\documentclass[11pt]{article}

\usepackage{mathrsfs}
\usepackage[T1]{fontenc}
\usepackage{mathpazo}
\usepackage{setspace}
\usepackage{amsfonts}
\usepackage{amssymb}
\usepackage{amsmath}
\usepackage{epsfig}
\usepackage{latexsym}
\usepackage{color}
\usepackage{graphicx}
\usepackage{nicefrac}
\usepackage[latin1]{inputenc}
\usepackage{slashed}
\usepackage{multirow}
\usepackage{framed}

\def\hybrid{\topmargin -30pt    \oddsidemargin 0pt 
        \headheight 0pt \headsep 0pt
        \textwidth 6.25in       
        \textheight 9.5in       
        \marginparwidth .875in
        \parskip 5pt plus 1pt   \jot = 1.5ex}

\hybrid

\def\baselinestretch{1.2}

\catcode`\@=11

\def\marginnote#1{}
\newcount\hour
\newcount\minute
\newtoks\amorpm
\hour=\time\divide\hour by60
\minute=\time{\multiply\hour by60 \global\advance\minute by-\hour}
\edef\standardtime{{\ifnum\hour<12 \global\amorpm={am}%
        \else\global\amorpm={pm}\advance\hour by-12 \fi
        \ifnum\hour=0 \hour=12 \fi
        \number\hour:\ifnum\minute<10 0\fi\number\minute\the\amorpm}}
\edef\militarytime{\number\hour:\ifnum\minute<10 0\fi\number\minute}

\def\draftlabel#1{{\@bsphack\if@filesw {\let\thepage\relax
   \xdef\@gtempa{\write\@auxout{\string
      \newlabel{#1}{{\@currentlabel}{\thepage}}}}}\@gtempa
   \if@nobreak \ifvmode\nobreak\fi\fi\fi\@esphack}
        \gdef\@eqnlabel{#1}}
\def\@eqnlabel{}
\def\@vacuum{}
\def\draftmarginnote#1{\marginpar{\raggedright\scriptsize\tt#1}}

\def\draft{\oddsidemargin -.5truein
        \def\@oddfoot{\sl preliminary draft \hfil
        \rm\thepage\hfil\sl\today\quad\militarytime}
        \let\@evenfoot\@oddfoot \overfullrule 3pt
        \let\label=\draftlabel
        \let\marginnote=\draftmarginnote
   \def\@eqnnum{(\theequation)\rlap{\kern\marginparsep\tt\@eqnlabel}%
\global\let\@eqnlabel\@vacuum}  }

\def\draft2{
        \def\@oddfoot{\sl preliminary draft \hfil
        \rm\thepage\hfil\sl\today\quad\militarytime}
        \let\@evenfoot\@oddfoot \overfullrule 3pt
        \let\label=\draftlabel
        \let\marginnote=\draftmarginnote
   \def\@eqnnum{(\theequation)\rlap{\kern\marginparsep\tt\@eqnlabel}%
\global\let\@eqnlabel\@vacuum}  }

\def\preprint{\twocolumn\sloppy\flushbottom\parindent 2em
        \leftmargini 2em\leftmarginv .5em\leftmarginvi .5em
        \oddsidemargin -.5in    \evensidemargin -.5in
        \columnsep .4in \footheight 0pt
        \textwidth 10.in        \topmargin  -.4in
        \headheight 12pt \topskip .4in
        \textheight 6.9in \footskip 0pt
        \def\@oddhead{\thepage\hfil\addtocounter{page}{1}\thepage}
        \let\@evenhead\@oddhead \def\@oddfoot{} \def\@evenfoot{} }

\def\numberbysection{\@addtoreset{equation}{section}
        \def\theequation{\thesection.\arabic{equation}}}

\def\underline#1{\relax\ifmmode\@@underline#1\else
        $\@@underline{\hbox{#1}}$\relax\fi}

\def\titlepage{\@restonecolfalse\if@twocolumn\@restonecoltrue\onecolumn
     \else \newpage \fi \thispagestyle{empty}\c@page\z@
        \def\thefootnote{\fnsymbol{footnote}} }

\def\endtitlepage{\if@restonecol\twocolumn \else \newpage \fi
        \def\thefootnote{\arabic{footnote}}
        \setcounter{footnote}{0}}  

\catcode`@=12
\relax

\def\pl#1#2#3{ {\sl Phys. Lett.\/} {\bf#1}, #2(#3)}

\def\figcap{\section*{Figure Captions\markboth
        {FIGURECAPTIONS}{FIGURECAPTIONS}}\list
        {Figure \arabic{enumi}:\hfill}{\settowidth\labelwidth{Figure
999:}
        \leftmargin\labelwidth
        \advance\leftmargin\labelsep\usecounter{enumi}}}
 \relax
\def\tablecap{\section*{Table Captions\markboth
        {TABLECAPTIONS}{TABLECAPTIONS}}\list
        {Table \arabic{enumi}:\hfill}{\settowidth\labelwidth{Table
999:}
        \leftmargin\labelwidth
        \advance\leftmargin\labelsep\usecounter{enumi}}}
 \relax
\def\reflist{\section*{References\markboth
        {REFLIST}{REFLIST}}\list
        {[\arabic{enumi}]\hfill}{\settowidth\labelwidth{[999]}
        \leftmargin\labelwidth
        \advance\leftmargin\labelsep\usecounter{enumi}}}
 \relax
\makeatletter
\newcounter{pubctr}
\def\publist{\@ifnextchar[{\@publist}{\@@publist}}
\def\@publist[#1]{\list
        {[\arabic{pubctr}]\hfill}{\settowidth\labelwidth{[999]}
        \leftmargin\labelwidth
        \advance\leftmargin\labelsep
        \@nmbrlisttrue\def\@listctr{pubctr}
        \setcounter{pubctr}{#1}\addtocounter{pubctr}{-1}}}
\def\@@publist{\list
        {[\arabic{pubctr}]\hfill}{\settowidth\labelwidth{[999]}
        \leftmargin\labelwidth
        \advance\leftmargin\labelsep
        \@nmbrlisttrue\def\@listctr{pubctr}}}
 \relax
\makeatother

\def\be{\begin{equation}}
\def\ee{\end{equation}}
\def\ba{\begin{eqnarray}}
\def\ea{\end{eqnarray}}

\def\del{\partial}
\def\pl{\partial}

\def\k{\kappa}
\def\r{\rho}
\def\a{\alpha}
\def\A{\Alpha}
\def\b{\beta}
\def\B{\Beta}
\def\g{\gamma}
\def\G{\Gamma}
\def\d{\delta}

\def\e{\epsilon}

\def\th{\theta}

\def\m{\mu}
\def\n{\nu}

\def\s{\sigma}

\def\cH{{\cal H}}
\def\cL{{\cal L}}

\def\cM{{\cal M}}
\def\cN{{\cal N}}
\def\cO{{\cal O}}
\def\cP{{\cal P}}

\def\no{\noindent}

\def\qq{\qquad}

\def\IR{\relax{\rm I\kern-.18em R}}

\def\alp{\alpha^\prime}
\def\nn{\nonumber}

\def\inv{^{\raise.0ex\hbox{${\scriptscriptstyle -}$}\kern-.05em 1}}

\def \A { {\bar A} }
\def \B {{\bar B}}

\def \ha {{\frac{1}{2}}}

\def \ov {\over}

\def\diag{{\rm diag}}

 \def\EDIT#1{#1}

\begin{document}

\renewcommand{\theequation}{\arabic{equation}}
\renewcommand{\theequation}{\thesection.\arabic{equation}}

\renewcommand{\theequation}{\thesection.\arabic{equation}}
\csname @addtoreset\endcsname{equation}{section}

\begin{titlepage}
\begin{center}


\phantom{xx}
\vskip 0.5in

{\large \bf Duality Symmetric String and M-Theory}

\vskip 0.45in

  {\bf David S. Berman}${}^{1}\,$\footnote{{\tt d.s.berman@qmul.ac.uk} }  and\ {\bf Daniel~C.~Thompson}${}^{2}\,$\footnote{{\tt dthompson@tena4.vub.ac.be}}

\vskip 0.2in

${}^1$ Queen Mary University of London, Centre for Research in String Theory,\\
School of Physics, Mile End Road, London, E1 4NS, England \
 \vskip .1in
 
${}^2$  Theoretische Natuurkunde, Vrije Universiteit Brussel, \\
and The International Solvay Institutes, \\
Pleinlaan 2, B-1050 Brussels, Belgium. \

 \vskip .2in
\end{center}

\vskip .4in

\centerline{\bf Abstract}

\no
We review recent developments in duality symmetric string theory.  We begin with the world-sheet doubled formalism  which describes  strings  in an extended spacetime with extra coordinates conjugate to winding modes.   This formalism is T-duality symmetric and can accommodate non-geometric T-fold  backgrounds which are beyond the scope of Riemannian geometry. Vanishing of the conformal anomaly of this theory can be interpreted as a set of spacetime equations for the background fields.  These equations follow from an action principle that has been dubbed Double Field Theory (DFT).  We review the  aspects of generalised geometry relevant for DFT.  We outline recent extensions of DFT and explain how, by relaxing the so-called strong constraint with a Scherk-Schwarz ansatz, one can obtain backgrounds that simultaneously depend on both the regular and T-dual coordinates. This provides a purely geometric higher dimensional origin to gauged supergravities that arise from non-geometric compactification.  We then turn to M-theory and describe recent progress in formulating an $E_{n(n)}$ U-duality covariant description of the dynamics.  We describe how spacetime may be extended to accommodate coordinates conjugate to brane wrapping modes and the construction of generalised metrics in this extended space that unite the bosonic fields of supergravity into a single object.  We review the action principles for these theories and their novel gauge symmetries.   We also describe how a  Scherk-Schwarz reduction can be applied in the  M-theory context and the resulting relationship to the embedding tensor formulation of maximal gauged supergravities. 

\end{titlepage}
\vfill
\eject

\def\baselinestretch{1.2}
\baselineskip 10 pt
\noindent

\tableofcontents

\def\baselinestretch{1.2}
\baselineskip 20 pt
\no

\newcommand{\eqn}[1]{(\ref{#1})}

 \def\beq{\be}
 \def\eeq{\ee}
 \def\bea{\ba}
 \def\eea{\ea}
 
 \section{Introduction}

One of the great successes of string theory is that its low energy effective description produces a gravitational theory. As such one is tempted to think that at low energies supergravity and string theory are one and the same. However, this is not quite right. String theory  possesses duality symmetries which equate seemingly different supergravity backgrounds. This duality, known as T-duality, has its origin in the extended nature of the string world-sheet and the possibility that closed strings have to wrap around nontrivial one cycles in spacetime.  The string spectrum contains  {\it{winding modes}}  which ``see'' spacetime in a different way to the momentum modes of the string. T-duality is a symmetry built on the exchange of the winding and momentum modes whilst changing the gravity background to render the physics invariant.  

  T-duality tells us that strings experience geometry in a rather different way to point particles and teases the question of whether there is a more appropriate geometrical language with which to understand string theory.  More precisely, we may consider keeping both the winding and momentum modes of the closed string.  It is  then sensible to seek a unified description of spacetime incorporating both the winding mode and the momentum mode perspectives on an equal footing.  In such a formulation T-duality  would be a manifest symmetry.  A number of related formulations \cite{Duff:1989tf,Tseytlin:1990nb,Tseytlin:1990va,Maharana:1992my,Cremmer:1997ct,Hull:2004in,Hull:2006va} of this   duality symmetric string  have been constructed as a string $\sigma$-model in a spacetime with double the number of dimensions.      The extended nature of this target space allows for a symmetric inclusion of both a geometry and its T-dual.    
 
What then is the geometry of this doubled target space?  The description of the background equations of motion for the doubled target space goes by the name of {\it{Double Field Theory}} (DFT). It realises T-duality as a manifest symmetry and incorporates the momentum modes and windings modes on the same footing.  Recently, using string field theory techniques, Hull and Zwiebach proposed such a spacetime duality symmetric theory   \cite{Hull:2009mi} which is related to earlier pioneering works of Siegel \cite{Siegel:1993th,Siegel:1993xq} and Tseytlin \cite{Tseytlin:1990va}.   This theory has a tangent space which is \EDIT{closely related} to the generalised geometry developed by Hitchin \cite{Hitchin:2004ut}.   The dynamical field content of this theory is encoded in a {\em{generalised metric}} on the doubled space.  This generalised metric has the advantage that it elegantly combines both the regular spacetime metric and the NS-NS two-form potential into a single object.   The Double Field Theory action is then written in terms of this generalised metric and a single (shifted) dilaton.  Thus not only does this approach promote T-duality to the level of a manifest symmetry, it can be said to geometrise the NS-NS two-form by combining it with a metric into a single quantity.   This is much more in the spirit of world-sheet string theory where both these fields come from the level 2 mode of the closed string as opposed to the traditional supergravity description where the two-form field is viewed as living in a  spacetime background given by the metric.

At this point let us examine the energy scales of the different modes. To this end we introduce the string tension:
\beq
T= \frac{1}{\EDIT{2\pi} {\alpha}'}
\eeq
and consider a spacetime which has a circle of length, $R$. 
The energy of the string winding around the circle will be simply the length times tension:
\beq
E_{winding} {\EDIT \approx} \frac{R}{ {\alpha}'}  \, .
\eeq
On the other hand, because we consider a compact space, the energies associated to momentum modes are  quantised in units of inverse length and are independent of the string tension i.e.
\beq
E_{momentum} {\EDIT \approx} \frac{1}{R}  \, .
\eeq 
It is convenient for the purposes of considering limits to work with dimensionless quantities and so we look at the ratio of the winding mode energy to the momentum mode energy 
\beq
E_{winding}/ E_{momentum} {\EDIT \approx} \frac{R^2}{ \alpha'} \, .
\eeq
For $R^2\ll \alpha'$ the   winding modes are light and should dominate but for the converse, $R^2 \gg \alpha'$ the momentum mode are lighter and dominate the low energy physics.  The energy of the oscillator modes of the string are proportional to the tension, $T$ but independent of $R$. So let us see which modes are light in various regimes. For $R\gg \sqrt{\alpha'}$, the momentum modes are lighter than  both winding and oscillator modes and supergravity is a consistent low energy effective theory. For $R\ll \sqrt{\alpha'}$ the winding modes are lighter than the other modes and these modes again give supergravity description albeit in  T-dual  variables.   At around the string scale where $R\approx \sqrt{\alpha'}$   we must keep all modes (oscillators, winding and momentum modes) and at other scales there is a hierarchy  with a low energy effective action (a supergravity action) for the lightest modes.  Thus we see that there is not a regime where one should keep both winding and momentum modes but throw away the oscillators of the string.

So Double Field Theory is not a low energy effective action. It is \EDIT{perhaps} best thought of as a truncation of the string spectrum whose  {\em post-hoc} justification will be that  it does not exhibit any pathologies. Formulating the string $\sigma$-model in this doubled background and demonstrating all the usual quantum tests such as modular invariance and the vanishing of the $\beta$-function are passed is   a crucial consistency check that we must make.  The theory can also be made supersymmetric (with the maximal number of supercharges) which is also an indicator that the theory is more consistent than it has a right to be.  \EDIT{That DFT has passed these checks supports the view that it is a consistent theory  even though it is not a true low energy effective description.}

In DFT T-duality is promoted from a hidden symmetry to a global symmetry of the action. When the background possesses $d$ commuting isometries the T-duality group is described by $O(d,d)$ and is realised linearly in the doubled spacetime. There are several local symmetries of DFT.  There is a local $O(d) \times O(d)$ symmetry which becomes the equivalent of the Lorentz group for the doubled space\footnote{\EDIT{In DFT one may formally double all directions, including the time direction, in which case the local symmetry group will be $O(d-1,1)\times O(1,d-1)$ reflecting the Lorentzian signature.}}.
In addition, there are the diffeomorphisms from general coordinate transformations and the gauge transformations of the NS-NS two-form which are described locally by a one-form. These local symmetries do not commute and together they form a Courant algebra.  In DFT the diffeomorphisms and gauge transformations are combined into a single geometric local symmetry whose infinitesimal action is generated by a generalised Lie derivative.  

Throughout there will be a tightly constraining interplay between manifesting all these symmetries: the global $O(d,d)$ symmetry of T-duality; the local tangent space group $O(d) \times O(d)$; and the local symmetry combining diffeomorphisms and p-form gauge transformations. 

A key property of string theory is that T-duality is a perturbative symmetry  present and visible from the perturbative world-sheet description. However, string theory also possesses other nonperturbative symmetries which are not accessible from string perturbation theory. In particular, the type IIB string has S-duality which changes the NS-NS sector into the RR sector and inverts the string coupling. This symmetry is very much like the S-duality of $N=4$ Yang-Mills theory in that it exchanges perturbative and nonperturbative states and relates strong to weak coupling. S-duality does not commute with T-duality so composing them provides   a new symmetry group for the string.
 
This combination of T and S-duality goes by the name of U-duality. It was the presence of U-duality that gave rise to the idea of M-theory where different perturbative string theories were realised as different backgrounds of  a single underlying theory \cite{Hull:1994ys,Witten:1995ex}. One may say that U-duality is the T-duality of M-theory. It is the symmetry that relates naively different M-theory backgrounds.

The winding/momentum mode analysis of T-duality then becomes much more complicated as one now has windings of branes of differing world-volume dimensions. All the possible brane windings become important and the resulting symmetry group becomes highly dimensionally dependent.

The U-duality group $G$, associated with reduction on a $d$ dimensional torus is given by the exceptional group $E_d$ (with the cases for $d<6$ given by the obvious identifications of the corresponding Dynkin diagrams). There is a local {\it{Lorentz}} group $H$, just as before which is the maximal compact subgroup of $E_d$, and also the local symmetries combining diffeomorphisms and p-form gauge transformations. The appearance of these symmetries was noted many years before M-theory  when  \cite{Julia} observed that eleven-dimensional supergravity dimensionally reduced on a torus has scalar fields that inhabit the $G/H$ coset.   This is essentially a consequence of eleven-dimensional supergravity being the low energy effective description of M-theory.\footnote{\EDIT{At this stage we would like to warn the reader of a potentially confusing usage of language that we make in this report.  Strictly speaking the term ``U-duality''  should refer to the discrete M-theory duality group denoted $E_{n(n)}(\mathbb{Z})$ formed by intertwining $S$ and $T$ dualities that was conjectured  by Hull and Townsend \cite{Hull:1994ys} to remain an exact quantum symmetry of M-theory.  This is a subgroup of the continuous group $E_{n(n)}(\mathbb{R})$ which are the ``hidden'' symmetries displayed by eleven-dimension supergravity reduced on a torus.  In this report, in chapter 5 especially, we will be abusive in our use of the term  U-duality;  we shall frequently  use it to mean the continuous supergravity symmetry group.  Since we are working entirely in the supergravity limit in chapter 5 the intended meaning will be clear from the context. }}

Again, as with the doubled formalism of the string, one would wish for a formalism where all these symmetries are manifest.  Now we see that in order for the global symmetry to be manifest   requires much more than just doubling the space. To realise the $E_d$ linearly we need a space with the same dimension as the relevant representation of $E_d$ and this is always bigger than $2d$ for $d>3$.  The M-theory version of Double Field Theory calls for an exceptional generalised geometry which extends the space to linearly realise the exceptional algebra \cite{Hull:2007zu,Pacheco:2008ps}. One can, as in Double Field Theory, then form an action entirely from a (generalised) metric on this extended space \cite{Berman:2010is,Berman:2011pe,Berman:2011jh}. The realisation of the local gauge symmetries   becomes more involved in the M-theory version since they must contain not just the two-form gauge symmetries of string theory but simultaneously the three-form and six-form gauge symmetries of eleven-dimensional supergravity. Again, there is a generalised Lie derivative that generates these local transformations as a single entity.  The closure of the gauge algebra will place constraints on the theory. It is remarkable that the structure of Double Field Theory could be repeated for the U-duality groups of M-theory.

Given that the theory is now defined in an extended spacetime one finds the need to restrict the dynamics to end up with the correct physical degrees of freedom and a consistent theory.  This is achieved with a  {\it{physical section}} condition specifying locally a subspace of the extended space which is to be viewed, in a given duality frame, as physical.  For DFT this amounts to picking a maximal isotropic  subspace \EDIT{(a $d$-dimensional subspace that is null with respect to the invariant inner product of the T-duality group $O(d,d)$)}.    In the M-theory scenario the {\it{physical section}} condition must become more complicated in that its solutions should project onto a comparatively smaller subspace.   

 One way \EDIT{to} satisfy the section condition is to enforce from the very outset that fields and gauge parameters only ever depend on the regular spacetime coordinates (that is they have no dependence on the extra coordinates of the extended space conjugate to winding charges). \EDIT{At this point in the discussion we must take care to specify the difference in DFT between the so called weak and strong constraints. What is being described here is known in DFT as the strong constraint. It is essentially,
\be
\eta^{MN} \partial_M \phi_1 \partial_N \phi_2=0  \, ,
\ee
for any fields $\phi_1, \phi_2$ \EDIT{where $\eta$ is the invariant inner product of T-duality group}. This is solved this by demanding fields be independent of either a coordinate or the conjugate dual coordinate. The so called weak constraint,
\be
\eta^{MN} \partial_M \partial_N \phi =0 \, ,
\ee
comes directly in string field theory from the level matching condition and one imagines DFT can be formulated with only the weak constraint being required. These issues are discussed in more detail later in the review.}

  Demanding that fields are independent of dual coordinates  means that one arrives in a formulation  \cite{Coimbra:2011ky,Coimbra:2011nw,Coimbra:2012af} where tangent space is extended but the underlying space itself is not.  One ends up with a theory whose structure is precisely the generalised geometry of Hitchin for the case of strings and its exceptional counterpart for M-theory.  Obeying  the section condition globally in this fashion    ultimately means that \EDIT{there} is no new physical content in the theory beyond supergravity although it does provide an extremely elegant and potentially powerful reformulation making the duality symmetry manifest. 

One of the key developments has been to determine the ways in which one could consider relaxing the physical section condition. Of course, having a completely unconstrained theory seems incorrect. \EDIT{However some relaxations of the section conditions do seem possible.} We will see how   a generalised version of the Scherk-Schwarz ansatz   allows explicit dependence on all coordinates of the extended spacetime. \EDIT{The} extra dimensions of the extended spacetime augment M-theory in that previously isolated gauged supergravities now appear from geometric compactifications of a single theory.

In all of this, global questions will be crucial with so called {\it{nongeometric}} backgrounds appearing from allowing holonomies in the global group $G$. There is then a subtle interplay between the patching together of p-form potentials with gauge transformations on overlapping patches and the holonomy of the space under $G$ or $H$. This type of holonomy allows the construction of so called exotic branes and T-folds. Thus, string and M-theory have backgrounds that are described by more than just supergravity because their symmetries allow novel solutions relying on the additional duality structures in the theory.

\subsection{Outline} 

The outline of the main body of this review is as follows.  In section \ref{sec:Background} we begin by introducing T-duality. We first consider it in the simplest context of the $S^1$ radial inversion duality and then introduce the  extension  to toriodal backgrounds and the corresponding   $O(d,d, \mathbb{Z})$ T-duality group.  Also in section \ref{sec:Background} we introduce the concepts of non-geometric backgrounds, T-folds and exotic branes.   

In section \ref{sec:worldsheetdoubled} we introduce the string world-sheet doubled formalism.     We explain the basic features of this T-duality symmetric approach and discuss some of its quantum mechanical properties.  In particular we discuss the conditions for conformal invariance of the doubled string; these give rise to a set of  background field equations that the doubled spacetime geometry should obey.  Understanding these equations from a spacetime perspective becomes the focus of the second half of this report. 

In section \ref{sec:DFT} we turn our attention to the spacetime T-duality symmetric theory that has been dubbed Double  Field Theory (DFT). We first show how the symmetries of the NS sector of supergravity naturally motivate an algebra built around a Courant bracket rather than a Lie bracket.  This hints at the rather central role that generalised geometry plays in DFT.  We provide a review of the essential details of generalised geometry required to understand DFT and its M-theory analogue.   After this somewhat mathematical interlude we introduce DFT itself.   We spend some time discussing the gauge symmetries of DFT and the emerging new geometrical concepts that it invokes.  We close section \ref{sec:DFT} with a presentation of some of the extensions of DFT including the inclusion of RR fields, fermions, heterotic constructions and its relationship with gauged supergravities.   

Section \ref{sec:Mtheory} concerns the M-theory extension of these ideas.  We begin by providing an overview of M-theory, its low energy limit of eleven-dimensional supergravity and its hidden symmetries.  We proceed by reviewing the construction of  exceptional generalised geometries, that is the extended tangent bundle that supports a natural action of the U-duality groups.  We discuss the construction of the U-duality symmetric M-theory extensions of DFT and illustrate this with the specific example of the $SL(5)$ U-duality group.  We discuss the section condition required to reduce the theory to a physical subspace and the gauge structure   paying careful attention to the ghost structure of gauge transformations.  We also outline how group theoretic techniques allow for the constructions of generalised metrics that encode the field content of  these theories. 

  We close section \ref{sec:Mtheory} by considering the dimensional reduction of the U-duality invariant theory.  First we show how it can be reduced on an appropriate torus in the extended spacetime to give rise to DFT.  Then we go on to detail the linkage between the Scherk-Schwarz reduction of the U-duality invariant theory and maximal gauged supergravities in lower dimensions and how this allows for a consistent relaxation of the section condition. 
  
We conclude the report in section \ref{sec:Concs} where we give our own, subjective,  outlook of this subject.   \EDIT{In the appendix we provide the reader with a short tool-kit to understand some of the features of chiral bosons} in two dimensions which underpin much of the construction of the world-sheet doubled theory of  section \ref{sec:worldsheetdoubled}.
 
 This report comes complete with a lengthy bibliography however now is
an opportune moment to extend an inevitable apology to the authors of
works that through oversight or ignorance we have unintentionally
failed to refer.

 \section{Dualities and Symmetries}\label{sec:Background}

\subsection{Path Integral (Buscher) Approach to T-duality} 

The history of T-duality is long and can be traced back to observations made in \cite{Kikkawa:1984cp,Sakai:1985cs} and a review of early   developments and perspectives can be found in \cite{Giveon:1994fu}.  Here we present the path integral perspective of T-duality which originates from the work of Buscher  \cite{Buscher:1987qj,Buscher:1987sk}.

We consider bosonic string theory whose target space admits a $U(1)$ isometry generated by a Killing vector $K$ acting by the Lie derivative such that
\be\label{eq:isometry}
L_K G = L_K H  = L_K \Phi = 0 \ ,
\ee 
 where $G$ is the metric of the target spacetime, $H=dB$ is the field strength of the NS two-form and $
\Phi$ the dilaton.  We work with coordinates adapted to this isometry, \EDIT{$\{X^I\} = \{\theta, x^i \}   $,} such that the Killing vector becomes $K = \partial_\theta$.   \EDIT{In the interests of pedagogy, we shall begin with a stronger assumption than eq.~\eqref{eq:isometry}; we impose that  $L_K B = 0$ (this is not needed for the $\sigma$-model to be invariant and in general, such a choice of potential may not be globally possible -- we shall revisit this shortly).}  We choose to work with dimensionless radii and coordinates such that $\theta \sim  \theta + 2 \pi$ and $G_{\theta \theta}= R^2$.  In conformal gauge the string $\sigma$-model is\footnote{\EDIT{We work with lightcone coordinates $\sigma^\pm = \frac{1}{2} \left( \tau \pm \sigma \right)$  such that $\partial_\pm =   \partial_\tau \pm \partial_\sigma $.  Here we work in Lorentzian signature but when we discuss string perturbation theory effects a Euclidean worldsheet should be understood. }}  
\bea
\label{ungauged}
S_0 
  &=&\frac{1}{4\pi} \int d^2 \sigma \, (G + B)_{ij}\partial_{+}X^i \partial_{-}X^j \nn \\
 &=& \frac{1}{4\pi} \int d^2 \sigma \,  R^2 \partial_+ \theta \partial_- \theta + E_{i\theta}\partial_+x^i \partial_- \theta + E_{\theta i}\partial_+ \theta \partial_- x^i  +     E_{ij}\partial_{+}x^i \partial_{-}x^j \, , \quad
\eea  
where we have defined $E = G + B$.   The isometric condition implies that the background fields do not depend on $\theta$ but may have some dependence on the $x^i$ spectator fields. \EDIT{In particular, let us emphasise that the radius $R$ need not be constant but can be a function of the spectator coordinates i.e. $R= R(x)$.}

A powerful approach to T-duality is the Buscher procedure \cite{Buscher:1987qj,Buscher:1987sk}.  This is a three step recipe for obtaining dual $\sigma$-models:
\begin{enumerate}
\item Gauge the isometry.
\item Invoke a flat connection for this gauge field with a Lagrange multiplier.
\item Integrate by parts and solve for the non-propagating gauge field.
\end{enumerate}
This approach may be readily generalised, for instance one can consider applying it to non-Abelian isometry groups \EDIT{\cite{delaossa:1992vc}} or even to isometries in fermionic directions \cite{Berkovits:2008ic}.  Both of these generalisation have been studied in recent years and have led to some surprising insights especially within the context of the AdS/CFT correspondence;  non-Abelian T-dualities have been used to generate interesting new supergravity backgrounds  \cite{Lozano:2011kb,Sfetsos:2010uq,Itsios:2012zv,Itsios:2012dc,Lozano:2012au,Jeong:2013jfc,Itsios:2013wd} and self-duality of the $AdS_5\times S^5$ background under a combination of bosonic and fermionic T-dualities \cite{Alday:2007hr,Berkovits:2008ic} has been used to give an explanation of the Wilson loop/scattering amplitude duality and associated dual superconformal symmetry at strong coupling in ${\cal N}=4$  SYM.  Reviews of recent developments in fermionic T-duality can be found in \cite{OColgain:2012si,Bakhmatov:2011ab}. One caveat with these generalisations is that they are not expected to hold   to all orders in string genus perturbation theory  for reasons that we shall return to shortly.  Nonetheless they are interesting spin-offs of this basic approach and can certainly be employed as solution generating symmetries of supergravity.   

We now apply this recipe to the case at hand.  We gauge the global $U(1)$ symmetry which acts as $\delta \theta = \omega$ by introducing a $U(1)$ valued gauge field $A = A_+ d\sigma^+ + A_-d\sigma^-$ which has a gauge transformation rule $\delta A = -d \omega$ and invariant field strength $F= dA$.  The covariant derivative is
\be
D \theta = d \theta + A  \, .
\ee 
Additionally we constrain the gauge connection to be flat using a Lagrange multiplier. The gauged $\sigma$-model is given by
\be 
\label{gauged}
S_{\mbox{\footnotesize gauged}}=\frac{1}{4\pi} \int d^2 \sigma \,  R^2 D_+ \theta D_- \theta + E_{i\theta}\partial_+x^i D_- \theta + E_{\theta i}D_+ \theta \partial_- x^i  +     E_{ij}\partial_{+}x^i \partial_{-}x^j + \lambda F_{+-}  \, . 
\ee 
Integrating out the Lagrange multipliers invokes the constraint $F_{+-} = 0$ which can be solved locally by  $  A_\pm = \partial_\pm \phi $. 
  Substituting this pure gauge connection back into eq. (\ref{gauged}) one finds an action equivalent to eq. (\ref{ungauged}) after a trivial field redefinition.  There are, of course, some global concerns that we must treat carefully when considering string perturbation theory defined on \EDIT{Euclidean} world-sheets of arbitrary genus -- we will return to this point shortly.  

We continue by integrating-by-parts the Lagrange multiplier term to obtain
 \be
S_{\mbox{\footnotesize gauged}}=\frac{1}{4\pi} \int d^2 \sigma \, R^2 D_+ \theta D_- \theta + E_{i\theta}\partial_+x^i D_- \theta + E_{\theta i}D_+ \theta \partial_- x^i       +  \partial_- \lambda A_+ - \partial_+ \lambda A_-   +     E_{ij}\partial_{+}x^i \partial_{-}x^j\,.
\ee 
In this form the gauge fields are auxiliary; they have an algebraic equation of motion  and can be eliminated. This procedure is made simplest by fixing the gauge in which $\theta = 0$  where the relevant equations of motion become
\be
\label{Aeqm}
  0 = R^2 A_+ +  E_{ i\theta} \partial_\EDIT{+} x^i - \partial_+ \lambda \ , \quad 0 =  R^2 A_- + E_{\theta i} \partial_- x^i + \partial_- \lambda \, .
\ee 
Replacing $A_\pm$ by these expressions results in an action given by
\be
\label{dual}
S_{\mbox{\footnotesize dual}}= \frac{1}{4 \pi} \int d^2 \sigma \,  \frac{1}{R^2}  \partial_+ \lambda \partial_- \lambda - \frac{1}{R^2} E_{i\theta}\partial_+x^i \partial_- \lambda +\frac{1}{R^2} E_{\theta i}\partial_+ \lambda \partial_- x^i  +     \left( E_{ij} - \frac{E_{i\theta}E_{\theta j}}{R^2} \right)\partial_{+}x^i \partial_{-}x^j \, .
\ee
This dual action is of the same form as the initial $\sigma$-model eq. (\ref{ungauged}) but with the following redefinitions
\be 
\label{buscher}
G_{\theta \theta }  \rightarrow  \frac{1}{G_{\theta \theta} }  , \quad 
E_{\theta i}   \rightarrow  \frac{1}{G_{\theta \theta}} E_{\theta i}  \, , \quad
E_{i \theta}   \rightarrow   - E_{i \theta}  \frac{1}{G_{\theta \theta}}   \, ,\quad 
E_{ij}  \rightarrow  E_{ij} - E_{i\theta}  \frac{1}{G_{\theta \theta}}  E_{\theta j}\, .
\ee 
When the procedure is done in a path integral  with the inclusion of a Fradkin-Tseytlin dilaton coupling  \cite{Fradkin:1984pq,Fradkin:1985ys} one obtains, in addition to the above transformation rules, a shift in the dilaton given by \cite{Buscher:1987qj,Buscher:1987sk}
\be\label{eq:dilshift}
\Phi \rightarrow  \Phi - \frac{1}{2} \ln G_{\th \th}  \ . 
\ee

This derivation is, in essence, due to Buscher  \cite{Buscher:1987qj,Buscher:1987sk} and the above transformations and their generalisations are known as the Buscher rules.  Since the actions $S_0 $ and $S_{\mbox{\footnotesize dual}}$ are both equivalent  to a common action $S_{\mbox{\footnotesize gauged}}$ we say that these are dual descriptions of the same physics.  It is from this perspective that we can consider this a non-perturbative duality on the world-sheet by regarding $R$ as the coupling constant.   In fact, this derivation is somewhat similar to that used in showing S-duality in the Lagrangian of $\cN=2$ gauge theories in four dimensions wherein a Lagrange multiplier is used to enforce the Bianchi identity for the field strength; if instead of integrating out the Lagrange multiplier one integrates out the gauge field itself one finds the S-dual action \cite{Seiberg:1994aj, Seiberg:1994rs}.   Another rewarding pay-off from this Buscher procedure is \EDIT{to} provide a constructive derivation of \EDIT{bosonisation} as duality (both Abelian and non-Abelian) \cite{Burgess:1993np,Burgess:1994np}.

\EDIT{As is well known, and we shall come back to this essential point later, the string theory defined by the $\sigma$-model in eq.~\eqref{ungauged} is conformal when, to first loop order in $\alpha'$, the background fields obey the equations of motion of the common NS sector of type II supergravity.  When the background fields entering eq.~\eqref{ungauged} do obey these conditions, then it can be explicitly shown that so too will the T-dual geometry defined by the Buscher rules eqs.~\eqref{buscher} and \eqref{eq:dilshift}.  It is believed, and arguments have been proposed dating back to Sen \cite{Sen:1991zi} and Meissner and Veneziano \cite{Meissner:1991ge,Meissner:1991zj} that T-duality is present to all orders in $\alpha'$ and the T-duality map can be expanded as a perturbative series in  the inverse string tension.  It is of course an extremely difficult task to explicitly calculate the $\alpha'$ corrections to the T-duality map but this has been done at two-loops in \cite{Kaloper:1997ux}.  Whilst this represents conventional wisdom, a recent proposal \cite{Hohm:2013jaa} has been made in which, in the context of Double Field Theory, T-duality remains uncorrected in $\alpha'$ however the price to pay is that the gauge structure of the theory receives corrections. }

A further useful observation on T-duality comes from looking at how world-sheet derivatives \EDIT{transform}. Making use of eq. (\ref{Aeqm})  \EDIT{together with the pure gauge condition on the gauge field} one finds that 
\be
\label{wsderivsa}
    \partial_+ \theta  = R^{-2}( \partial_+ \lambda  - E_{ i\theta} \partial_\EDIT{+} x^i   )  \ , \quad  \partial_- \theta  = -  R^{-2}(\partial_- \lambda+ E_{\theta i} \partial_- x^i   ) \, .
\ee 
The implication of this is rather profound; under T-duality left and right movers have different transformation properties.  This can be seen very clearly when the spectator fields are ignored by setting   $E_{ i\theta} =    E_{\theta i }=0$ in  \EDIT{eq. (\ref{wsderivsa})} and for simplicity setting $E_{ij}=\delta_{ij}$.  In that case the T-duality simply acts by
\be
\label{wsderivsb}
    \partial_+ \theta  = R^{-2} \partial_+ \lambda     \ , \quad  \partial_- \theta  = -  R^{-2} \partial_- \lambda  \, ,
\ee 
i.e. as a reflection on the right movers.   A nice property is that the relations in  \EDIT{eq. (\ref{wsderivsa})}  define a canonical transformation in phase space between the original and dual $\sigma$-models \cite{Alvarez:1994wj}. 

The knowledge of how world-sheet derivatives \EDIT{transform} in eq. \EDIT{(\ref{wsderivsa})} also allows one to \EDIT{determine the action of T-duality} on the RR sector.  Let us again take the case where we ignore spectator fields.  In that case, after T-dualisation the right movers define the pull-back   of a set of frame fields for the target space geometry, 
\be
\hat{e}_{(+)}^\lambda = R^{-1} d \lambda   \ , \qquad \hat{e}_{(+)}^i = d x^i ,
\ee
and left movers define a different set of pulled back frame fields 
\be
\hat{e}_{(-)}^\lambda = -R^{-1} d \lambda   \ , \qquad \hat{e}_{(-)}^i = dx^i . 
\ee
The left and right moving frames define the same geometry and are thus related by a Lorentz rotation 
\be
\hat{e}_{(+)} = \Lambda \hat{e}_{(-)}  \ ,
\ee
 which for the simple case at hand is given by
 \be
 \Lambda =  \diag(-1, 1, \dots 1) \ . 
 \ee
 \EDIT{This is evidently not a proper Lorentz rotation; it has $\det \Lambda = -1$. } This induces an action on spinors under T-duality given by a matrix $\Omega$ obtained through the invariance properties of the gamma matrices,
 \be
 \Omega^{-1} \Gamma^i  \Omega = \Lambda{}^i{}_j  \Gamma^j \ .
 \ee
 For the case at hand $\Omega = \Gamma^{11} \Gamma^\lambda$ -- notice the presence of a chirality changing $\Gamma^{11}$, this is because a single T-duality takes one from type IIA to type IIB string theory and vice-versa.

 The spinorial counterpart of this Lorentz rotation generates transformations of RR flux fields in the following way. {\EDIT{In the democratic formalism the RR field strengths are combined into a sum of p-forms $F=\sum_{n=0}^5 F_{2n}$ for (massive) IIA and $F= \sum_{n=0}^4 F_{2n+1}$ in IIB. This sum of forms is known is the literature as a polyform. It is a somewhat formal object since one is {\it{adding}}
 objects with different tensor properties. The polyforms $F$ obey the Bianchi identity $(d -H\wedge) F = 0$ and can be expressed in terms of potentials as $F=(d -H\wedge)C+F_0 e^B$ where the last term should be included only for massive IIA. }}   \EDIT{The} RR fields are combined with their Hodge duals to form a bispinor defined as
\be
{\rm IIB}:\ P  = {e^{\Phi}\ov 2} \sum_{n=0}^4 \slashed{F}_{2n+1}\ ,
\qq
{\rm IIA}:\ \hat{P} ={e^{\hat \Phi}\ov 2} \sum_{n=0}^5 \slashed{\hat F}_{2n}\ , 
\ee
where $\displaystyle \slashed{F}_{p}={1\ov p!}\G^{\m_1\dots \mu_p}F_{\m_1\m_2\dots \mu_p}$.
The higher $p$-forms are related to the lower ones by 
\be
F_p = (-1)^{int\left[ \frac{p}{2}\right]} \star F_{10-p}\  .  
\label{demoll1}
\ee 
The T-dual RR fluxes are simply obtained by 
multiplication with $\Omega^{-1}$. If, for instance, the \EDIT{duality maps} 
from type-IIB to   type-IIA  the
transformation rules for the RR-fluxes are given 
by comparing the two sides of  
\be
\hat{P} =  P \cdot \Omega^{-1} \ .
\label{ppom}
\ee
For the case of a single T-duality in the simplified scenario this corresponds to the intuitive rule of thumb that the effect of dualisation is to delete legs of flux in the direction of the isometry and add them where they were not present before i.e.  we have 
\be
F^{p} = {\cal{F}}^{p-1}\wedge e^\theta +{ \cal{F}}^p \longrightarrow F^{p-1} =  {\cal{F}}^{p-1} \ , \quad F^{p+1} =  {\cal{F}}^p \wedge e^\lambda .
\ee 

The transformations rules for RR fields were derived in   \cite{Bergshoeff:1995as} by comparing the dimensionally reduced supergravities but the above explanation   originates from the   work of Hassan \cite{Hassan:1999mm}  (see also \cite{Cvetic:1999zs} for a world-sheet derivation in the Green-Schwarz approach and \cite{Benichou:2008it} for one in the pure spinor approach).  \EDIT{An alternative approach to the transformation of RR fields as a Fourier-Mukai transformation was proposed by Hori in \cite{Hori:1999me}.}

\subsubsection{\EDIT{Some Comments on Global Issues}}
\EDIT{There are a number of subtle global issues that we now give some further details on.}

\EDIT{ In the preceding discussion we made a rather strong assumption that the NS two-form was invariant under the isometry used to T-dualise.  This is not necessary, and in general it may be not possible to define such a potential globally. We begin with a global and covariant treatment of T-duality for the case of a single isometry which was first given  by \cite{Alvarez:1993qi}.}
 
 \EDIT{ As a potential for $H$, the NS two-form $B$ need only be locally defined, that is to say $H$ must be closed but not necessarily exact. We choose a suitable open covering $\{ {\cal U}^\a\}$ of the target space where each patch ${\cal U}^\a$ is isomorphic to $R^n$ and each two fold intersection of patches, ${\cal U}^\alpha \cap {\cal U}^\beta$ is also isomorphic to $R^n$. In each patch $ {\cal U}^\a$ one has a choice of potential 
\begin{equation}
H = d B^{\alpha} \ ,
\end{equation}   
which differ only by a gauge transformation on overlaps i.e. $B^\alpha - B^\beta = d \Lambda^{\a \b}$ on ${\cal U}^\alpha \cap {\cal U}^\beta$ and obey the cocycle condition on triple overlaps.  The contribution to the string $\sigma$-model  of $H$ can be given by a Wess-Zumino term  
\begin{equation}
S_{WZ} = \frac{1}{8\pi} \int_{V} H \ ,
\end{equation}
in which $H$ should be understood to have been pulled-back to $V$, a three-dimensional manifold whose boundary is the world-sheet $\Sigma$ of the string.\footnote{\EDIT{In this discussion we are making heavy use of form notation, pull-backs are   implicit and we use frequently the Cartan identity $L_K = d \iota_K + \iota_K d $.}}  Although   $S_{WZ}$ depends on a choice of $V$, the difference between actions defined with different choices of $V$ is a topological number giving rise to only a phase in the (Euclidean) path integral.  This ambiguity will not effect the quantum theory providing this phase is unity -- this is the case when $H$ is a representative of an appropriately quantised cohomology class.}

\EDIT{
It is clear that under the action of the Killing vector $K$ the WZ term varies as 
\begin{equation}
\delta S_{WZ} = \frac{1}{8\pi} \int_{V} d \iota_K H = \frac{1}{8\pi} \int_{\Sigma}   \iota_K H \ . 
\end{equation}
Hence invariance then requires only that  
\begin{equation}
  \iota_K H = - d v \ ,
\end{equation} 
for a globally defined one-from $v$.  Thus it is sufficient that the potential is only invariant up to a gauge transformation; on a patch ${\cal U}^\alpha$ one requires
\begin{equation}
L_K B^\alpha = d \omega^\alpha \ , \quad \omega^\alpha = \iota_K B^\alpha - v \ . 
\end{equation}
The dualisation procedure in this more general context was performed in \cite{Alvarez:1993qi} and gives rise to a modification of the Buscher rules eq.~\eqref{buscher}
\begin{equation}
\tilde{G}_{\theta \theta }  =  \frac{1}{K^2 }  \, , \quad 
\tilde{G}_{\theta i}   =   \frac{v_i  }{K^2}      \, , \quad
\tilde{B}_{\theta i}   =   \frac{K_i  }{K^2}  \, ,  \quad
 \tilde{G}_{ij}= G_{ij} - \frac{K_i K_j - v_i v_j }{K^2}    \, ,\quad 
\tilde{B}_{ij}  =  B_{ij} - \frac{K_i v_j - v_i K_j }{K^2}  \ .
\end{equation}
Here the gauge symmetry of the $\sigma$-model has been fixed by using local adapted coordinates.  When $\omega =0$, $v_i = B_{\theta i}$ and using that $K_i = G_{\theta i }$  and  $K^2 = G_{\theta \theta}$ one recovers the earlier results eq.~\eqref{buscher}. 
}

\EDIT{
In \cite{Alvarez:1993qi} it was also suggested, again for the case of a single isometry, that one can perform the T-duality locally on patches and then glue together to reconstruct globally the T-dual background.  This may be unavoidable when adapted coordinates do not globally exist.  An explicit example is given by performing a T-duality on  $S^3$.  Locally $S^3$ looks like $S^2 \times S^1$ but this is of course not true globally.   For instance, on the round $S^3$, Euler angles provide a local adapted coordinate system in which a $U(1)$ isometry is realised by  shifts of one of the angles $\chi \rightarrow \chi + \epsilon$.   However since Euler angles are not global coordinates for $S^3$ one can not determine the full global properties of the T-dual geometry.  Instead, by considering the T-duality patchwise in the northern and southern hemispheres and carefully patching together the result, one can rebuild a global picture of the T-dual and conclude it has topology $S^2 \times S^1$.  This is a nice example of topology change from T-duality.  In   \cite{Alvarez:1993qi}  it was further argued that this patchwise approach can be used when $H$ is not exact. } 

\EDIT{
When there are several commuting isometries these global issues become even more delicate. For a start, the gauging of a $\sigma$-model used in the dualisation procedure can become subtle and afflicted by topological obstructions   \cite{Hull:1990ms,Hull:1989jk}.   However, the conditions for T-duality are weaker than those for gauging; this was examined for principle circle bundles in \cite{Bouwknegt:2004tr} and more generality in \cite{Hull:2006qs}.   An important perspective  of  \cite{Hull:2006qs} is that most of the topological obstructions to gauging can be eliminated by extended the sigma model with additional coordinates representing the T-dual torus -- because of this the doubled $\sigma$-model is apt to describing T-folds \cite{Hull:2006va}.  However, there remains one topological obstruction for a conventional T-duality; given $d$ Killing vectors $K^{(i)}$ generating a  free $U(1)^d$ action, conventional T-duality requires that $\iota_{i} \iota_{j} \iota_{k} H=0$ (i.e. the contraction of the three-form with any three of the Killing vectors should vanish).  Nonetheless it has been suggested in \cite{Dabholkar:2005ve} that even this condition may be relaxed and T-duality be possible though the result is not expected to be described by conventional geometry even locally.  
}

\EDIT{  A second class of global issues concerns the periodicities of T-dual coordinates.  In the Polyakov  path integral one is instructed to sum over all genera of (Euclidean) world-sheets.  One is thus led to ask whether the gauged $\sigma$-model is  equivalent to the ungauged $\sigma$-model for any genus and, in particular, what happens to the topological information contained in the gauge field? The winding modes of the Lagrange multiplier $\lambda$ serve as Lagrange multipliers for the holonomies of the gauge field.  It was outlined first for the case of a single radial inversion duality in   \cite{Rocek:1991ps} and then in more generality in \cite{Giveon:1991jj}, that by appropriately tuning the periodicity of $\lambda$ -- i.e. the periodicity of the T-dual coordinates -- these holonomies can be constrained to be the identity element of $U(1)$ and the duality related models become completely equivalent.  A further discussion of these global concerns can be found in \cite{Alvarez:1993qi,Giveon:1993ai}.  Let us illustrate this more explicitly.  }
 
\EDIT{ On a world-sheet Riemann surface of genus $g$ there are a set of $2g$ canonical homology one-cycles labelled $(A_i,B_i)$ and one must therefore keep track of the information contained in the holonomies of the connection around these cycle given by the path ordered exponential ${\cal P} \exp i \oint_\gamma A$.  For the two $\sigma$-models (\ref{gauged}) and (\ref{ungauged}) to be truly equivalent as conformal field theories we require that such holonomies are trivial.   To see how this works we look at the torus world-sheet and use the Lagrange multiplier term given by
\be
S_{\lambda}  
= \frac{1}{2\pi} \int_\Sigma d\lambda \wedge A  \, . 
\ee
In the path integral we should sum over all configurations for $\lambda$ including topological sectors.  We therefore write 
\be
d \lambda = d \hat{\lambda} + 2\pi L  p \alpha +2 \pi L q \beta\,, \quad p,q \in \mathbb{Z}
\ee
where  $\hat{\lambda}$ is single valued, and where $(\alpha, \beta)$ are the Poincar\'e dual one-forms to the cycles $(A,B)$ obeying $\oint_A \alpha = \oint_B \beta = 1$ and $\oint_A \beta= \oint_B \alpha = 0$.   $L$ is an, as yet undetermined, periodicity for the Lagrange multiplier.  In the winding sector of the path integral we have, making use of Riemann's bilinear identity (see e.g. \cite{Frak}),
\bea
\sum_{p,q} e^{\frac{i2\pi L}{2\pi}\int (p\a + q\b)\wedge A }&=&\sum_{p,q}  e^{\frac{i2\pi L}{2\pi}\left(p \oint_B A - q \oint_A A \right)} \nonumber \\
&=& \sum_{m,n} \delta\left(m - \frac{L}{2\pi} \oint_A A \right)\delta\left(n - \frac{L}{2\pi} \oint_B A \right)\, .  
\eea 
Then when the periodicity of the Lagrange multiplier is tuned such that $L=1$ we find that the holonomies of the gauge field are just the identity element of $U(1)$.  Thus,  integrating out the single valued piece of the Lagrange multiplier puts the gauge field in pure gauge as before and integrating out the winding modes of the gauge field makes the holonomies trivial \cite{Rocek:1991ps}. One can readily extend this argument to higher genus by simply adding in a sum over canonical pairs of homology cycles.  In a similar way one can show that the constant mode of the Lagrange multiplier enforces the constraint $\oint_\Sigma F=  0$ and hence restricts the curvature to a trivial class. These arguments ensure the full equivalence of the gauged and ungauged $\sigma$-models. }

\EDIT{    For the generalisation to non-Abelian or fermionic  isometries this argument fails to hold\footnote{\EDIT{An immediate obstruction is that in the non-Abelian case the holonomies require path ordering which is not captured in the above analysis \cite{Alvarez:1993qi,Giveon:1993ai}.}}
 and it is for this reason that the generalisations are not thought to be exact symmetries of string genus perturbation theory but rather as a mapping between related CFTs \cite{Alvarez:1993qi,Giveon:1993ai}.}

\subsection{Toroidal Compactification and the $O(d,d)$ Duality Group}
A full understanding of toroidal compactification defined by $d$ compact bosonic directions $X^i \simeq X^i + 2 \pi$ and the internal metric data $E_{ij} = G_{ij} + B_{ij}$ was first given by Narain et al.     \cite{Narain:1985jj,Narain:1986am}.\footnote{For simplicity we ignore spectator fields in this presentation.}   By using an ansatz  $X^i= \sigma n^i + q^i(\tau)$  for the zero modes,   one may deduce that the zero mode momenta must have the following form
\bea
p_{L \bar{i}} =  p_L^i G_{ij} e^i_{ \bar{i}} = (n_i + E_{ji}m^j) e^i_{ \bar{i}} \, , \quad p_{R \bar{i}} =  p_R^i G_{ij} e^i_{ \bar{i}} = (n_i + E_{ij}m^j) e^i_{ \bar{i}} \, ,
\eea
with $n_i$ and $m^i$ integer valued and in which we have introduced vielbeins defined such that
\be
G_{ij}  = e_i^{ \bar{i} } e_{j}^{\bar{j}} \delta_{\bar{i}\bar{j}}\, . 
\ee
 These momenta $(p_L, p_R)$ define a lattice $\Gamma_{d,d} \subset R^{2d}$ , and moreover, since 
\be
p_L^2 - p_R^2 = 2 m^in_i \in 2\mathbb{Z}
\ee 
this lattice is said to be even.  A more detailed investigation of partition functions shows that the lattice must also be self dual to ensure modular invariance \cite{Narain:1985jj}.   The contribution of these momenta to the Hamiltonian is given by 
\be
H_0 = \frac{1}{2}(p_L^2 + p_R^2)  = \frac{1}{2} \left(\begin{array}{c} m^i \\ n_j \end{array}\right) \left(\begin{array}{cc} (G - BG^{-1}B)_{ij}  & (BG^{-1})_{i}^{\ j} \\ - (G^{-1}B)^{i}_{\ j} & G^{ij} \end{array}\right)      \left(\begin{array}{cc} m^j & n_j \end{array}\right) \, .
\ee
These last expressions illustrate underlying moduli space structure of toroidal compactification which is \cite{Narain:1985jj}, 
\be
\label{modspace}
\cM_d =O(d,\mathbb{R} )\times O(d, \mathbb{R})/O(d,d, \mathbb{R})\backslash {\rm T-dualities}\, ,
\ee  
where we have indicated a duality group which we will shortly identify.  The dimension of this moduli space  is $d^2$ in accordance with the number of parameters in $E_{ij}$. The fact that the momenta reside in an even self dual lattice  gives rise to the numerator in (\ref{modspace}) and the invariance of Hamiltonian under  separate $O(d,\mathbb{R})$ rotations of $p_L$ and $p_R$ gives rise to the $O(d\EDIT{, \mathbb{R}}) \times O(d\EDIT{, \mathbb{R}})$ quotient. 

$O(d,d,\mathbb{R})$ is the group which consists of real matrices ${\cal O}$ that respect an inner product with $d$ positive and $d$ negative eigenvalues given, in a conventional frame, by  
\be
\label{oddmetric}
 \eta_{IJ} = \left(\begin{array}{cc} 0 & 1 \\ 1 & 0 \end{array}\right) 
\ee
such that
\be
{\cal O}^t \eta  {\cal O} = \eta \, . 
\ee
Note that the condition that the lattice $\Gamma_{d,d}$ is even is defined with respect to this inner product.  In what follows we shall frequently make use of the $O(d,d,\mathbb{R})/\left( O(d, \mathbb{R} ) \times  O(d, \mathbb{R} )  \right)$ coset representative which packages  the $d^2$ moduli fields :
\be\label{eq:Hdef}
\cH_{IJ} = \left(\begin{array}{cc} (G - BG^{-1}B)_{ij}  & (BG^{-1})_i^{\ j} \\ - (G^{-1}B)^{i}_{\ j} & G^{ij} \end{array}\right)  \, .  
\ee 
The action of an $O(d,d,  \mathbb{R})$ element
\be
\cO =  \left(\begin{array}{cc} a  & b \\ c & d \end{array}\right)\,  , \quad \cO^t \eta \cO = \eta\, ,
\ee
is transparent on the coset form
\be
\label{Hacts}
\cH^\prime = \cO^t \cH \cO\, ,
\ee
but the equivalent action on  $E=G+B$ is more complicated and is given by the fractional linear transformation 
\be
\label{Eacts}
E^\prime = (a E + b)(c E + d )^{-1}\, . 
\ee

For an $S^1$ compactification we saw that the moduli space was further acted on by a discrete duality group (in that case $\mathbb{Z}_2$).  We may ask what is the duality group for toroidal compactification, or what subgroup of $O(d,d, \mathbb{R})$ transformation leave\EDIT{s} the physics completely invariant.  

There are essentially three types of contributions, the first are large diffeomorphisms of the compact torus which preserve periodicities and correspond to the action of a $GL(d, \mathbb{Z})$ basis change. We refer to such transformations as being in the geometric subgroup of the duality group.   These correspond to $O(d,d)$ matrices of the form
\be
\cO_{A} = \left(\begin{array}{cc} A^T & 0 \\ 0 & A^{-1} \end{array} \right)\, , \quad A \in GL(d, \mathbb{Z})\, .
\ee  

The second transformation arises by considering constant shifts in the  $B$ field.  \EDIT{Such a shift in the $B$ field by an antisymmetric matrix with integer components} results in a shift in the action by $2 \pi  \mathbb{Z}$ and produces no change in the path integral.  For $d$ compact directions this allows us to consider the shift $B_{ij} \rightarrow B_{ij}+ \Omega_{ij}$ where $\Omega$ is an antisymmetric matrix with integer entries.   The $O(d,d)$ form of this transformation is 
\be
\cO_{\Omega} = \left(\begin{array}{cc} 1 & \Omega \\ 0 & 1 \end{array} \right)\,,  \quad \Omega_{ij}=-\Omega_{ji} \in \mathbb{Z} \, .
\ee  

The final set of dualities, sometimes called factorised dualities,  consist of things more akin to the radial inversion and have the form \cite{Giveon:1994fu}
\be\label{eq:factorisedduality}
\cO_{T} =  \left(\begin{array}{cc} 1 -e_i & e_i \\ e_i & 1-e_i \end{array} \right)\,, 
\ee  
where $(e_i)_{jk} = \delta_{ij}\delta_{ik}$.   All together these three transformations generate the duality group $O(d,d, \mathbb{Z})$ which henceforward shall be known as the T-duality group. These results above were first fully established in \cite{Giveon:1988tt,Shapere:1988zv}.  To calculate a T-dual geometry one simply performs the action (\ref{Hacts}) or (\ref{Eacts}) using an   $O(d,d, \mathbb{Z})$ transformation. The generalisation of the dilaton transformation for the toroidal case becomes
\be
\label{dilshift2}
\phi^\prime = \phi  + \frac{1}{4} \ln \frac{\det G^\prime}{\det G}   \, .
\ee
The calculation of this shift is rather subtle involving careful regularisation of determinants and is detailed in \cite{Tseytlin:1990va,Tseytlin:1991wr, Schwarz:1993fk}. 

\subsection{Example of $O(2,2, \mathbb{Z})$}
An illustrative example of this $O(d,d,\mathbb{Z})$ can be found by considering string theory whose target is a $T^2$ and this will be useful for our discussion of T-folds. The duality symmetry is already sufficiently  rich since the theory with a torus target space has four moduli encoded in the three components of $G_{ij}$ and the single component of $B_{ij}$. In general we may write the metric of a torus as 
\be
ds^2 = \frac{A}{\tau_2} | dx + \tau dy  |^2 \ . 
\ee
The three metric moduli are neatly encoded in the volume modulus $A$ and the complex structure modulus $\tau$.  The volume modulus and integrated B field can be combined to form a complexified K\"ahler modulus $\rho = B_{xy} + i A$. 

The geometric subgroup of $O(2,2,\mathbb{Z})$ is $GL(2,  \mathbb{Z})$ whose volume preserving subgroup  $SL(2,\mathbb{Z})$ acts as $\cH^\prime = \cO^t \cH \cO$ with generators
\be
\cO_T= \left(\begin{array}{cccc}
1&0&0&0\\
1&1&0&0\\
0&0&1&-1\\
0&0&0&1\\
\end{array}\right)\, , \quad \cO_S= \left(\begin{array}{cccc}
0&-1&0&0\\
1&0&0&0\\
0&0&0&-1\\
0&0&1&0\\
\end{array}\right)\, . \ee
These act to transform the complex structure  $\tau$ by the standard modular transformations
\be
\label{geosub}
T: \tau \rightarrow \tau + 1 \, \quad S: \tau \rightarrow - \frac{1}{\tau} \, . 
\ee    
\EDIT{We denote this subgroup by $SL(2, \mathbb{Z})_\tau$ and it has a geometric interpretation as large diffeomorphisms of the torus.   There is a second subgroup $SL(2, \mathbb{Z})_\rho$ which acts in a similar fashion on the K\"ahler modulus $\rho$.  Since the $SL(2,\mathbb{Z})_{\rho}$ mixes components of the metric with the B-field it is a fundamentally stringy feature.} 
A  further $\mathbb{Z}_{2}$  subgroup serves to interchange    K\"ahler and complex structure  $\tau \leftrightarrow \rho$ and is generated by transformations of the following type
\be
\cO_R= \left(\begin{array}{cccc}
0&0&1&0\\
0&1&0&0\\
1&0&0&0\\
0&0&0&1\\
\end{array}\right)\,. 
\ee 
For the case where $B=0$ and $G=diag(R_1^2 , R_2^2) $ this simply acts by inversion of the radius $R_1 \rightarrow 1/R_1$.  
In summary, the duality group  $O(2,2, \mathbb{Z})$ can be identified with  $(SL(2,\mathbb{Z})_{\tau}\times SL(2,\mathbb{Z})_{\rho})\times\mathbb{Z}_2 $.\footnote{A further discrete $\mathbb{Z}_2$ identification corresponds to world-sheet parity.}    

\subsection{T-folds, U-folds and Non-Geometric Backgrounds}
\subsubsection{\EDIT{T-folds}} 
\label{secT-folds}
 
 
The celebrated paper of Scherk and Schwarz  \cite{Scherk:1979zr} presents two different reduction mechanisms both of which are important  because, for instance, they give rise to potentials that may stabilise  moduli fields -- a long standing challenge in String compactifications.  The first type, a {\em duality-twisted-reduction} (keeping with the naming conventions in  \cite{Dabholkar:2002sy}), involves the reduction of a theory on a torus with fields picking up monodromies about each circle valued in the global duality group of the theory.  In the present context we could consider performing a duality-twisted-reduction with modondromies in the T-duality group as in \cite{Hull:1998vy,Dabholkar:2002sy,Hull:2004in,Hellerman:2004fk,Flournoy:2005kx,Gray:2005ea} -- the resultant backgrounds have been coined  T-folds  by  \cite{Hull:2004in}.  Though exotic sounding, T-folds are a rather inevitable part of any string theory landscape, for instance they may easily be produced by performing a T-duality on a background supported by NS flux as in  \cite{Kachru:2003uq,Lowe:2003qy}.  T-folds look locally like geometric patches but globally they can not be understood as manifolds since T-dualities are used as transition functions.   Providing a modified geometrical picture to describe T-folds was a key rational for Hull's introduction of the Doubled Formalism  \cite{Hull:2004in,Hull:2006va,Hull:2006qs}.  

The second sort of reduction, which we continue to call a Scherk-Schwarz reduction, is when the dependence  of all fields on the internal coordinates, $y^{i}$, is completely specified by  a matrix  $W^{m}_{i}(y)$.  The dependence on the internal coordinates drops out of the dimensionally reduced action provided the one-forms $ W^{m}_{i} dy^{i}$ obey a structure equation 
\begin{equation}
dW^{m }= -\frac{1}{2} f^{m}_{np} W^{n}\wedge W^{p} 
\end{equation}
such that the $f^{m}_{np}$ are constants and define the structure constants for a Lie group $K$.   In general, the reduction space can then be identified with a quotient $K/\Gamma$ where $\Gamma$ is a discrete subgroup of $K_{L}$, the left action of $K$ on itself  as explained in \cite{Hull:2005hk} (which also discusses the case where $K$ is non-compact).  Such a space is often called a twisted torus.   The theory of flux compactification on twisted tori is by now well developed  \cite{MyersOdd,Kachru:2003uq,Derendinger:2004jn,Villadoro:2005cu,Dall'Agata:2005ff,Camara:2005dc,Hull:2005hk,Aldazabal:2006up,Hull:2006tp,Grana:2006kf,ReidEdwards:2009nu} but lies outside the main thrust of this report (a helpful introduction can be found in the lectures of Wecht \cite{Wecht:2007wu}).
  
%
%
%
%
%
%
The T-dualisation of a compact space with H-flux can result in a twisted torus and subsequent T-dualisation of such a twisted torus can produce a T-fold.  These ideas can be well illustrated using the explicit example of a $T^3$ which has been studied in some detail   in the literature \cite{Hull:1998vy,Kachru:2003uq,Flournoy:2005kx,Shelton:2005cf,Shelton:2006fd,Hellerman:2004fk,Dabholkar:2005ve}.   We begin with a $T^3$ equipped with flux described by 
\be
\label{H-flux}
ds^2 = dx^2 + dy^2 + dz^2 \ , \quad  B_{xy}=  N z \, , \, N\in \mathbb{Z}\, .
\ee
This background may be thought of as a $T^2$ in the $(x, y)$ directions fibred over an $S^1$.  After performing a T-duality in the $x$ direction we find a slight surprise -- there is no longer any B-field and the resultant metric is given as
\be
\label{Geoflux}
ds^2 = (dx - N z dy)^2 + dy^2 + dz^2\, . 
\ee
This background is still a torus fibration over the $S^1$, however, it is non-trivial and topologically distinct from the starting $T^3$ since the complex structure  given by 
\be
\tau = -Nz + i, 
\ee
has a monodromy under a circulation of the base $S^1$
\be
\label{mon}
\tau \rightarrow \tau - N \,. 
\ee
This geometry is a twisted torus, in fact it is a nilmanifold -- the quotient of a nilpotent Lie group by a discrete subgroup.  It can be identified with the Heisenberg group of upper triangular $3\times 3$ matrices subject to the action 
\begin{equation}
\Gamma : (x,y,z) \sim (x+1,y,z) \sim ( x, y+1 , z) \sim (x + Ny, y, z+1)   . 
\end{equation}
The globally defined frame fields\footnote{If we parameterise   a group element by  the matrix  
\begin{equation}
h = \left(\begin{array}{ccc}
1& N z &  x \\
0 & 1&  y \\
0 & 0& 1
\end{array} \right) \ , 
\end{equation}
the  frame fields of \eqref{eq:twistedtorusframes} correspond to the left-invariant one-forms $h^{-1} d h = L_i T^i$ where the $T^i$ are (suitably normalized) generators of the Heisenberg algebra.  This was detailed explicitly in \cite{ReidEdwards:2009nu}.  }  
\begin{equation}
\label{eq:twistedtorusframes}
e^{x} = dx - N z dy \ , \quad e^{y} = dy \ ,  \quad  e^{z} = dz \ , 
\end{equation}
obey a  structure equation with the only non-trivial structure constant coming from
\begin{equation}
d e^{x} = N e^{y} \wedge e^{z} \ . 
\end{equation}
This background (\ref{Geoflux}) is sometimes described as having `geometric-flux'  in analogy to the H-flux of the background (\ref{H-flux}).
This space might technically also be thought of as a T-fold since, when viewed as a $T^{2}$ bundle over $S^{1}_{z}$ the monondromy upon circumnavigating the $z$ direction is one of the $O(2,2, \mathbb{Z})$ transformations.  However things are still reasonably geometric; the element of the duality group corresponding to the transformation (\ref{mon}) is contained in the geometric subgroup (\ref{geosub}).  

Things get even more exciting upon performing a second T-duality along the $y$ direction.\footnote{Here we suppress a subtlety: even though the geometry does not depend on the $y$ coordinate, the Killing vector $\partial_{y}$ is not globally defined. However a  T-duality can still be established by working in a covering space (this is nicely explained in \cite{ReidEdwards:2009nu}).  A  treatment of fibrewiseT-duality for non globally defined Killing vector fields can be found in section 7 of  \cite{Hull:2006qs}.   } The Buscher rules give the resultant geometry as
\be
\label{F-flux}
ds^2 = \frac{1}{1+ N^2 z^2} (dx^2 + dy^2) +  dz^2\, , \quad B_{xy} =\frac{Nz}{1+ N^2 z^2}\, . 
\ee  
In this case it is the K\"ahler modulus 
\be
\rho = \frac{Nz}{1+N^2 z^2} + i  \frac{1}{1+ N^2 z^2}
\ee
that has monodromy as $z$ goes from $ 0$ to $1$
\be
\rho \rightarrow \rho - \frac{i N}{N - i } \, ,
\ee
which can be written more compactly as $\rho^{-1} \rightarrow \rho^{-1} +N $.  
This gives us a non-geometric T-fold. In terms of $O(2,2, \mathbb{Z})$, this transformation mixes metric and B-field and hence does not lie in the geometric subgroup and is generated by the following $O(2,2)$ matrix
\be
\cO =  \left(
\begin{array}{llll}
 1 & 0 & 0 & 0 \\
 0 & 1 & 0 & 0 \\
 0 & -N & 1 & 0 \\
  N  & 0 & 0 & 1
\end{array}
\right)\, . 
\ee
This background is locally geometric but clearly no-longer globally geometric.  In the terminology of \cite{Shelton:2005cf,Wecht:2007wu} this background possesses non-geometric flux or  `Q-flux'.   It should be pointed out at this stage that the above example is really only a toy demonstration; it does not represent a complete string background satisfying the conformal invariance criteria of vanishing $\beta$-functionals.  It is somewhat tricky to find an explicit full string background with a twisted torus in its geometry   (although see e.g.  \cite{Hull:1998vy, Schulgin:2008fv,Grana:2006kf} for explicit realisations).

One may even go a step further and postulate that there should also be a notional third T-dual background obtained by T-dualising the $S^1$ corresponding to the $z$ coordinate.  However, at this point the Buscher procedure ceases to be of use since the background is not isometric in the $z$ direction.   Nonetheless there are reasons to believe that a T-dual should exist.  At special points in moduli space (where the scalar potential is minimised) and for certain duality twists, the   isometry is recovered and then one can construct a corresponding orbifold CFT which can be T-dualised  \cite{Dabholkar:2002sy}.  We expect then to be able to deform away from these points in moduli space in both the starting  orbifold CFT and its T-dual   \cite{Dabholkar:2002sy,Dabholkar:2005ve}.     The relationship between these sorts of non-geometric backgrounds and asymmetric orbifolds \cite{Narain:1986qm} (in which left and right moving sectors are subject to different orbifold actions) has been explored in \cite{Dabholkar:2002sy,Hellerman:2004fk,Flournoy:2005kx,Flournoy:2005xe,Dabholkar:2005ve,HackettJones:2006bp} and more recently in \cite{Condeescu:2012sp,Condeescu:2013yma}. 

Furthermore, it is expected that such a background may not even have a locally geometric interpretation since the description of physics may depend locally on both the physical coordinates and the T-dual coordinates \cite{Dabholkar:2005ve,Shelton:2006fd}.  A schematic argument for this given in  \cite{Shelton:2006fd} is that a D3 brane can not be wrapped on the geometry \eqref{H-flux} since the Bianchi identity for a world volume gauge field would be violated.  Performing T-dualities  in all three directions one reaches the conclusions that D0 branes can not be placed in the resultant background and thus it can't be probed by point like objects.  Such backgrounds are often described as having `R-flux' and have been explored in \cite{Dabholkar:2005ve,DallAgata:2007fk,Hull:2009sg}.    An extension to the doubled formalism that is capable of describing R-flux backgrounds was discussed in \cite{Hull:2007jy} and fully spelt out in \cite{ReidEdwards:2009nu}.

A significant amount of work has been done in understanding the construction of these non-geometric backgrounds and their implications for string compactifications.  One reason that such backgrounds have seen such interest is because they allow for moduli stablisation  \cite{Becker:2006ks,Flournoy:2005kx,Wecht:2007wu}.  There has been some suggestion in the literature that these non-geometric backgrounds may be as numerous as conventional backgrounds \cite{McOrist:2010jw}.  The idea of T-folds can be naturally extended to compactifications with S-duality twists (i.e. twisting by a symmetry of the equations of motion \cite{Hull:2003kr} )  or U-duality twists (U-folds) \cite{Hull:2004in,Hull:2006va,Cederwall:2007je} and, given the understanding of mirror symmetry as T-duality \cite{Strominger:1996it}, may allow mirror-folds for Calabi-Yau compactifications \EDIT{as first suggested in \cite{Hull:2004in} (this concept has been studied in the context of $K3$ target spaces in \cite{Cvetic:2007ju,Kawai:2007uq,Ganor:2012ek}).}

 \subsubsection{Exotic Branes, U-folds and Non-Geometric Backgrounds}  
 It is well known that string theory includes various extended objects such as D-branes which are non perturbative in nature: their tension depends on the string coupling as $1/g_s$ meaning that they become very heavy at weak coupling. There are also solitonic objects, for instance the NS5 brane, whose tension scales like $1/g_s^2$.   T-duality can be used to relate the D-branes to each other; a D$p$ brane wrapped around an $S^1$ gets T-dualised to a D$(p-1)$ brane that is unwrapped.   
 
Upon dimensional reduction one however encounters states whose higher dimensional origin is less-clear. Such objects are known as exotic states \EDIT{introduced in \cite{Elitzur:1997zn,Blau:1997du,Hull:1997kb,Obers:1997kk,Obers:1998fb}}.  That these exotic states can not be given an \EDIT{interpretation} in terms of ordinary branes is evident from the fact their tensions are typically  proportional to $1/g_s^3$ or $1/g_s^4$ and thus at weak coupling grow  more rapidly than either D-branes or the solitonic branes.  
 
An interesting recent proposal \cite{Boer:2010fk,deBoer:2012ma}  is that  such exotic branes, which can be obtained from performing T-duality transformations on co-dimension two branes in a dimensionally reduced theory,   can be identified with T-folds.   One might think that the logarithmic divergences associated to such codimensional two objects means that they should be rendered un-physical and ignored.  However it was argued in  \cite{Boer:2010fk,deBoer:2012ma}  that since  ordinary branes may spontaneously polarise to form exotic branes via a supertube effect    they are inevitable and fundamental in string theory. 

Let us give the most famous example of an exotic brane, the $5_2^2$ brane.  This can be obtained in the context of  type II string theory compactified on a $T^2$.  We let the coordinates $x^{8,9}$ label the torus and consider an unwrapped NS5 brane (that is its six world-volume {\EDIT{directions}} are {\EDIT{extended}} in time together with spatial directions $x^{3\dots 7}$).  Performing a T-duality in one of the directions in the torus gives the KK monopole and a subsequent T-dualisation about the second direction of the torus produces the $5_2^2$ brane.  \EDIT{The name of the brane reflects the fact that its mass is proportional to the squares of the radii of the two dimensions $x^{8}$ and $x^{9}$, i.e. the $T^2$ directions, linear in the five radii of the dimensions $x^{3\dots 7}$ and depends on $g_s^{-2}$.  } 

One can obtain a supergravity solution that corresponds to this $5_2^2$ brane. There are various ways to obtain this, either by performing the above T-dualisation chain to the NS5 brane for which one needs to smear the KK monopoles in order to perform the second dualisation or by appropriately dualising a D7 brane background (thus avoiding the smearing).   The result is the following geometry
\bea\label{eq:522branemetric}
& ds^2 = H(r) \left( dr^2 + r^2 d\theta^2\right) + K(r, \theta)^{-1} H(r) dx^2_{89}  + dx^2_{034567}  \, , \nonumber \\
& e^{2\Phi} = H K^{-1} \ , \quad B = - K^{-1} \theta \sigma dx^8 \wedge dx^9 \ ,   
\eea
where the functions $H, K$ are given by 
\be
K = H^2 + \sigma^2\theta^2 \ , \qquad H = h_0 + \sigma \log \frac{\mu}{r} 
\ee
in which $\s = R_8 R_9 / 2\pi\alpha^\prime$ encodes the dependance on the radii of the $T^2$ and $\mu$ enters in regulating and ``renormalising'' the logarithmic divergences associated to the co-dimension two object.  A careful analysis shows that this background preserves half the super symmetries which is to be expected since it was obtained from dualisation of a KK  monopole with that amount of supersymmetry. 

Notice that in this solution the components of the metric and NS two-form associated to the $T^2$ have a dependance on $\theta$.   This dependance is rather similar to the T-fold of the previous section. 
%
It is clear that upon performing a circumnavigation the moduli matrix $\cH(\theta)$ is not single valued, it obtains a monodromy
\be
{\cal H}(2\pi )= U^t {\cal H}(0) U
\ee
 where $U$ is the $O(2,2)$ matrix which acts to send the Kahler modulus $\rho \rightarrow 1/(-2\pi \sigma \rho + 1) $. 
It is in precisely this sense that exotic branes can be associated to non-geometric backgrounds.  \EDIT{Recently, the paper  \cite{Hassler:2013wsa} discusses $Q$ and $R$ branes as the source for the $Q$ and $R$ fluxes of the T-fold and locally non-geometric background respectively. It was shown that    these solutions can be described in the context of double geometry and by considering intersecting Q- and R-branes, together with NS 5-branes and KK monopoles, it was possible to construct 6D supersymmetric geometries with ($SU(3)$ and $SU(2)$ G-structures) giving $AdS_4$ vacua.}

Relevant recent work in this area includes the construction in   \cite{Kimura:2013fda}  of the gauged linear $\sigma$-model which in the IR describes the $5_{2}^{2}$ brane and the consideration in \cite{Kimura:2013zva} of how world-sheet instantons correct the geometry in analogue with the known corrections to the KK monopole  \cite{Gregory:1997te,Tong:2002rq,Harvey:2005ab}.  The construction of the NS5 brane and KK monopole in the T-duality symmetric  doubled formalism -- the topic of the next chapter -- was considered in  \cite{Jensen:2011jna}.

 \subsubsection{Non-Geometry and Monodromy} 
 Above we saw two examples of the idea of T-folds. In both cases there was a monodromy in the spacetime background that takes values in the T-duality or, more generally, U-duality group. Let us emphasise the subtle difference between them.  In the first example, the dualisation of $T^{3}$ with flux, the monodromy occurs as one circumnavigates a non-contractible cycle of a compact spacetime.  In the second case, the exotic branes, the U-duality is non-trivially fibered over a contractible circle in the non compact directions.  In this latter case one may think of the monodromy as a generalised charged associated to a point like object in the non-compact directions. The effect of performing a U-duality is then to take a given charge (monodromy) and conjugate it with a U-duality element.  In this way one can obtain duality orbits of charges in the same conjugacy class of the U-duality group. 
 
 Let us close this section with a short comment on F-theory \cite{Vafa:1996xn}.   The D7 brane of type IIB is a co-dimensional two solution and can readily be considered as an exotic brane with some similarities to the $5^{2}_{2}$ brane above. The axio-dilaton combination $\tau = C_{0} + i e^{\phi}$ displays an $SL(2,\mathbb{Z})$  mondoromy as one encircles the position of the brane in the transverse two space.  A  [p,q]-7 brane (the hypersurface on which open (p,q) strings may end) can be obtained by $SL(2,\mathbb{Z})$ conjugation of the monodromy associated with   the D7.   A crucial idea in F-theory is to geometrise this monodromy. That is one considers the axio-dilation, $\tau$, as a complex structure of an auxillary elliptic curve fibered over the transverse space.  In this sense F-theory introduces an extended (12 dimensional) spacetime in which the $SL(2, \mathbb{Z})$ duality of IIB becomes promoted to a geometric symmetry.  The basis of this idea is, in spirit at least, extremely similar to the ideas that follow in the rest of this report. Recasting F-theory in exactly the same language as used for the duality symmetric M-theory has not yet been carried out in detail, but it is in principle exactly the same idea. The space is extended to allow a realisation of the duality group, in F-theory this is just $SL(2, \mathbb{Z})$, in what follows we will look at the higher rank duality groups and extend the space further. The equivalent of the F-theory D7 branes will then produce the exotic branes with monodromies in the U-duality group. As such F-theory is the precursor to a duality invariant    formulation and its exotic objects i.e. D7 branes, are just the tip of the U-duality group valued exotic objects. Extending F-theory to bigger U-duality groups was already considered a while ago \cite{Kumar:1996zx} but here as opposed to F-theory we actually have a full Lagrangian description of the theory.

\section{The Worldsheet Doubled Formalism}\label{sec:worldsheetdoubled}

The T-fold backgrounds introduced above are examples where string theory disregards familiar notions of geometry.  Understanding string theory on such backgrounds is an important question for two main reasons. Firstly, T-fold compactifications represent an interesting and novel corner of any string landscape and so should be better understood in their own right.  The second reason is that by considering scenarios where the picture of a string propagating in a geometric target space no longer makes sense, we may hope to uncover some clues as to the deeper nature of string theory. 

Parallel to the need to understand non-geometric string backgrounds is a long standing desire to make T-duality a more manifest symmetry of the string $\sigma$-model.   This idea of a duality symmetric  formulation of string theory for Abelian T-duality has a long history with previous approaches including  the formulation of Tseytlin \cite{Tseytlin:1990va, Tseytlin:1990nb}, the work of Schwarz and Maharana \cite{Schwarz:1993vs, Maharana:1992my} and the earlier work by Duff \cite{Duff:1989tf}.   In this section we introduce the duality symmetric theory known as the Doubled Formalism recently championed by  Hull in \cite{Hull:2004in,Hull:2006va,Hull:2006qs,Hull:2007jy,Hull:2009sg} and which has its roots in the early approaches of \cite{Duff:1989tf,Maharana:1992my,Schwarz:1993vs} and the subsequent work of \cite{Cremmer:1997ct}.  (See also \cite{Maharana:2013uvy} for a recent review of a related perspectives from the world-sheet ).

The Doubled Formalism \cite{Hull:2004in,Hull:2006va} is an alternate description of string theory on target spaces that are locally $T^n$ bundles over a base $N$.   
%
The essence of the Doubled Formalism is to consider adding an additional $n$ coordinates so that the fibre is doubled to be a $T^{2n}$.  Because the number of coordinates have been doubled it is necessary to supplement the $\sigma$-model with a  constraint to achieve the correct degree of freedom count.   This constraint takes the form of a self duality (or chirality) constraint.  The final step of the Doubled Formalism is to define a patch-wise splitting $T^{2n} \rightarrow  T^n \oplus \tilde{T}^n$ which specifics a physical $T^n$ subspace and its dual torus $\tilde{T}^n$.    
 
  In this doubled fibration the T-duality group $O(n,n, \mathbb{Z})$ appears naturally as a subgroup of the $GL(2n, \mathbb{Z})$ large diffeomorphisms of the doubled torus.  Because of this, geometric and non-geometric backgrounds are given equal footing in the Doubled Formalism.  They are distinguished only according to whether the splitting (also known as polarisation)  can be globally extended or not. 
  
 \subsection{Lagrangian and Constraint}
 \def\X{{\mathbb{X}}}
  \def\H{{\cal{H}}}
  \def\Ab{{\bar{A}}}
  \def\Bb{{\bar{B}}}
  \def\openone{\mathbb{1}} 
 \EDIT{ In this section we introduce the $\sigma$-model for the Doubled Formalism given by Hull in \cite{Hull:2004in}.}
  
 To describe the $\sigma$-model we first introduce some local coordinates on the $T^{2n}$ doubled torus which we denote by $\X^I$ and coordinates on the base $N$ denoted $Y^a$.   The world-sheet of the $\sigma$-model is mapped into the doubled torus by $\X^I(\sigma)$.  This allows us to define a world-sheet one-form\footnote{Following  \cite{Hull:2004in,Hull:2006va} we will work with world-sheet forms on a Lorentzian world-sheet and will slightly abuse notation by not indicating the pull back to the world-sheet explicitly.}
 \be
{ \cal P}^I = d\X^I \, . 
 \ee
 Additionally we introduce a connection in the bundle given by the  one-form 
\be\label{eq:doubleconnection}
{\cal A}^I ={ \cal A}^I_a dY^a\, , 
\ee
and covariant momenta
\be
\hat{{ \cal P}}^I = d\X^I  + {\cal A}^I  \, . 
\ee
 As before, the $n^2$ moduli fields on the fibre are packaged into the coset form 
 \be
 \label{fibremetric}
 \cH_{IJ}(Y) = \left(\begin{array}{cc} (G - BG^{-1}B)_{ij}  & (BG^{-1})_{i}^{\ j} \\ - (G^{-1}B)^{i}_{\ j} & G^{ij} \end{array}\right) \, ,
\ee
but note the moduli are allowed to depend on the coordinates of the base.    

The starting point for the doubled $\sigma$-model is the Lagrangian 
\be
\label{doubledlag}
\mathcal{L} = \frac{1}{4} \H_{IJ}(Y)\hat{{ \cal P}}^I\wedge\ast \hat{{ \cal P}}^J - \frac{1}{2} \eta_{IJ}{ \cal P}^I \wedge {\cal A}^J + \mathcal{L}_{base}(Y) + \mathcal{L}_{top}(\X)\, , 
\ee
where  $\mathcal{L}_{base}(Y)$ is a standard $\sigma$-model on the base $N$ given (in these conventions) by 
\be
\label{baselag}
 \mathcal{L}_{base}(Y) = \frac{1}{2}g_{ab}dY^a \wedge \ast dY^b +  \frac{1}{2}b_{ab}dY^a \wedge  dY^b \, .
 \ee 
The final term in (\ref{doubledlag}) is a topological term which we will examine in a little more detail when considering the equivalence with the standard formulation of string theory.    Note that the kinetic term of (\ref{doubledlag}) is a factor of a half down compared with the Lagrangian on the base (\ref{baselag}) and that an overall factor of $2\pi$ in the normalisations of these Lagrangians is omitted for convenience. 

This Lagrangian is to be supplemented by the constraint 
\be
\label{constraint}
\hat{{ \cal P}}^I = \eta^{IK}\H_{KJ} \ast \hat{{ \cal P}}^J\, ,
\ee
where $\eta_{IJ}$ is the invariant $O(d,d)$ metric introduced in (\ref{oddmetric}). For this constraint to be consistent we must require that 
\be
 \eta^{IK}\H_{KJ}\eta^{JL}\H_{LM} = \delta_{M}^I
\ee
which is indeed true for the $O(d,d)$ coset form of the doubled fibre metric (\ref{fibremetric}).  

For the remainder, we shall make two simplifying assumptions to aid our presentation, first that the fibration is trivial in the sense that the connection ${\cal A}^I$ can be set to zero.  In terms of the physical $T^n$ fibration over $N$ this assumption corresponds to demanding that the off-diagonal components of the background \EDIT{metric and two-form field} with one index taking values in the torus and the other in the base are set to zero (i.e. $E_{ai}=E_{ia}=0$).  The second is that we shall assume that on the base the two-form $b_{ab}$ vanishes.   With these assumptions the Lagrangian is simply 
\be
\label{doubledlag2}
\mathcal{L} = \frac{1}{4} \H_{IJ}(Y)d\X^I\wedge\ast d\X^J + \mathcal{L}_{base}(Y) + \mathcal{L}_{top}(\X) \, , 
\ee
 and the constraint is 
 \be
 \label{constraint2}
 d\X^I = \eta^{IJ}\H_{JK}\ast d \X^K\, .
 \ee
  To understand this constraint it is helpful to introduce a vielbein to allow a change to a chiral frame (denoted by over-bars on indices) where:
\bea
\H_{\Ab\Bb}(y) = \left( \begin{array}{cc}
\openone &0\\
0 & \openone
\end{array}\right), &  \eta_{\Ab\Bb} = \left( \begin{array}{cc}
\openone &0\\
0 & -\openone
\end{array}\right).
\eea
In this frame the constraint (\ref{constraint2}) is a chirality
constraint ensuring that half the $\X^\Ab$ are chiral Bosons and half
are anti-chiral Bosons.  Such a frame can be reached by introducing a vielbein for the fibre metric $\H_{IJ}$ given by (\ref{fibremetric}) 
\bea
\H_{IJ} =  V^{\hat{I}}_I \delta_{\hat{I}\hat{J}} V^{\hat{J}}_{ J}\, ,
\eea
with 
\bea
V^{\hat{A}}_A = \left( \begin{array}{cc}
e^{t} & 0 \\ -e^{-1}B& e^{-1}
\end{array} \right)\, ,
\eea 
and $G_{ij} =e_i^{\ \hat{i}} e_j^{\ \hat{j}} \delta_{\hat{i}\hat{j}}$.   
This allows the definition of an intermediate frame in which 
\bea
\eta_{\hat{I}\hat{J}} = \left(\begin{array}{cc}
	0 & 1 \\ 1 & 0 
\end{array} \right)\, , &  \H_{\hat{I}\hat{I}} = \left(\begin{array}{cc}
	1 & 0 \\ 0 & 1
\end{array} \right)\,
\eea 
Supplementing this with a basis transformation defined by
\bea
O_{\bar{A}}^{\, \hat{I}} = \frac{1}{\sqrt{2}} \left(\begin{array}{cc}
	1 & 1 \\ 1 & -1 
\end{array} \right)
\eea
brings $\H$ and $\eta$ into the required form.   In this chiral basis the one-forms ${\cal P}^{\bar{A}}= (P^{\bar{i}}, Q_{\bar{i}})$ obey simple chirality constraints 
\bea
 P^{\bar{i} }_+ = 0\, ,  &  Q_{\bar{i} - }=0 \, .
\eea

\subsection{Implementing the Constraint}
\def\d{{\partial}}
\EDIT{ There are, of course, a variety of ways in which this chirality constraint  eq.~\eqref{constraint2} can be implemented. The literature on chiral bosons is vast and we don't intend to survey it here, but even within the context of the duality symmetric string several approaches have been used:  The constraint has been imposed by gauging a suitable current as suggested in \cite{Hull:2004in} and detailed in \cite{Hull:2006va}.   The constraint can be treated as a second-class constraint and canonical quantisation can proceed with Dirac brackets as was shown in \cite{HackettJones:2006bp}.  The calculation of the partition function of the duality symmetric string in \cite{Berman:2007vi} uses the holomorphic factorisation approach of  \cite{Witten:1991mm,Witten:1996hc}.  In this section we will consider how, at the classical level at least, the constraint can be implemented at the level of the action using the approach of  Pasti, Sorokin and Tonin (PST) \cite{Pasti:1996vs,Pasti:1995us,Pasti:1997mj}.\footnote{Although there are some reasons to believe the PST approach is well behaved quantum mechanically  \cite{Pasti:1996vs,Lechner:1998ga}, it may not in fact be   so straightforward to quantise the PST action using a Faddeev-Popov procedure.  One reason is that PST symmetry is potentially anomalous and a second is the non-polynomial form of the action.  These issues have not yet been completely resolved in the literature, see \cite{Sevrin:2013nca} for a discussion.}  }

To understand the constraint more clearly we consider the simplest example, the one-dimensional target space of a circle, with
constant radius $R$. The doubled action on the fibre is
\be\label{eq:S1doubled}
S_{d}=\frac{1}{4} R^2 \int dX\wedge \ast dX + \frac{1}{4} R^{-2} \int d\tilde X \wedge \ast d\tilde X .
\ee
We first change to a basis in which the fields are chiral by defining
\bea
\X_+= RX+R^{-1}\tilde{X}, &  \d_-\X_+ =0\, , \nonumber \\ \X_-= RX-R^{-1}\tilde{X}, & \d_+\X_- =0\, . 
\eea 
In this basis the action becomes
\be
S_{d}= \frac{1}{8}\int d\X_+\wedge \ast d\X_+ + \frac{1}{8} \int d  \X_- \wedge \ast d \X_-\, .
\ee

One may then incorporate the constraints into the action using the
method of Pasti, Sorokin and Tonin (PST) \cite{Pasti:1996vs,Pasti:1995us,Pasti:1997mj}. \EDIT{The first application of the PST formalism  in the context of duality symmetric string theory is by Cherkis and Schwarz in \cite{Cherkis:1997bx} in which the PST approach was used to covariantise the duality symmetric heterotic string written in Tseytlin form \cite{Tseytlin:1990nb,Tseytlin:1990va}.   Here we go the other way round, as was advocated in \cite{Hull:2006tp} and carried out in \cite{Berman:2007xn}, 
    we start with an unconstrained action and implement a chirality constraint using the PST formalism and after gauge fixing recover a non-covariant form given by Tseytlin \cite{Tseytlin:1990nb,Tseytlin:1990va}.}

We define one-forms
\bea
\mathcal{P}=d\X_+-\ast d\X_+ ,& \mathcal{Q}=d\X_-+\ast d\X_-\, ,\eea
which vanish on the constraint.  These allow us to incorporate the constraint into the action via the introduction of two auxiliary closed one-forms $u$ and $v$ as follows: 
\be\label{eq:PSTform}
S_{PST}= \frac{1}{8}\int d\X_+\wedge \ast d\X_+ + \frac{1}{8} \int d  \X_- \wedge \ast d \X_-   -\frac{1}{8}\int d^2\sigma\left( \frac{(\mathcal{P}_m u^m)^2}{u^2} +\frac{(\mathcal{Q}_m v^m)^2}{v^2}\right).
\ee

As explained in the appendix, the PST action works by essentially introducing a new gauge symmetry, the PST symmetry, that allows the gauging away of fields that do not
obey the chiral constraints. Thus only the fields obeying the chiral
constraints are physical.

One may now consider gauge fixing the PST-style action.  It is possible to choose a physical gauge that completely fixes this invariance but which breaks manifest Lorentz invariance.  An alternative is to quantise within the PST framework whilst maintaining covariance and introduce ghosts to deal
with the PST gauge symmetry. Here we choose the non-covariant
option and immediately gauge fix to give a Floreanini-Jackiw \cite{Floreanini:1987as}
style action. Picking the auxiliary PST fields ($u$ and $v$) to be
time-like produces two copies of the FJ action (one chiral and one anti-chiral)
\be\label{FJL}
S = \frac{1}{4}\int d^2\s ( \d_1\X_+\d_{-}\X_+ -  \d_{1}\X_- \d_+\X_-).
\ee
If we re-expand this in the non-chiral basis we find an action
\be
\label{eqTseyaction}
 S= \frac{1}{2}\int d^2\sigma\left[ -(R\d_1 X)^2 - ( R^{-1}\d_1 \tilde{X})^2 + \partial_0X\partial_1\tilde{X} + \partial_1 X \partial_0 \tilde{X} \right] \, ,
\ee
which may be recognised as Tseytlin's  duality symmetric formulation \cite{Tseytlin:1990nb,Tseytlin:1990va}.   
The constraints 
\be
\d_0 \tilde{X} = R^2 \d_1 X \, ,  \qquad \d_0 X = R^{-2} \d_1 \tilde{X} \, ,
\ee
 then follow after integrating the equations of motion and the string wave equation for the physical coordinate $X$ is implied by combining the constraint equations. \EDIT{At first it might seem that one needs to resort to boundary conditions to remove a function $f(\tau)$ that arises when integrating up the equations of motion in this way.  In fact, this is not quite true, we fix this arbitrary function of $\tau$  by
   observing that (\ref{eqTseyaction}) has $\delta{X} = f(\tau)$ gauge
   invariance (in the sense that it is a symmetry with vanishing corresponding Noether charge).   }
   
   \EDIT{Let us comment briefly on what happens when the PST procedure is used in a more general scenario.  When $R$ is not constant but instead varies as a function of the base coordinate $R=R(y)$ one might be worried that the PST gauge symmetry will no longer leave the action invariant.  This is in fact true, however what saves the day (as is often the case in the Doubled Formalism) is that the chiral fields enter the action paired with anti-chiral partners.  When $R=R(y)$ one finds that the  PST approach can be used   but that the auxiliary fields $u$ and $v$ that enter the action eq.~\eqref{eq:PSTform} can no longer be considered as independent fields and only a single PST symmetry is preserved.  This can be readily extended to a general $T^{2n}$ doubled torus with generalised metric ${\cal H}_{IJ}(Y)$ to give the results found in  \cite{Cherkis:1997bx}.   Let us now review the covariant PST approach leading to the results in \cite{Cherkis:1997bx}.  }
  \def\Dp{\partial_{+}} 
  \def\Dm{\partial_{-}}
  \def\cH{{\cal H}}
    \def\cS{{\cal S }}
  \def\X{{\mathbb{X}}}
  \def\Pp{{\cal P}^{(+)}} 
   \def\Pm{{\cal P}^{(-)}} 
      \def\Ppm{{\cal P}^{(\pm)}} 
            \def\Pmp{{\cal P}^{(\mp)}} 
            \def\ep{\epsilon^{+}}
            \def\Em{\epsilon^{-}}
            \def\hpp{h_{++}} 
            \def\hmm{h_{--}}

The $O(d,d)$ structure of the Doubled Formalism allows us  to  define some projectors 
  \begin{equation} 
(\Ppm)^I{}_J  = \frac{1}{2} \left( \delta^I{}_J \pm \cS^I{}_J \right)  \ , \quad \Ppm\Ppm = \Ppm \ , \quad \Ppm\Pmp = 0  \ , 
   \end{equation} 
   where $\cS^I{}_J= \cH^{IK} \eta_{KJ}$. 
In terms of these the desired chirality constraint  eq.~\eqref{constraint2} becomes
    \begin{equation}  \label{eq:Constraint2}
 (\Pp)^{I}{}_{J} \Dm \X^{J} = 0  , \quad (\Pm)^{I}{}_{J} \Dp \X^{J} = 0  \ .
\end{equation} 
The following action 
\begin{equation}     
  \begin{aligned}
      S  =   \frac{1}{2} \int d^2 \sigma (2 \cH_{MN}(Y) \partial_+ \X^N \partial_- \X^M + &   \frac{\partial_- A}{\partial_+A} \eta_{MN}(\Pm \partial_+ \X)^M (\Pm \partial_+ \X)^N\\
     & -  \frac{\partial_+ A}{\partial_- A} \eta_{MN}(\Pp \partial_- \X)^M (\Pp \partial_- \X)^N + {\cal L}_{base}  )
      \end{aligned}
   \end{equation}
  has a PST local symmetry that acts as
 \begin{equation} 
 \delta \X^M = \zeta \left( \frac{1}{\partial_+ A} \Pm \partial_+\X^M  + \frac{1}{\partial_- A} \Pp \partial_- \X^M   \right)  \ , \quad \delta A = \zeta\ , 
   \end{equation}
   where $\zeta =\zeta(\sigma, \tau)$ is a scalar function on the world-sheet.  A second local invariance is given by
   \begin{equation} \label{eq:localdelta}    
 \delta \X^M = f^M(A) \ , \quad \delta A = 0 \ . 
    \end{equation}  
The first gauge symmetry   ensures that the degrees of freedom we wish to set to zero via the constraint  eq.~\eqref{eq:Constraint2} are rendered pure gauge. Indeed,  the equations of motion for $\X$  imply the desired constraint equation.  An additional unwanted solution to the equations of motion is removed by the second local invariance eq.~\eqref{eq:localdelta} \cite{Pasti:1996vs, Cherkis:1997bx,Lechner:1998ga}.


 \EDIT{The first gauge freedom can be fixed by choosing $A= \tau$ in which the action reduces to the non-covariant duality symmetric action of Tseytlin
\begin{equation}
 S  = \frac{1}{2} \int d^2 \sigma \,\left(  - \cH_{IJ}(y) \partial_{1} \X^{I}\partial_{1}\X^{J} + \eta_{IJ} \partial_{0}\X^{I} \partial_{1 }\X^{J } + {\cal L}_{base}   \right) \ .  \label{TseytlinA}
  \end{equation}  
}
For the fibre coordinates we have the equation of motion 
\bea
\d_1 \left( \H \d_1 \X \right) = \eta \d_1\d_0 \X,
\eea
which integrates \EDIT{-- again using what remains after fixing $A= \tau$ of the second local invariance eq.~\eqref{eq:localdelta} --} to give the constraint (\ref{constraint2}).   

 \EDIT{Let us mention, before moving on, that the reinstatement of the connection given in eq.~\eqref{eq:doubleconnection} can also be easily done in the PST approach.\footnote{\EDIT{We thank A. Sevrin for discussions on this point. }} }

\subsection{Action of $O(n,n,\mathbb{Z})$ and Polarisation}
 The Lagrangian (\ref{doubledlag2})  has a global symmetry of $GL(2n, \mathbb{R})$ acting as 
\be
\cH^\prime  = \cO^t \cH \cO\, ,  \quad  \X^\prime = \cO^{-1} \X \, , \quad \cO \in GL(2n, \mathbb{R})\,.
\ee
However,  to preserve the periodicity of the coordinates $\X$ this symmetry is broken down to the discrete subgroup  $GL(2n, \mathbb{Z})$.  Under these transformations the constraint (\ref{constraint2}) becomes
\be
\cO^{-1} d\X=\eta^{-1} \cO^{t} \H \ast d \X  \, .
\ee
Hence, to preserve the constraint we require that 
\be
\cO^t \eta \cO = \eta 
\ee 
and therefore that the global symmetry is reduced to $O(n,n,\mathbb{Z}) \subset GL(2n ,\mathbb{R})$.  It is now clear that the T-duality group has been promoted to the role of a manifest symmetry.   

To complete the description of the Doubled Formalism one needs to specify the physical subspace which defines a splitting  of the coordinates $\X^I = (X^i, \tilde{X}_i)$.  Formally this was done \EDIT{in \cite{Hull:2004in}}  by introducing projectors $\Pi$ and $\tilde{\Pi}$ such that $X^i = \Pi^i_{\ I}\X^I$    and   $\tilde{X}_i = \tilde{\Pi}_{i I}\X^I$ .   These projectors are required to pick out subspaces that are maximally isotropic in the sense that $\Pi \eta^{-1} \Pi^t = 0$ (a more conventional definition of a maximal isotropic can be found in section 4.2.1).    These projectors are sometimes combined into a  `polarisation vielbein' given by $\Theta = (\Pi , \tilde{\Pi})$.   

This is a somewhat formal way of saying that we are picking out a preferred basis choice for the $\sigma$-model but has use when describing T-duality.    Indeed, the formulas for the coset matrix $\H_{IJ}$ and the invariant metric $\eta_{IJ}$  given by (\ref{fibremetric}) and (\ref{oddmetric}) should be understood to be in the basis defined by this splitting.  The components of $\H$ are given by acting with projectors so that for example
\be
G^{ij} = \Pi^{i I} \Pi^{j J} \cH_{IJ} \, ,
\ee   
in which we have raised the indices on the projectors with the invariant metric $\eta^{IJ}$.  

\EDIT{T-duality can now be viewed in two ways, as either an active or a passive transformation \cite{Hull:2004in}.  In the active viewpoint the} polarisation is fixed but the geometry transforms according to the usual transformation law
\be 
\cH^\prime = \cO^t \cH \cO \, . 
\ee  
The second, passive, approach is to consider the geometry as fixed but that the polarisation vielbein changes according to
\be
\Theta^\prime = \Theta \cO \, . 
\ee
In this view T-duality amounts to picking a different choice of polarisation.  It is clear that all T-dual backgrounds may be treated on an equal footing in the doubled formalism.

\subsection{Geometric vs. Non-Geometric} 

To discuss the difference between geometric and non-geometric backgrounds one must examine topological issues and in particular the nature of the transition functions between patches of the base.  That is we need to see how T-duality may be non-trivially fibered.  For two patches of the base ${\cal U}_1$ and ${\cal U}_2$, on the overlap  ${\cal U}_1 \cap {\cal U}_2$ the coordinates are related by a transition function
\be
\X_1 = g_{21} \X_2 \, , \quad  g_{21} \in O(n,n , \mathbb{Z}) \, .
\ee
In the active view point, where polarisation $\Theta$ is fixed but the geometry is transformed,  the condition for this patching to be geometric is that the physical coordinates, obtained from projecting with the polarisation, are glued only to physical coordinates.  More formally we have 
\be
\Theta \X_1 = (X_1, \tilde{X}_1) \, , \quad   \Theta \X_2 = (X_2, \tilde{X}_2)\, , 
\ee
 and therefore 
 \be
 \Theta \X_2 = \Theta g_{12} \X_1= \Theta g_{12} \Theta^{-1} \Theta \X_1 = \hat{g}_{12} \Theta \X_1\, . 
\ee
Then the physical coordinates are glued according to 
\be
X^i_2 =  (\hat{g}_{12})^{i}_{\ j} X_1^j + (\hat{g}_{12})^{ij} \tilde{X}_{1j}  \, . 
\ee
Thus, for the physical coordinates to be glued only to physical coordinates, all transition functions must be of the form
\be
(\hat{g})^I_{\ J} = 
\left(
\begin{array}{cc}
 g^i_{\ j}  & 0     \\
  g_{ij }  &  g_i^{\ j}  
\end{array}
\right)
\ee
which are the $O(n,n, \mathbb{Z})$ elements formed by the semi direct product of $Gl(n, \mathbb{Z})$ large diffeomorphisms of the physical torus and integer shifts in the components of B-field.    Otherwise the background is said to be non-geometric.\footnote{One can tighten the definition of geometric to exclude the B-field shifts. This restriction is equivalent to demanding that the non-physical T-dual coordinates also only get glued amongst themselves.}

\subsection{Recovering the Standard $\sigma$-model}
 The equations of motion arising from the doubled Lagrangian (\ref{doubledlag2}) are, for the fibre coordinates,
\be
\label{eqm1}
0= \H_{IJ} \partial^2 \X^J + \partial_a \H_{IJ} \partial_\alpha Y^a \partial^\a \X^J \, ,
\ee
and for the base coordinates
\be
\label{eqm2}
0= \frac{1}{4} \partial_c \H_{IJ} \partial_\a \X^I \partial^\a \X^J - g_{ac}\left( \partial^2 Y^a +\hat{\Gamma}^a_{de} \partial_\a Y^d \partial^\a Y^e  \right) \, , 
\ee
where $ \hat{\Gamma}^a_{de}$ is the  symbol constructed from the base metric. 

After choosing a polarisation, the constraint can be written as
\be
 \partial^\a \tilde{X}_j = G_{jk} \e^{\a\b} \partial_\beta X^k +B_{jk}\partial^\a X^k \, , 
\ee
and that this can be used at the level of equations of motion to replace all occurrences of  the dual coordinates $\tilde{X}_i$ with the physical coordinates $X^i$.    After expanding out in the $(X,\tilde{X})$ basis one finds that the equation of motion (\ref{eqm1}) can be written as
\be
0 = g_{ij} \partial^2 X^j + \partial_a g_{ij} \partial_\a Y^a \partial^\a X^j + \partial_a b_{ij} \e^{\a\b} \partial_\a Y^a \partial_\b X^j 
\ee
and (\ref{eqm2}) as 
\be
0=   \frac{1}{2}  \partial_a g_{ij} \partial_\a X^i \partial^\a X^j + \frac{1}{2} \partial_a b_{ij} \e^{\a\b} \partial_\a X^i \partial_\b X^j 
 -      g_{ac}\left( \partial^2 Y^a +\hat{\Gamma}^a_{de} \partial_\a Y^d \partial^\a Y^e  \right)\, . 
\ee
The latter two equations correspond to the equations of motion obtained from the standard $\sigma$-model of the form 
\be
\label{standardsigma}
\cL = \frac{1}{2} g_{ij}(Y)  d X^i \wedge \ast dX^j + \frac{1}{2} b_{ij}(Y) dX^i\wedge dX^j + \frac{1}{2} g_{ab}(Y) dY^a \wedge \ast dY^b \, . 
\ee
This demonstrates that the doubled $\sigma$-model together with the constraint is equivalent, at the classical level of equations of motion, to the standard string $\sigma$-model. 

\subsection{Extensions to the Doubled Formalism}
We now outline a few advances and extensions to the above formalism. 
\subsubsection{Dilaton}
Alongside the metric and B-field a background in string theory is equipped with a scalar field, the dilaton.  One is forced to ask  how should this scalar field be included in the Doubled Formalism.  A first guess is to include a scalar field $d$ with the same Fradkin-Tseytlin coupling \cite{Fradkin:1985ys,Fradkin:1984pq} as for the standard string dilaton namely
\be
S_{dil}= \frac{1}{4\pi} \int d^2 \sigma \sqrt{\gamma} d(Y)  R^{(2)}
\ee 
where $R^{(2)}$ is the scalar curvature of the world-sheet metric $\g$ and, in keeping with the ansatz for metric and B-field, the scalar only depends on the base coordinates.  This `doubled dilaton' is duality invariant under $O(n,n,\mathbb{Z})$ and related to the standard dilaton field $\phi$ by \cite{Hull:2006va}
\be
d = \phi - \frac{1}{4}\ln \det G
\ee
so that from the invariance 
\be 
d^\prime = d  \Rightarrow \phi^\prime - \frac{1}{4}\ln \det G^\prime = \phi - \frac{1}{4}\ln \det G
\ee
we recover the dilaton transformation rule (\ref{dilshift2}).    The introduction of this duality invariant dilaton corresponds to the well know feature that in supergravity the string frame measure $\sqrt{|G|} e^{-2\Phi}$ is a T-duality invariant. 
\subsubsection{Branes}
In the original work of Hull \cite{Hull:2004in}, the extension of the Doubled Formalism to open strings and their  D-branes interpretation was suggested.  Since T-duality swaps  boundary conditions,  of the $\X$ coordinates exactly half should obey Neumann boundary conditions and half Dirichlet.   This splitting is, in general, different from the splitting induced by polarisation choice.   If, for a given polarisation, exactly N of the physical coordinates obey Neumann conditions then there is a Brane wrapping N of the compact dimensions.  The dimensionality of the brane then changes according to the usual  T-duality rules under a change in polarisation.  The projectors that specify Neumann and Dirichlet conditions are required to obey some consistency conditions detailed in   \cite{Lawrence:2006ma,Albertsson:2008gq} and extended further in \cite{Albertsson:2011ux}. Further related work on the role of D-branes in non-geometric backgrounds can be found in \cite{Kawai:2007qd}.
\subsubsection{Supersymmetry}
A world-sheet $\cN=1$ supersymmetric extension of the formalism has been considered in  
\cite{HackettJones:2006bp,Hull:2006va}.  To achieve this one simply promotes coordinates into superfields and world-sheet derivatives into super-covariant derivatives.  The constraint   (\ref{constraint2}) then is generalised to a supersymmetric version which includes a chirality constraint on fermions.   Preliminary findings of an $\cN=2$ formulation directly in superspace have been reported in \cite{Sevrin:2013oca}. 
\subsubsection{Interacting Chiral Bosons: Doubled Everything}
\label{doubledeverything}
A slightly displeasing feature of this formalism is the need to, from the outset, seperate the base of the fibration which remains undoubled with the $T^d$ fibre that becomes doubled.   A natural question to ask is whether there is a more democratic approach in which everything becomes doubled.   Let us consider the most general doubled action, in Tseytlin form, 
\be
S = \int d^2 \sigma \, {\cal H}( {\mathbb{X}})_{IJ} \pl_1 \mathbb{X}^I  \pl_1 \mathbb{X}^J - {\cal C}( {\mathbb{X}})_{IJ}     \pl_0 \mathbb{X}^I  \pl_1 \mathbb{X}^J \ . 
\ee 
Now all coordinates are doubled and the matrices $  {\cal H}$ and  $ {\cal C}$  have arbitrary coordinate dependence and $  {\cal C}$ need not be be symmetric.  As it stands this action is too general; one should implement some constraints such that the theory has: a) first order equations of motion allowing half the degrees of freedom to be eliminated; b) emergent on-shell Lorentz invariance and c) an off-shell invariance under a set of modified Lorentz transformations.  Let us remark on two ways to solve these constraints, further discussion and derivation can be found in \cite{DallAgata:2008qz,Avramis:2009xi,Sfetsos:2009vt,Thompson:2010sr}.

\begin{itemize}
\item {\bf DFT Constraint: Depending on half the coordinates}  
\end{itemize}

 Setting ${\cal C}( {\mathbb{X}})  = \eta$, the constant $O(d,d)$ invariant metric,  and allowing   the matrix $  {\cal H}(\mathbb{X})$   to depend on only half of the coordinates provides a consistent model (as noticed long ago by Tseytlin in \cite{Tseytlin:1990va}).  This requirement, which can be thought of as an off-shell requirement, will be closest to what follows in the DFT approach   which introduces the so called strong constraint  that $\eta^{IJ} \partial_{I}\partial_{J}\phi_{1}$ {\em and} $\eta^{IJ}\partial_I \phi_1 \partial_J \phi_2$ evaluates to zero for all fields $\phi_1,\phi_2$. As a consequence of the strong constraint the fields only depend on half the coordinates as shown explicitly in \cite{Hohm:2010jy}.

\begin{itemize}
\item   {\bf Scherk-Schwarz ansatz}  
\end{itemize}

One can in fact allow for further coordinate dependence  by assuming the presence of some underlying group structure and allowing a Scherk-Schwarz \cite{Scherk:1979zr}   ansatz for the background fields.  This  allows for geometries that depend simultaneous on the regular and T-dual coordinates.    As we shall see in the sequel this has   profound implications and paves the way to the relaxation of the strong constraint in Double Field Theory. 

Introducing some left-invariant one-forms for a $2d$ dimensional group manifold $L^{\bar{A}}(\X) = L_I^{\bar{A}}(\X) d \mathbb{X}^I$ we impose that 
\be
\label{twisted1}
{\cal C}_{(IJ)}(\X) = L_I^{\bar{A}}(\X) \eta_{{\bar{A} \bar{B}}}  L_J^{\bar{B}}(\X)  \ , \qquad  {\cal H}_{IJ}(\X) = L_I^{\bar{A}}(\X)  {\cal H}_{{\bar{A} \bar{B}}} L_J^{\bar{B}}(\X) \ , 
\ee
such that all the coordinate dependence on $\X$ is contained in the $L_I^{\bar{A}}$.  The matrix $\eta$ is a inner product on the underlying algebra and the matrix ${\cal H}$ and $\eta$ must be related by the orthogonality condition 
\be
\label{twisted2}
{\cal H}^T \eta {\cal H} = \eta  \ . 
\ee
Finally the anti-symmetric piece of ${\cal C}$ serves as a potential for a torsion that equals the spin connection on the group manifold.  In summary the result is rather reminiscent of a WZW model:
\be
\label{doubleWZW}
S = \frac{1}{4} \int_\Sigma d^2 \sigma \, {\cal H}( {\mathbb{X}})_{IJ} \pl_1 \mathbb{X}^I  \pl_1 \mathbb{X}^J - \eta( \mathbb{X})_{IJ}     \pl_0 \mathbb{X}^I  \pl_1 \mathbb{X}^J +  \frac{1}{12}  \int_{{\cal B}} f_{\bar{A}\bar{B}\bar{C}} L^{\bar A} \wedge L^{\bar B} \wedge L^{\bar C} \ ,
\ee 
where $f_{\bar{A}\bar{B}\bar{C}}$ are the structure constants, the three-manifold is such that  $\partial {\cal B}= \Sigma$ and pull-backs are implicit in the final term.   When ${\cal H}_{{\bar{A} \bar{B}}}  = \eta_{{\bar{A} \bar{B}}}   = \delta_{\bar{A} \bar{B}}$ then this   corresponds to the chiral WZW described by Sonnenschein in   \cite{Sonnenschein:1988ug}.  Such models were also considered in the original work of Tseytlin \cite{Tseytlin:1990va}. 

Such an approach was directly taken by Hull and Reid-Edwards \cite{Hull:2007jy,Hull:2009sg,ReidEdwards:2006vu,ReidEdwards:2009nu,ReidEdwards:2010vp} in proposing a doubled $\sigma$-model capable of describing {\em locally non-geometric backgrounds} (e.g. the postulated R-flux backgrounds that may be obtained by performing a T-duality when Buscher rules no longer apply).          For the case that the group manifold is a Drinfeld Double then this form corresponds exactly to the Poisson-Lie duality symmetric action of \cite{Klimcik:1995dy,Klimcik:1995kw,Klimcik:1995ux} which contains as a sub-case non-Abelian T-duality.  That is to say that the above $\s$-model also provides a doubled formalism for non-Abelian T-duality.
  
 A remarkable feature, and a theme that we shall return to, is that by allowing dependance on both the dual and regular coordinates (albeit in a way constrained by the group structure) we find a $\s$-model for non-geometric backgrounds and moreover that this theory will  turn out to provide a world-sheet description of gauged supergravities!

\subsection{Quantum Aspects of the Doubled Formalism}
 
Thus far our discussion has been rather classical but there has been  progress  made on understanding the quantum aspects of the Doubled Formalism which we shall now review. There is by now good evidence that the doubled formalism is quantum mechanically sound and its equivalence to the standard formalism extends beyond the classical level.

\subsubsection{Canoncial Quantisation}
The canonical quantisation of the doubled formalism was performed in \cite{HackettJones:2006bp}. The idea is to view the theory as constrained system with second class constraints.  One may then proceed in quantisation via Dirac brackets.  In this approach a central role is played by the constraint  eq.~\eqref{constraint2}  which in phase space reads 
 \be
 \chi_{I} = \pi_{I} - \eta_{IJ} \partial_{\sigma} \X^{J} \approx 0  \ , 
 \ee
 where $\pi_{I}$ is the momenta conjugate to $\X^{I}$ with Poisson bracket $\{ \X^{I}(\sigma) , \pi_{J}(\sigma^{\prime}) \} = \delta_{J}^{I} \delta(\sigma - \sigma^{\prime})$.   The constraint has a non-vanishing Poisson bracket with itself but generates no extra constraints since its time evolution vanishes weakly
\be 
\{ \chi_{I}(\sigma) , \chi_{J}(\sigma^{\prime})  \} =  -2\eta_{IJ}  \delta'(\sigma - \sigma^{\prime})  \equiv \Delta_{IJ}(\sigma - \sigma^{\prime}) \ , \quad \{ \chi_{I} (\sigma) , H\} = \partial_{\sigma}(- \cH_{I}{}^{J} \chi_{J} )   \approx 0 \ .
\ee
This means the constraint is second class (it is not associated to a gauge symmetry) and that the dynamics may be restricted to the constraint surface $\chi_{I} = 0$ by introducing Dirac brackets 
\be
\{ f, g \}_{D}= \{ f, g\} - \int \{ f, \chi_{I} (\sigma) \} ( \Delta(\sigma - \sigma^{\prime})^{-1})^{IJ}  \{ \chi_{J} (\sigma') , g \} \ . 
\ee
The Dirac brackets of the fields are then given by
\be
\begin{aligned}
&\{  \X^{I}(\sigma), \X^{J} (\sigma') \}_{D} = -\frac{1}{4} \left(\theta( \sigma - \sigma^{\prime}) - \theta( \sigma^{\prime} - \sigma) \right)\ ,  \\
 & \{ \X^{I}(\sigma) , \pi_{I}(\sigma^{\prime}) \}_{D} = \frac{1}{2}\delta_{I}^{J} \delta(\sigma - \sigma^{\prime}) \ ,  \\ 
 & \{ \pi_{I}(\sigma) , \pi_{J}(\sigma^{\prime}) \}_{D} = \frac{1}{2}\eta_{IJ} \delta(\sigma - \sigma^{\prime})\ .  \\ 
\end{aligned}
\ee
Quantisation may now proceed by hatting i.e. promoting these Dirac brackets to quantum commutators. One rather nice feature of this approach is that no choice of polarisation is either assumed or needed.  An explicit example was provided in \cite{HackettJones:2006bp} using this approach to quantise in a T-fold background rather    like the one described in \ref{secT-folds}.  In the non-doubled language, the background considered was an  $S^{1}$ fibered over a base, also an $S^{1}$, such that upon circumnavigation of the base the radius of the fibre is inverted i.e. it has a T-duality monodromy. One might think that in the doubled approach this lifts to a simple geometric configuration and can be quantised without much thought.  However it was explained in   \cite{HackettJones:2006bp} that even in the doubled formalism one has an orbifold which requires keeping track of two different twisted sectors.  Nonetheless, it was shown using the results of   \cite{Hellerman:2006tx}, how to  construct the appropriate Hilbert space for this theory and that it \EDIT{gives rise to a modular invariant partition function. }

\subsubsection{Gauging the Current} 

An alternative to this above canonical quantisation was proposed by Hull in \cite{Hull:2006va}.  In this approach the conserved current associated to shift symmetries in the fibre together with the trivial Bianchi identity for fibre momenta can be used \EDIT{to form} a conserved current
 \be
 J_{I} = \H_{IJ} d\X^J - L_{IJ} \ast d\X^J \, . 
 \ee 
 The constraint equation (\ref{constraint2}) is equivalent to demanding that the polarisation projected component $J^i = 0$.    The symmetry giving rise to this current can be consistently gauged by introducing a world-sheet gauge field. However to ensure that the action is completely gauge invariant, including large gauge transformations, it was found in  \cite{Hull:2006va}  that a topological term 
 \be
 \cL_{top} = \frac{1}{2} d\tilde{X}_i \wedge dX^i \, 
 \ee 
 must also be included to supplement the minimal coupling.   As well as ensuring gauge invariance this topological term also ensures that the winding contributions from the dual coordinate $\tilde{X}$ can   be gauged away.  This gauged theory can be seen to be equivalent to the standard $\sigma$-model (\ref{standardsigma}) and, upon fixing a gauge, equivalent to the Doubled Formalism. 
 
\EDIT{ This can be extended to the quantum level using a BRST argument given in \cite{Hull:2006va} and   expanded in section 4.6 of \cite{Hull:2009sg}.  Physical states are cohomology classes of the BRST charge operator Q of ghost number zero.  As a consequence of this one has that on physical states $J_{+}^{i} | \psi\rangle_{phys} = 0$ whilst $J_{-}^{i} =0$ can be imposed by a Lagrange multiplier. }

\subsubsection{The Doubled Partition Function}
Let us first consider the one-loop partition function for a single compact boson of radius R and mention how T-duality acts.  \EDIT{We work on the Euclidean theory on a torus worldsheet $\Sigma$ whose modulus we denote by $\tau$. } We include   both single valued and winding contributions in $H^{1}(\Sigma , \mathbb{Z})$ by introducing the field $L = d X + n \a + m \beta$ where $n,m \in \mathbb{Z}$ and $\alpha,\beta$ are the canonical one-cycles of the torus.  The Lagrangian  is given by 
\be
L =  - \pi R^{2 } L \wedge \ast L \, .
\ee    
The partition function then receives both instanton contributions -- meaning sums over $n,m$, the integers corresponding to fields valued in the first cohomology and oscillatory contributions -- meaning integrating over the field, $X$
\be
Z  = Z_{osc} Z_{inst} \ . 
\ee 
The Gaussian integral giving the oscillator contribution may be evaluated using $\zeta$-function regularisation to be
\be
Z_{osc} = \frac{R}{\sqrt{2}} (\det{}'\Box)^{-1/2} = \frac{R}{\sqrt{2 \tau_{2}} |\eta|^{2}} 
\ee
in which the Dedekind $\eta$-function appears. (The prime above the \EDIT{$\det$} denotes omission of the zero mode which is accounted for by the $\tau_2$ factor.) The instanton contribution can be directly evaluated as a sum over the two integers labelling the windings.  After performing a Poisson resummation on one of these integers one finds that
\be\label{eq:Zinst}
Z_{inst} = \frac{\sqrt{\tau_{2}}}{R} \sum_{m, \omega } \exp\left[ \frac{i \pi}{2} \tau p_{L}^{2 } -  \frac{i \pi}{2} \bar \tau p_{R}^{2 }     \right] \ , 
\ee
where $p_{L/R} = R n \pm \frac{\omega}{R}$.   The full partition function is invariant under the T-duality inversion $R \rightarrow  1 /  R $.\footnote{On an arbitrary fixed genus the partition function is not invariant under the radial inversion but acquires an extra factor of  R to a power depending on the genus.   In the Polyakov sum over genera this is accommodated by shift in the  dilaton under T-duality.}   

Now we consider the doubled string  in this context.   Since we are considering a theory with chiral bosons one must exercise care.  The strategy followed in \cite{Berman:2007vi} is based on the approach of Witten in dealing with chiral bosons, known as holomorphic factorisation \cite{Witten:1991mm,Witten:1996hc}.  The rough idea is that to calculate the partition function of a chiral boson one first calculates  that of a non-chiral theory into which the chiral theory can be embedded.  Then one finds that after a variety of manipulations and resummations,  the result may be written as a product of a holomorphic and an anti-holomorphic piece.  One then identifies the holomorphic piece with the chiral   partition function.  

The Lagrangian of this theory was given by  eq.~\eqref{eq:S1doubled} but we now should accommodate winding modes for the dual coordinate by defining $\tilde{L} = d \tilde{X} + \tilde{n} \alpha + \tilde{m} \beta$. In calculating the partition function it becomes vital to include the topological term which in this case reads  
\be
L_{top} =  \pi L \wedge \tilde{L} \ . 
\ee
Although this is a total derivative and does not effect the classical equations of motion, it contributes to the instanton sum and is needed to make the quantum theory equivalent to the un-doubled one.  A careful calculation reveals that instanton sum   factorises as 
\be
Z^{doubled}_{inst} = Z_{f} \times \bar{Z}_{f}
\ee
where $Z_{f}$ is, up to a factor that will cancel with the doubled oscillatory contributions, the instanton part of the partition function for a standard boson given in  eq.~\eqref{eq:Zinst}.   One may now go-ahead and proceed by taking the holomorphic factor and identifying it with the partition function of the doubled string once the chirality condition has been imposed.  In this way one demonstrates the equivalence between the doubled approach and the standard.   The fact that the chiral bosons appear in pairs of opposite chiralities in the doubled approach is essential.  In the above we have skipped over some of the subtleties, for instance the treatment of spin structures, and refer the reader to  \cite{Berman:2007vi} for details. This was generalised to the ${\cal N}=1$ supersymmetric partition function in \cite{Chowdhury:2007ba}.

We should also state that when embedded into a full critical background this equivalence shows that the doubled formalism will be modular invariant and thus represents an important consistency check of the approach. 

\subsubsection{The Doubled $\beta$-functions: Towards a Spacetime Doubled Formalism}  
To conclude this section we take our first steps to a spacetime interpretation of the duality invariant framework.  We recall a fundamental result in string theory that the low energy effective dynamics of strings can be described by (super)gravity.  We consider a string in a curved spacetime described by the non-linear $\sigma$-model
\be
S =  \frac{1}{2 \pi\alpha^\prime} \int d\s^+d\s^-   G_{ij}(X) \partial_+ X^i \partial_- X^j  \ . 
\ee
This theory is not quantum mechanically conformal, the couplings contained in $G$ run and have an associated beta-function which at one-loop is given by
\be
\beta_{ij}^G  \sim \mu\frac{ \partial G }{\partial \mu}  = \alpha^\prime R_{ij} \ . 
\ee
One may think of this equation as saying how geometry changes with scale; the metric undergoes Ricci flow.  In string theory we require that conformal invariance persists at the quantum level and therefore that this beta-function vanishes.  The target space must then be Ricci flat or more generally, once the additional massless fields in the string spectrum are incorporated, be a solution of (super)gravity whose NS sector is given by 
\be
\label{gravNS}
S_{\rm grav}   =   {1\ov 2\kappa^2}\int d^n x  \sqrt{g} e^{-2\Phi}
\left(R + 4 (\del\Phi)^2 -   \frac{1}{12} {H^2} \right)    \ , 
\ee
where $n$ is the appropriate critical dimension of the string theory in question. Classic references explaining this connection include \cite{AlvarezGaume:1981hn,Callan:1985ia}. There are of course other ways to arrive at the same effective action, for instance by considering graviton scattering, but in what follows we find this approach most helpful. 

What then are the effective actions governing the duality symmetric string?   The calculation of the $\beta$-function was performed in \cite{Berman:2007xn,Berman:2007yf} working with the Tseytlin action of eq. (\ref{TseytlinA}).   The calculation is somewhat involved for two main reasons; firstly the lack of manifest world-sheet Lorentz invariance and secondly the lack of an obvious background field expansion that respects the underlying geometry of the doubled formalism.    

The approach is to construct an effective action by  considering  quantum fluctuations about  a classical background.  The one-loop effective action, from which the one-loop $\beta$-functions can be established, is obtained by considering all one-loop diagrams with only classical background fields as external legs.   For want of a  good covariant approach one  may choose to perform a linear splitting defining the quantum fluctuation about a classical background  $\X^I = \X^I_{cl} + \xi^I$.\footnote{Although the original paper \cite{Berman:2007xn} used a  background field method covariant with respect to the metric $\cH$, one may see that it is just as efficient to use this linear splitting. When the classical background is on-shell the one-loop effective actions obtained will be equivalent in either approach.}   The complexities caused by the non-Lorentz invariant structure can be readily seen in the Euclidean two-point functions of the quantum fluctuation which read \cite{Tseytlin:1990va,Berman:2007xn}
\be
\langle \xi^I(z) \xi^J(0) \rangle \approx  \cH^{IJ} \log|z|^2  + \eta^{IJ} i \arg z   \ . 
\ee
When $z\rightarrow 0$, the term proportional to the generalised metric $\cH$ in the above gives a familiar UV divergence which can be evaluated in dimensional regularisation to give a $\frac{1}{\epsilon}$ pole.  The $\beta$-functions for $\cH$ can be read off as the coefficient of this pole. However, the term proportional to $\eta$ shows that the two point function is sensitive to the angle with which we take $z\rightarrow 0$.   It is suggested in  \cite{Tseytlin:1990va} that any angular dependence in the effective action represents a global Lorentz anomaly.  In principle just as demanding a cancellation of the Weyl anomaly constrains the background fields, one should demand that the effective action does not suffer from any angular dependance or global Lorentz anomaly\footnote{Anomaly is perhaps an abusive term here since the classical action is non-manifestly Lorentz invariant in any case.  Nonetheless, the point stands that one should like all this angular dependance to cancel.}.  

  Despite these added complexities, one can indeed carry out the calculation at one-loop as was done in  \cite{Berman:2007xn}.  The first result -- and an important consistency test of the theory -- is that the effective action does not suffer from any Lorentz anomaly; no additional constraints need to be placed on the background for this to be true.  The second, and somewhat related, result is that the $O(d,d)$ invariant inner product $\eta_{IJ}$ does not get renormalised.

The $\beta$-functions for ${\cal H}$, the doubled metric on the fibre, $g_{ab}$, the metric on the base, and $\d$ the doubled dilaton can be established to be
\be
\begin{aligned}
\label{betafinal}
 \beta[\cH_{IJ}] & =  -\frac{1}{2} \hat{\nabla}^2 \cH_{IJ} + \frac{1}{2} \hat{\nabla}_a \cH_{IL} \cH^{LM}\hat{\nabla}^a \H_{MJ} + \hat{\nabla}_a \cH_{IJ} \hat{\nabla}^a d  \\ 
 \beta[g_{ab}] & =  \hat{R}_{ab} + \frac{1}{8}\mbox{tr}  (\hat{\nabla}_b \H^{-1} \hat{\nabla}_a \H ) + 2 \hat{\nabla}_a \hat{\nabla}_b d 
  \\ \beta[d] & =  -\frac{\alp}{2} \left( \hat{\nabla}^2 d -  (\hat{\nabla}d)^2 + \frac{1}{2} \hat{R} + \frac{1}{16}  \mbox{tr}  (\hat{\nabla}_b \H^{-1} \hat{\nabla}^a\H ) \right)
  \end{aligned}
\ee
where $\hat{\nabla}$ are covariant with respect to the metric on the base whose curvature is $\hat{R}$.  The vanishing of these $\beta$-functions are the equations of motion of the following gravity theory 
\be
\label{effa}
S_{26-d}= \frac{ vol(T^d)}{2 \kappa^2}  \int d^{26-d}y \sqrt{-g} e^{-2d}\bigl\{ \hat{R}(g) +4 \left(\hat{\nabla} d \right)^2 
 + \frac{1}{8}\mbox{tr} \left(\eta \hat{\nabla}_a \H \eta \hat{\nabla}^a \H\right) \bigr\}\, .
\ee
which may be obtained by dimensional reduction of eq. (\ref{gravNS}).    This result makes clear now the linkage between the doubled formalism and the spacetime theory and helps establish the full quantum validity of the doubled formalism.    However it is still somewhat limited in that we assumed from the outset no dependence on the internal coordinates \EDIT{(that is $\cH= \cH(Y)$ depends only on the coordinates of the base).}  

One might wonder what the situation is for the general theories of interacting chiral bosons described in section \ref{doubledeverything} in which all coordinates are doubled.  For the case that ${\cal H}$ obeys the strong constraint and depends only on half of the coordinates   one obtains the following one-loop $\beta$-function \cite{Copland:2011wx,Copland:2011yh}
\be
\beta[{\cal H}]_{IJ} = {\cal R}_{MN} 
\ee
where ${\cal R}_{MN}  = \frac{1}{2} \left( {\cal K}_{MN} -  \cH_M{}^P {\cal K}_{PQ} \cH^{Q}{}_N  \right)$ and 
\begin{align}
\label{kis}
\begin{split}  
{\cal K} _{MN}={}&\frac{1}{8}\, \partial_{M}{\cal H}^{KL}
  \,\partial_{N}{\cal H}_{KL}
  -\frac{1}{4}(\partial_L - 2 (\partial_L d) ) 
  ({\cal H}^{LK} \partial_K {\cal H}_{MN})
 \\&+2 \,\partial_{M}\partial_N d\,  -\frac{1}{2} \partial_{(M}{\cal H}^{KL}\,\partial_{L}
  {\cal H}_{N)K}
  \\
  &+ \frac{1}{2} (\partial_L - 2 (\partial_L d) )  \bigl({\cal H}^{KL} \partial_{(M}
   {\cal H}_{N)K}
  + {\cal H}^K{}_{(M}  \partial_K {\cal H}^L{}_{N)}  \bigr) \,.
   \end{split}
\end{align}
 Understanding the spacetime interpretation of this result will be the topic of the next section -- but the vanishing of this $\beta$-function will be equivalent to the equations of motion of Double Field Theory.
 
For the case in which the weak constraint is violated by means of an underlying group structure as in eq. (\ref{twisted1})-(\ref{doubleWZW}) one obtains \cite{Sfetsos:2009vt,Avramis:2009xi}

\be
\beta[{\cal H}]_{AB} =\frac{1}{4}
(\cH_{AC}\cH_{BF}-\eta_{AC}\eta_{BF})
(\cH^{KD}\cH^{HE}-\eta^{KD}\eta^{HE})
f_{KH}{}^Cf_{DE}{}^F\ ,
\ee
in which the $f_{AB}{}^C$ are the structure constants of the group.   This result has two interpretations.  On  the one hand it shows how Poisson-Lie or non-Abelian T-duality  holds at one-loop (for instance this equation encapsulates the equivalent running of both of the pair of T-dual $\sigma$-models)  \cite{Sfetsos:2009vt}.  A second remarkable interpretation is the following  \cite{Avramis:2009xi}:   the vanishing of this $\beta$-function can be understood as the equations of motion for the scalars in electric gaugings of four-dimensional ${\cal N}=4$ supergravities.   It is known that not all such supergravities can be given a conventional geometric higher dimensional origin.   However in the doubled formalism, the Scherk-Schwarz twisted doubled torus provides the relevant group manifold to support the gauge algebra of {\em all} these gauged supergravities. The embedding tensor that defines the supergravity theories becomes a purely geometric flux in the doubled space \cite{DallAgata:2007fk}.   What is perhaps all the more surprising about this result is that, given we started with a purely bosonic construction,   it has anything to do with supersymmetry at all.  It appears that secretly the doubled approach already knows something of supersymmetry.   

We shall see more of this deep connection between gauged supergravities and the doubled formalism in the sequel.   
  
\section{The Spacetime T-duality Invariant Theory: Double Field Theory } \label{sec:DFT}

We have seen that strings on $T^d$ toriodal backgrounds exhibit an $O(d,d , \mathbb{Z})$ T-duality   and that this could be promoted to a manifest symmetry on the worldsheet in the doubled formalism. This introduced $d$ extra coordinates, extending the fibration to a $T^{2d}$.  In the case that the metric on the doubled fibre ${\cal H}$ only depended on the base coordinates we saw that the vanishing of the $\beta$-functions corresponded to the equations of motion of toriodally compactificed gravity.    A generalisation, that dispenses with the base/fibre distinction, in which everything is doubled is possible providing that a supplementary constraints on ${\cal H}$ are  imposed.  We now describe the spacetime interpretation of this. 

The goal of this section is to present a manifestly $O(d,d  )$ invariant  theory for the metric  ${\cal H}_{IJ}$  on an extended spacetime with coordinates $\mathbb{X}^{I}= (x^i, \tilde{x}_j)$ that contains as a subcase the standard NS sector of supergravity.     Such a proposal was recently made by Hull and Zwiebach \cite{Hull:2009mi} dubbed "Doubled Field Theory" (DFT) though has  older heritage in the pioneering  works of Siegel \cite{Siegel:1993th,Siegel:1993xq} and Tseytlin before that \cite{Tseytlin:1990va}.  The approach of Hull and Zwiebach  was motivated by considering string field theory on a toroidal background \cite{Kugo:1992md}.   String field theory treats momenta and winding symmetrically and consequentially the components of the string field depend on {\emph{both}}  momenta and winding numbers.  Whilst Fourier transforming the momenta gives components with dependance on regular coordinates (the $x^i$), Fourier transforming the winding numbers  gives dependance on the $\tilde{x}_i$ coordinates of an extended spacetime.   This represents an example of geometrising charges, i.e.  the replacement of a charge associated to an extended object, in this case string winding, with extra dimensions of spacetime.  

For a generic perturbative string state given by
\be
 | \Psi \rangle = \sum_{I}  \int dk \sum_{p, \, \omega}    \phi_I ( k, p_i , \omega^i)  {\cal O}^I | k, p, \omega \rangle 
\ee
the component fields have Fourier transforms $\phi_I ( y^\mu, x^i , \tilde{x}_i)$ that depend on the coordinates of the base (the $y^\mu$)  and the coordinates of the doubled torus.  Physical states must obey the level matching condition
\be
(L_0 - \bar{L}_0)  | \Psi \rangle  = (N - \bar{N} + p_i \omega^i ) | \Psi \rangle = 0 \ , 
\ee
which for a component field $\phi_I$ after Fourier transformation to position space reads
\be\label{eq:levelmatch}
 (N_I - \bar{N}_I) \phi_I   =   -  \partial_i \tilde{\partial}^i  \phi_I   \ .
 \ee
 The free field equation can be set up as a BRST cohomology  $Q   | \Psi \rangle =0 $  modulo gauge transformations $\delta  | \Psi \rangle  = Q | \Lambda \rangle$.  Then just as the string field depends on both $x$ and $\tilde x$, the gauge parameters will depend on the coordinates of the doubled torus. 

Of course, the full closed string field theory is a huge challenge and the work of Hull and Zwiebach focuses on a tractable truncation to the massless sector and in particular to the $N= \bar{N} = 1$ sector. One needs to be rather careful about what `massless' means in this context;  in \cite{Hull:2009mi} the action is restricted to the massless fields of the decompactified theory rather than the compactified theory, which are not the same.  Moreover, as explained in the introduction,  one should not think of DFT as an `effective theory'.      Even this truncation might seem ambitious; one must still work order by order in fields and it is not at all obvious from the outset how background independence should arise in this approach.      

Thus in this truncation there are three component fields
\be\label{eq:dftfields}
h_{ij} ( x^\mu, x^i , \tilde{x}_i) \ , \quad  b_{ij} ( x^\mu, x^i , \tilde{x}_i) \ , \quad d  ( x^\mu, x^i , \tilde{x}_i) \ . 
\ee 
By virtue of the level matching condition eq.~\eqref{eq:levelmatch} the coordinate dependence of these fields must be constrained; the Kleinian operator of the doubled torus must annihilate them.  This constraint is known as the {\it weak} constraint.    

The aim of Double Field Theory is to provide a spacetime action to describe the dynamics of these fields.   Since this action should accommodate dependance of both the regular coordinates and the T-dual coordinates it must necessarily be rather novel.  This action must also possess a rather novel gauge symmetry since the gauge parameters depend on the dual coordinates also.  This can already be seen at the linearised level where in addition to the regular diffeomorphism symmetry 
\be
\delta h_{ij} = \partial_i \epsilon_j  +  \partial_j \epsilon_i \ , 
\ee
there are also "dual diffeomorphisms" of the form 
\be
\tilde\delta h_{ij} = \tilde\partial_i \tilde\epsilon_j  +  \tilde\partial_j \tilde\epsilon_i \ . 
\ee
\EDIT{In order to close the gauge algebra in the correct manner and to ensure that the action is gauge invariant the {\it weak} constraint is not sufficient.  Instead the {\it strong}   constraint (also called, interchangeably, the {\em physical section condition}) must be enforced \cite{Hull:2009zb}.  This says that in addition to the Kleinian annihilating fields, it must also annihilate products of fields (and gauge parameters). }

The first works of Hull and Zwiebach \cite{Hull:2009mi,Hull:2009zb} involved  a   {\em{tour-de-force}} calculation in string field theory to obtain first quadratic and cubic actions of these field defined in eq.~\eqref{eq:dftfields}, and demonstrated that the resultant action is  gauge invariant and also  T-duality invariant.   Shortly after the initial work, Hohm, Hull and Zwiebach \cite{Hohm:2010jy} extended this work to present first a background independent formalism (i.e. for the full metric and two-form and not just the fluctuation around a given background) and  then in \cite{Hohm:2010pp} a formulation involving the $O(d,d)$ coset representative ${\cal H}$.   It is this later formulation that makes the most contact with the world-sheet approach described previous and we shall focus on this.   Rather than replicate the laborious calculational details of  these papers required to arrive at this answer, we shall instead give a presentation of the results and place more emphasis on the gauge symmetries and the relation to generalised geometry \cite{Hitchin:2004ut,Gualtieri:2003dx}.   We begin with a shorter primer on generalised geometry that will help inform the remainder of this section. 
 
\subsection{Gauge Symmetry of Strings} 
To perform a T-duality via the Buscher procedure a requirement is the existence of an isometry as detailed in the very first equation of the body of this report eq.~\eqref{eq:isometry}.  It is important to keep in mind that it is only the NS three-form $H=db$ that needs to be invariant under the action of the Killing vector generating the isometry $L_X H=0$ and  in general the two-form potential $b$ need not be. However locally one can make a gauge transformation $b' = b + d\zeta'$ such that $L_X b' = 0$.   Then the condition that one can do a T-dualisation is that  
\be\label{eq:genkillingcond}
L_X g = 0 \ , \qquad L_{X} b   - d\zeta   = 0  \ , 
\ee 
where the one-form $\zeta$ can be related to the above gauge transformation  by $\zeta = - \iota_X  d\zeta' + d  f$.   This highlights that for T-duality one should consider the action of not just a vector but also of a one-form and indeed these are precisely the parameters that generate diffeomorphisms and gauge transformations of the NS sector of supergravity
\be
\delta_{X+\xi}  g = L_X g \ , \quad \delta_{X+ \xi} b = L_X b + d\xi \ . 
\ee
So we see the gauge parameters are sections of the bundle $E= TM \oplus T^\ast M$ which will play a prominent role in what follows.  Now let us consider the commutator of two successive gauge transformations.  The  action on the metric is trivial, it is just the Lie derivative taken along the commutator of the two vector fields.  Acting on the two-form is less trivial due to the mixing between the diffeomorphism and the two-form gauge transformations.   One finds that the closure of two gauge variations may be written (up to an overall exact piece that may be ignored)
\be\label{eq:comm}
[ \delta_{X+\xi}, \delta_{Y+ \eta}  ]  = \delta_{ [X,Y] +  L_X \eta - L_Y \xi } \ . 
\ee
One would like to interpret the  right hand side of this as the gauge variation along the "bracket"  of the two gauge parameters: 
\be
[ \delta_{X+\xi}, \delta_{Y+ \eta}  ]  = \delta_{  [[ X+\xi ,  Y+ \eta   ]]  } \ . 
\ee 
   Of course there is some ambiguity here since one may augment the right hand side of eq.~\eqref{eq:comm} with the action of an arbitrary exact form which generates a trivial (zero) transformation.   This ambiguity allows the choice of bracket 
\be
[[ \delta_{X+\xi}, \delta_{Y+ \eta}   ]]  \equiv [X,Y] +  L_X \eta - L_Y \xi  - \frac{1}{2} d \left( \iota_X \eta - \iota_Y \xi  \right) \ . 
\ee
This bracket is known as the Courant bracket and will play a key role in what follows.

So in summary we see that both the bundle $E= TM \oplus T^\ast M$ and the Courant bracket arise naturally and this motivates the consideration of generalised geometry. 

\subsection{A Generalised Geometry Primer}
\label{secgengeom}
In the mid 2000's Hitchin \cite{Hitchin:2004ut} and Gualtieri \cite{Gualtieri:2003dx} proposed the study of structures on the bundle 
\be
E= TM \oplus T^\ast M
\ee
whose sections consist of the formal sum of vectors and one-forms on an $n$-dimensional manifold $M$.  Although the approach we take later is different,  since we actually double the dimensionality of the manifold itself  by introducing extra coordinates, we shall find generalised geometry to be a powerful guiding tool.  The principal reason why this generalised geometry construction is relevant is that $E$ comes equipped with a natural $O(n,n)$ product given by
\be
\langle  X+ \xi, Y + \eta \rangle = \frac{1}{2} (\iota_X \eta + \iota_Y \xi ) \ . 
\ee
Indeed, things may become even clearer by working in local coordinates such that $X = X^i \partial_i $ and $\xi = \xi_i dx^i$ etc.  so that 
\be
\langle  X+ \xi, Y + \eta \rangle = \frac{1}{2} (X^i \eta_i + Y^i \xi_i ) \ . 
\ee
Or even more suggestively if we combine the components in to  $\X^I = (X^i , \xi_i)$ we have 
\def\Y{{\mathbb{Y}}}
\be
\langle  \X, \Y  \rangle =  \eta_{IJ} \X^I \Y^J  \ , 
\ee
in which $\eta$ is the $O(n,n)$ invariant metric introduced earlier. 
\subsubsection{Inner Product, Dorfman and Courant}
We may introduce a derivative on sections of this bundle know as the Dorfman derivative 
\be
\label{Dorf}
\X \bullet \Y \equiv (X+\xi) \bullet (Y + \eta) = L_X (Y+\eta ) - \iota_Y d\xi \ . 
\ee
A useful property is that symmetrisation of the Dorfman obeys 
\be
((\X , \Y  )) \equiv (( X+\xi, Y + \eta   )) = \frac{1}{2} \left(  (X+\xi) \bullet (Y + \eta)  + (Y+\eta) \bullet (X + \xi)  \right) = d ( \langle  X+ \xi, Y + \eta \rangle     ) \ . 
\ee
Even more important is the antisymmetrisation 
\ba
\label{Courant}
[[\X , \Y  ]] \equiv [[ X+\xi, Y + \eta   ]]  &=& \frac{1}{2} \left(  (X+\xi) \bullet (Y + \eta)  - (Y+\eta) \bullet (X + \xi)  \right)  \nonumber \\
&=& [X,Y] + L_X\eta - L_Y \xi - \frac{1}{2}  d ( \iota_X \eta - \iota_Y \xi ) \ . 
\ea
This skew product is known as the Courant bracket and plays the role of the Lie bracket in generalised geometry. 

As well as the obvious $Diff(M)$ symmetry the Courant bracket is also preserved by a further subgroup of $O(n,n)$ isomorphic to closed two-forms $\Omega_{cl}^2(M)$ which acts by 
\be
[[ {\cal O}_\omega \X  , {\cal O}_\omega \Y]] = {\cal O}_\omega [[ \X, \Y]] \ ,
\ee
 where
 \be
 {\cal O}_\omega : \X = X+ \xi \rightarrow X+ \xi + \iota_X \omega \ . 
 \ee 
 The full symmetries of the Courant bracket are then $Diff(M)\ltimes \Omega_{cl}^2(M)$ \ .   
 
 However, unlike the Lie bracket, this Courant bracket does not obey the Jacobi identity but it fails to do so in very particular way
 \be\label{eq:JacAnom}
 Jac (\X, \Y, \mathbb{Z} ) =  d ( Nij (\X, \Y, \mathbb{Z} )  )  \ , 
 \ee 
 where $Nij$ is the Nijenhuis operator 
 \be
3 Nij (\X, \Y, \mathbb{Z} )  =    \langle  [[\X , \Y ]] , \mathbb{Z} \rangle  + \mathrm{cyclic} \ .
\ee

\EDIT{ Let us introduce some useful concepts at this point: isotropics, involutivity and Dirac structures. 
 A subspace on which the inner product vanishes (i.e. one consisting of mutually orthogonal vectors) is known as an {\em isotropic} and if   such a space  has the maximal dimension, which is $n$, then it is said to be a {\em maximal isotropic}.  A subspace is said to be {\em involutive} if it is closed under the Courant bracket (that is  the  bracket of any two vectors in the subspace gives rise to a third vector that is also a member of the subspace).   With these definitions it is clear that on a subspace $L$ of the bundle $E= T M \oplus T^\ast M$ that is both involutive and isotropic,  the anomalous term in the Jacobi identity in eq.~\eqref{eq:JacAnom}  vanishes.  If  such an $L$ has maximal dimension, which is $n$, then  it is known as a {\em Dirac structure}.   A rather trivial but important example of a Dirac structure is just $TM \subset TM \oplus T^\ast M$.}

\subsubsection{Pure Spinors and Maximal Isotropics}  
From the above consideration it is clearly important to understand how to \EDIT{construct} isotropics and to do so let us first consider spinors in generalised geometry.   A spinor in generalised geometry is associated with a {\em polyform} $\lambda \in \Lambda^\bullet T^\ast M$ \EDIT{  (where $\Lambda^\bullet T^\ast M$ means  $\bigoplus_{i=1}^n \Lambda^i T^\ast M$ i.e. the formal sum of forms of different degrees).   This idea has already been introduced when discussing the action of T-duality on RR fields in section 2.1. Here we develop it a little more precisely. }  The action of a generalised vector on a polyform is given by 
\be
\label{cliffact}
(X+ \xi )\circ \lambda = (\iota_X + \xi \wedge) \lambda  
\ee
where $\iota_X$ denotes the interior product (contraction) of $\lambda$ with the vector $X$.  This action 
provides a representation of the Clifford algebra $CL(n,n)$ 
\be
(X+ \xi )^2 \circ \lambda  = \langle  X  + \xi , X + \xi \rangle  \lambda  \ . 
\ee
Thus  polyforms  provide a module for the Clifford algebra and they can then be considered as equivalent to spinors.  The notion of chirality of the spinor is determined by whether the polyform is the sum of even or of odd rank forms (which we denote as $\Lambda^{E/O} T^\ast $ respectively).  To be more precise, by considering how spinors transform under the $GL(n)$ subgroup the detailed isomorphism between spinors and polyforms is defined by
\be
\label{iso}
S^\pm \cong   \Lambda^{E/O} T^\ast M \times  | det T^\ast M |^{-\frac{1}{2}} \ , 
\ee
\EDIT{where $det T^\ast M$ refers to the line bundle obtained by taking the top exterior product $\Lambda^n T^\ast M$ and its presence reflects that fact that the isomorphism between polyforms and spinors depends on a choice of volume form.\footnote{Another way to understand   the need for this factor   is that the natural inner product on polyforms, known as the Mukai pairing, maps two polyforms to a top-form rather than a scalar, further discussion of this point can be found e.g. in section 3.3 of the review article  \cite{Koerber:2010bx}.}  }

To any spinor $\lambda$ one can associate an annihilator $L_\lambda$ given by 
\be
L_\lambda = \{ \X \in  E  | \X\circ \lambda = 0  \} \ . 
\ee 
One can immediately see, from the definition of the Clifford action eq. (\ref{cliffact}),   that this annihilator is an isotropic since
\be
0 =  \X \circ (\Y \circ \lambda) +    \Y \circ (\X \circ \lambda) = \langle \X, \Y \rangle  \lambda
\ee
for all $\X, \Y \in L_\lambda$.   If  $L_\lambda$ has dimension $n$ then it is a maximal isotropic and the corresponding spinor is said to be pure. This definition corresponds to the pure spinors encountered for instance in the covariant pure spinor approach to quantisation of the superstring.\footnote{To a physicist the following definition in $2n$  dimensions of a pure spinor of $Spin(2n)$ might be more familiar:
\beq
\lambda^\a \g_{\a\b}^{m_1\dots m_j} \lambda^\b = 0 \ , \quad  0 \leq j < n \ . 
\eeq
This implies that
\beq
\lambda^\a \lambda^\b \propto \g^{\a \b}_{m_1\dots m_n}  ( \lambda \g^{m_1\dots m_n} \lambda ) 
\eeq 
and so $\lambda \g^{m_1\dots m_n} \lambda$ defines a $n$ dimensional plane.  Here the spinors are chiral ($\a = 1 \dots 2^{(n-1)}$) and the $\gamma$ can be thought of as the off diagonal blocks of Dirac gamma matrices in the Weyl basis.  In Euclidean signature these spinors may be complex however in split signature they are rendered real. }   \EDIT{Above we introduced Dirac structures as maximal isotropics that are involutive under the Courant bracket.  Since  for $\X, \Y \in L_\lambda$ 
\be
[[\X, \Y]]\circ\lambda = \X\circ\Y\circ d\lambda , 
\ee
demanding that the annihilator $L_\lambda$ is inovlutive translates to a constraint on the pure spinor
\be
d\lambda = \X \circ \lambda 
\ee
 for some suitable $\X$  (a proof can be found in theorem 3.38 of Gualtieri's thesis \cite{Gualtieri:2003dx}).    }

\subsection{T-duality and Generalised Geometry} 
Let us now comment on how T-duality can be understood more precisely in the language of generalised geometry following \cite{Grana:2008yw}.   Generalised geometry is evidently well adapted to T-duality and the $O(d,d)$ coset representative $\cal H$ defined in eq.~\eqref{eq:Hdef} provides a second inner product (in addition to $\eta$) with which one can contract two sections of $E= TM \oplus T^\ast M$ to form a scalar $\cal H(\X , \Y)$.  It is in this sense that we sometimes refer to $\cal H$ as a generalised metric.   One can extend the definition of the Dorfman derivative acting on $\cal H$ in a coordinate free way according to 
\be
(\mathbb{Z} \bullet \cH)( \X, \Y) = (\mathbb{Z} \bullet \cH( \X, \Y) ) -  \cH( \mathbb{Z} \bullet \X, \Y) -  \cH(  \X, \mathbb{Z} \bullet \Y)
\ee
In this formulation the criteria for a T-duality eq.~\eqref{eq:genkillingcond} around a vector and one-form $ \X =X+ \xi$  may be expressed as the vanishing of the Dorfman derivative of the generalised metric
\be
 \mathbb{X} \bullet \cH  = 0  \ . 
\ee 
In this sense the pair $ \X =X+ \xi$  can be thought of as defining a generalised Killing vector for the generalised metric.   From this vector one can build an 
 $O(d,d)$ element \cite{Grana:2008yw} \be{\cal O}_{\X} = \mathbb{1} - 2 \X \otimes \X^T\eta  \, .\ee     Then the action of T-duality is simply to conjugate with this element $\H \rightarrow{\cal O}_{\X}^T \H {\cal O}_{\X} $.  In coordinates adapted to the isometry and  with a judicious choice of basis, one can express $\X = \hat{e}_i   + e^i$ where $\hat{e}_i$ is the vector dual to the frame field $e^i$  and in which case  the corresponding ${\cal O}_{\X}$ is   of form given in eq.~\eqref{eq:factorisedduality} (sometimes called factorised duality).

In fact the connections between generalised geometry and supergravity run far deeper than this.  One important and beautiful line of work led by Waldram and collaborators has been to harness generalised geometry to reformulate supergravity \cite{Coimbra:2011nw,Coimbra:2012yy}.  In these approaches we emphasise that spacetime itself remains completely conventional, it is only the tangent bundle that is extended. 

\subsection{ Extending Spacetime}
Now we take a leap of faith; we extend not just the tangent bundle but spacetime itself by introducing extra coordinates.  That is to say all quantities now depend on 2d  coordinates $ \X^I = (x^i  , \tilde{x}_i) $ of an extended spacetime. We shall equally double the gauge transformations to include `dual' diffeomorphisms and dual two-form gauge transformations.   

Let us consider a generalised gauge parameter and derivatives in the extended spacetime
\be
\xi^M = \left( \begin{array}{c} \xi^i \\  \tilde{\xi}_i  \end{array} \right) \ , \quad \partial_M = \left( \begin{array}{c} \partial_i  \\  \tilde{\partial}^i  \end{array} \right)  \ . 
\ee  
With these we may define a generalised Dorfman derivative, or D-derivative, 
\be\label{eq:genL1}
{ {\cal L}_\xi A_M = \xi^P \partial_P A_M + ( \partial_M \xi^P - \partial^P \xi_M) A_P \  .}
\ee
In this derivative the $O(d,d)$ invariant metric $\eta$ enters explicitly when indices are raised and lowered out of their natural positions. We will demand that all fields obey the strong constraint including the generalised vector field $\xi$ that generates the transformations.

The $O(d,d)$ invariant metric itself is preserved by this derivative (once one uses the strong constraint)
\be
{\cal L}_\xi \eta_{IJ} = 0  \ . 
\ee
Because of this structure, the derivative actually is reducible in the sense that for $\xi^M = \partial^M \chi$ 
\be
{\cal L}_{\partial \chi }  \equiv  0 \ . 
\ee

From this D-derivative we may define a generalised Courant bracket, which we shall call a C-bracket, by
\be
[[ \xi_1 , \xi_2]]_C  = \frac{1}{2} \left( {\cal L}_{\xi_1} \xi_2 -   {\cal L}_{\xi_2} \xi_1 \right)
\ee

Let us now investigate the closure of the D-derivative.  A short calculation reveals that
\be
[{\cal L}_{\xi_1} , {\cal L}_{\xi_2} ] V^M = {\cal L}_{ [[ \xi_1 , \xi_2]]_C} V^M - F^M(\xi_1, \xi_2 , V) \ , 
\ee
where
\be
F^M  = \xi_1^Q \partial^P \xi_{2Q} \partial_P V^M + 2 \partial_P \xi_{1Q} \partial^P \xi^M_2 V^Q  - \xi_1 \leftrightarrow \xi_2 \ . 
\ee
\EDIT{ Notice that the derivatives entering this anomalous contribution $F^{M}$ have their indices contracted together but act  on different fields.  By demanding that $\eta^{IJ} \partial_{I} \bullet \partial_{J} \bullet \equiv 0$, where the bullets denote any two fields or gauge parameters of the theory, this term can be made to vanish.  This is the {\em strong} constraint.  
Note that although {\it a priori} the dependance on the coordinates is subject only to the {\it weak} constraint, when the strong constraint is imposed then the commutator of D-derivatives closes onto the C-bracket.   When the strong constraint is solved by setting $\tilde{\partial}_i\equiv 0$, then the D-derivative and C-bracket reduce to the Dorfman derivative and Courant bracket respectively.   }

\subsubsection{Group Structure of the Generalised Derivative}
Note that if we define generators of $O(d,d)$ in the fundamental as 
\beq
(T_{IJ})^{M}_P = \delta_I^M \eta_{JP} - \delta_J^M \eta_{IP} 
\eeq
then we may recast the derivative as 
\beq
\hat{{\cal L}}_\xi V^M = \xi^P \pl_P V^M  + \frac{1}{2} \eta^{IK} \eta^{JL} (T_{IJ})^{M}_Q (T_{KL})^N_P  \partial_N \xi^P V^Q \ . 
\eeq
Since the Killing form is $  \k_{IJ , KL} \propto \eta_{I[K} \eta_{L]J}$ we can view the second term in the above as being the adjoint projection acting on the indices $N,P$ (see \cite{Coimbra:2011nw}). 

With a mind to later M-theory generalisations,  let us be even more systematic.  We consider the most general form of a derivative
\beq\label{eq:genL2}
{\cal L}_U V^M = L_U V^M + Y^{MN}{}_{PQ} \pl_N U^P V^Q
\eeq
where $L_U$ is the standard Lie derivative and $Y^{MN}{}_{PQ}$ is an invariant tensor of $O(d,d)$. We may build it out of the projectors (acting on the downstairs indices i.e. acting on the product of two fundamentals),
\be 
(P_{1})^{MN}{}_{PQ} =  \frac{1}{D} \eta_{PQ} \eta^{MN}  \ ,  \quad 
(P_{ST })^{MN}{}_{PQ} = \delta^M_{(P} \delta^N_{Q)}  - \frac{1}{D} \eta_{PQ} \eta^{MN}  \ , \quad
(P_{A})^{MN}{}_{PQ} =  \delta^M_{[P} \delta^N_{Q]}   \ . 
\ee 
Then the general choice can be written as 
\beq
Y^{MN}{}_{PQ}  = a_1 (P_{1})^{MN}{}_{PQ} + a_{ST} (P_{ST })^{MN}{}_{PQ} + a_{A}(P_{A})^{MN}{}_{PQ}  \, .
\eeq
Then the algebra closes providing that
\bea
Y^{MN}{}_{PQ} \pl_M \otimes \pl_N = 0 \ , \\
 (Y^{MN}{}_{TQ}Y^{TP}{}_{RS}-Y^{MN}{}_{RS} \delta^P_Q ) \pl_{(N} \otimes \pl_{M)} = 0  \ . 
\eea
Now the strong constraint (or section condition) says that to satisfy the first we must set $a_{ST} = a_A =0$ and then the second just gives $a_1 =D$.   This is the systematic way of establishing the structure of generalised derivatives. 

Instead one could decompose into projectors on the upstairs and downstair pair of indices (i.e. on the product of a fundamental and anti-fundamental) and one finds 
\beq\label{eq:Yadjoint}
Y^{MN}{}_{PQ} = - 2 (P_{adj})^M{}_Q{}^N{}_P +  \delta^M_P \delta^N_Q \ , 
\eeq
where the adjoint projector is 
\beq
 (P_{adj})^M{}_Q{}^N{}_P = -\frac{1}{4} \eta^{IK}\eta^{JL} (T_{IJ})^M_Q  (T_{KL})^N_P \ .  
 \eeq
 The second term in eq. \eqref{eq:Yadjoint} cancels with the same contribution coming from the standard Lie derivative in eq. \eqref{eq:genL2} leaving the result of  eq. \eqref{eq:genL1}.

 \subsection{An Action for DFT} 
 Having embraced an extended spacetime and proposed a generalised gauge structure let us now cut to the chase and present the key result of DFT which is  a manifestly $O(d,d)$ invariant action for the NS sector fields 
  \be
 \cH_{IJ}(\X ) = \left(\begin{array}{cc} (G - BG^{-1}B)_{ij}  & (BG^{-1})_{i}^{\ j} \\ - (G^{-1}B)^{i}_{\ j} & G^{ij} \end{array}\right) \, , \quad  e^{-2d(\X)}  = \sqrt{|G|} e^{-2\phi}  \ ,  
\ee
 where all fields are subject to the {\it weak} and {\it strong} constraints.  This action should obey the gauge symmetry generated by the D-derivative
\be
\delta_\xi \cH = {\cal L} _\xi \cH \ , \quad \delta_\xi d  =  {\cal L} _\xi d = \xi^M \partial_M d - \frac{1}{2} \partial_M \xi^M \   .
\ee
 The derivation from first principles of such an action is of course lengthy and here let us just present the result  \cite{Hohm:2010pp}:
  \be
\label{DFT}
{ S_{DFT}= \int dx d\tilde{x} e^{-2 d} \left(\frac{1}{8} \H^{ M N } \partial_{M} \H^{KL} \partial_{N} \H_{KL} - \frac{1}{2} \H^{M N}\partial_{M}\H^{KL} \partial_{L} \H_{KN} -2 \partial_{M} d \partial_{N} \H^{MN}
+ 4\H^{MN} \partial_{M} d \partial_{N} d  \right)\ . } 
\ee  
When setting $\tilde{\partial}=0$ one finds, after integration by parts, that this action then reduces to the standard action for the common NS sector of supergravity: 
\be
S_{DFT}  \xrightarrow{\tilde{\partial}=0} \int dx \sqrt{G}  e^{-2\Phi}
\left(R + 4 (\del\Phi)^2 -\frac{1}{12}{H^2} \right)  \ . 
\ee

To gain intuition for this action, consider  the case that the NS two-form and (standard) dilaton are turned off one may see that this Double Field Theory action gives
\be
S_{DFT}  =   \int dx d\tilde{x} \left[    R(G, \partial) + R(G^{-1}, \tilde\partial)  \right] \ , 
\ee
which clearly displays an invariance under the T-duality inversion $G \leftrightarrow \tilde{G} = G^{-1}$, $\partial \leftrightarrow \tilde\partial$.  Indeed exactly this result was obtained in the work of Tseytlin \cite{Tseytlin:1990va}.  The introduction of the NS field will serve to produce terms with mixing $x$ and $\tilde x$ derivatives. 

\subsection{Towards a Geometry for DFT}
\label{towardsgeom}
One should like very much to understand the action for DFT of eq.~(\ref{DFT}) in a more geometric manner.  We may cast the action in an Einstein-Hilbert like form 
\be
\label{DFTEinHil}
S_{DFT} = \int dx d\tilde{x} e^{-2 d} {\cal R} 
\ee 
where ${\cal R}$ is an $O(d,d)$ scalar and a gauge scalar and so it is natural to view it as  curvature scalar.  In a similar fashion, upon variation with respect to $\cH_{MN}$ one finds a candidate curvature tensor ${\cal R}_{MN}$.  One is then prompted to seek a geometrical construction of these objects based on the introduction of an appropriate connection {\it{ \`a la}} general relativity.    In fact a frame like formulation with a local $GL(d)\times GL(d)$ symmetry dates back to the pioneering work of Siegel \cite{Siegel:1993th,Siegel:1993xq}.   A number of recent works \cite{Hohm:2010xe,Hohm:2012mf,Hohm:2011si,Jeon:2011cn,Jeon:2010rw,Geissbuhler:2011mx,Berman:2013uda} have further developed the geometrical concepts predominantly in a Christoffel-like approach. 
 
The basic idea is to introduce an appropriate connection to covariantise all derivatives.  That is to say,  for a vector transforming under the gauge symmetry 
\be
\delta_\xi V_M  = {\cal L}_\xi V_M \ , 
\ee
one introduces the connection
\be
\nabla_N V_M = \partial_N V_M - \Gamma_{NM}{}^K V_K \ ,
\ee
transforming such that 
\be
\delta_\xi  (\nabla_N V_M) = {\cal L}_\xi   (\nabla_N V_M)    \ . 
\ee
One encounters a departure from conventional GR already at this stage, the connection $ \Gamma_{NM}{}^K $ cannot be chosen to be symmetric in its lower indices. One can proceed in a conventional way and try to define a curvature and torsion through the commutator of this connection:
\be
[\nabla_M \nabla_N] V_K = - R_{MNK}{}^L V_L - T_{MN}{}^L \nabla_L A_K \ . 
\ee
However neither of these objects are \EDIT{tensorial} under the gauge symmetry.  One may easily, albeit in an {\it { ad hoc}} manner, construct a tensorial four index object
\be
{\cal R}_{MNKL} = R_{MNKL}+ R_{KLMN} + \Gamma_{QMN} \Gamma^Q{}_{KL}  \ ,
\ee
in which indices have been raised an lowered with $\eta$.  For the torsion, a covariant definition is obtained by considering the change in the action of the generalised Lie derivative when all partial derivatives are replaced by covariant derivatives.  The resultant covariant generalised torsion is given by
\be
{\cal T}_{MNP} = T_{MNP} + \Gamma_{MNP} \ . 
\ee
One would now like to impose some constraints to determine the connection in terms of the physical fields $\cH$ and $d$.  The first constraint -- which as we shall see shortly need not be enforced --  is to impose the covariant torsion ${\cal T}_{MNP}$ vanishes.  A second constraint is to impose that the connection is metric compatible with {\it{both}} $\cH$ and the $O(d,d)$ invariant inner product $\eta$
\be
\nabla_{M} \cH_{NP}  = 0 \ , \qquad \nabla_{M} \eta_{NP}  = 0  \ . 
\ee
A final condition is that covariant derivative respects the presence of the dilaton factor in the integration measure of eq.~(\ref{DFTEinHil}) in the sense that 
\be
\int dx d\tilde{x} e^{-2 d}  V \nabla_M V^M = - \int dx d\tilde{x} e^{-2 d}  V^M   \nabla_M V \ .
\ee

An immediate puzzle, and one of the most pressing issues of this approach, is that unlike in GR these constraints \underline{do not} determine the connection, and consequently curvatures, uniquely in terms of the physical field data. A careful counting reveals that there are $\frac{2}{3} d (d+2 )(d-2)  $ undetermined components of connection \cite{Hohm:2011si,Coimbra:2011nw}.   The part of the connection that is completely determined is given, upon lowering one index with $\eta$ by \cite{Jeon:2011cn,Jeon:2010rw}
\be
\begin{aligned}\label{eq:Gammadetermined}
\hat{\Gamma}_{IJK} = &\frac{1}{2} \cH_{KQ} \partial_I \cH^Q{}_J + \frac{1}{2} \left( \delta_{[J}^P \cH_{K]}{}^Q + \cH_{[J}{}^P \delta_{K]}^Q \right) \partial_P \cH_{IQ}  \\ 
& + \frac{2 }{ d-1 } \left( \eta_{I[J}\delta_{K]}^Q + \cH_{I[J}\cH_{K]}{}^Q \right) \left( \partial_Q d  + \frac{1}{4} \cH^{PS} \partial_S \cH_{PQ} \right) \ .
\end{aligned}
\ee

One may continue anyhow and there are two options, the first is to work solely with the determined connection $\hat{\Gamma}$ and accept that derivatives $\nabla = \partial + \hat\Gamma$ are no longer completely covariant and curvatures defined with them are necessarily no longer completely tensorial.  This is the premise of the semi-covariant approach of \cite{Jeon:2011cn,Jeon:2010rw} in which a central role is played by the covariantly constant projector operators 
\be\label{eq:projects}
P_M{}^N = \frac{1}{2} \left(  \delta_M{}^N + \cH_M{}^N  \right)  \ , \qquad \bar{P}_M{}^N = \frac{1}{2} \left(  \delta_M{}^N - \cH_M{}^N  \right)  \ . 
\ee
Although the derivatives are not covariant, upon appropriate contraction with these projectors covariant objects can be recovered.   The alternative (and probably ultimately physically equivalent) philosophy is to insist on retaining the undetermined components of connection but to ensure that they do not enter in the final physical theory.  Indeed whilst the four index Riemann curvature contains undetermined parts of the connection one can form contractions such as 
\be
{\cal R} = {\cal R}_{MNPQ} P^{MP} P^{NQ} \ ,
\ee
in which all undetermined components drop out.  It is the scalar curvature so defined that enters into the action eq.~(\ref{DFTEinHil}).  Actually, the obvious scalar constructed as 
\be
{\cal R}_{MNPQ}  \eta^{MP} \eta^{NQ} = 0   \ .
\ee
The problem with this philosophy is that whilst it is evidently fine at leading order one might well encounter difficulties when attempting to extend the theory to higher derivatives. For instance, it is well known that the first correction to the bosonic and heterotic string effective actions consist of Riemann tensor squared.  It is currently unclear how to build, covariantly, such terms using the geometry introduced above. The resolution to this important question maybe somewhat subtle due to the possibilities of field redefinitions.  Hope is given by the work of Meissner \cite{Meissner:1996sa} in which the first $\alpha^\prime$ correction to the NS sector dimensionally reduced to one temporal dimension (with no dependence on the internal coordinates) are phrased in terms of $\cH$ and $d$. \EDIT{A recent proposal \cite{Hohm:2013jaa} has been made in which, in the context of Double Field Theory, T-duality remains uncorrected in $\alpha'$ however the price to pay is that the gauge structure of the theory receives corrections.}

\subsubsection{The Torsion Formalism} 
%

Of course, the preceding discussion does not absolutely preclude a different approach in which some extra or different constraints are imposed such that the connection is  uniquely determined.  In fact  such a formulation can be achieved but the price to pay is to accept torsion  - see \cite{Berman:2013uda} for a detailed discussion.   Dropping the requirement of vanishing torsion,  one can find a natural connection known as the Weitzenb\"ock connection for the generalised vielbein. This connection is metric compatible and $O(d,d)$ compatible, it is not however invariant under the local $O(d)\times O(d)$ symmetry of Double Field Theory. 

Its significant ``disadvantage" is that it has vanishing curvature; there is only torsion. Remarkably though one can still write down an action in terms of the torsion whose equations of motion reproduce the the equations of motion of Double Field Theory. This means the geometry of Double Field Theory can be turned entirely into a flat-torsionful geometry. The action constructed out of the torsion is then  invariant under the local $O(d)\times O(d)$ symmetry, even though the the individual torsion pieces it is constructed from are not.  (This is reminiscent of what occurred with the generalised Lie derivative. The action was invariant even though the building blocks for the action were not.)  We now present some of the details.

First one introduces a generalised vielbein for the generalised metric. We denote this by ${\cal E}^{\A}_{I}$.  In this section,   flat indices are barred and $I,J$ are the curved indices. The Weitzenb\"ock connection is given by
\begin{equation} \label{eq:Wbock}
 \Gamma^I_{JK} = {\cal{E}}_\A{}^I \partial_J {\cal{E}}^\A{}_K  \,  .
\end{equation}
We want covariant derivatives formed with this connection to be tensorial under the DFT gauge transformation generated by the generalised Lie derivative. Under a generalised Lie derivative generated
by $U^I$,
\begin{equation} \label{eq:VarCovDev}
 \delta_U \nabla_I V^J = \mathcal{L}_U \partial_I V^J + Y^{KL}{}_{IM} \partial_L U^M \partial_K V^J + \left( Y^{JL}{}_{KM} \partial_I \partial_L U^M - \partial_I \partial_K U^J \right) V^c + \delta_U ( \Gamma^J_{IK} V^K )\,.
\end{equation}
Thus the covariant derivative transforms \emph{up to section condition} as a generalised tensor if
\begin{equation} \label{eq:VarConn}
 \delta_U \Gamma^I_{JK} \approx \mathcal{L}_U \Gamma^I_{JK} +  \partial_J \partial_K U^I - Y^{IL}{}_{KM} \partial_J \partial_L U^M \,.
\end{equation}
One may have wanted to define a connection that gives a  covariant derivative without using the section condition.
\EDIT{This is not possible in general}: \eqref{eq:VarCovDev} will only ever give a transformation up to section condition because the second term on the right-hand side can only vanish when the section condition is used.

The Weitzenb\"ock connection obeys the transformation law \eqref{eq:VarConn} up to section condition and so defines a suitable connection for Double Field Theory. This connection is compatible with the generalised metric ${\cal H}_{IJ}$ and the $O(d,d)$ inner product $\eta_{IJ}$. It has zero curvature but non-zero torsion. The absence of curvature may be viewed as a problem for describing the dynamics of the theory: naively, one would expect the action to be constructed from the curvature. In fact, as we show the torsion of this connection alone is sufficient to construct the DFT action.

Let us now briefly verify the important properties of the Weitzenb\"ock connection.
Recalling the definition of the
generalised metric in terms of the vielbein, one can easily show that it is a metric
connection,
\begin{equation}
 \nabla_I {\cal{H}}_{JK} = 0\, .
\end{equation}
We can also ask for the covariant derivative of the vielbein to vanish. This is used to define
the spin connection $\omega_I{}^{\A\B}$.
The covariant derivative of the vielbein is, just like in general relativity, defined by
\beq
\nabla_I {\cal{E}}^\A{}_J = \partial_I {\cal{E}}^\A{}_J - \Gamma^K_{IJ} {\cal{E}}^\A{}_K - \omega_I{}^\A{}_\B {\cal{E}}^\B{}_J = 0\,.
\eeq
We now see that substituting in the Weitzenb\"ock connection yields a vanishing spin
connection.

The usual Riemann curvature tensor of a connection is defined by
\begin{equation}
 R^I{}_{IKL} = \partial_K \Gamma^I_{LJ} - \partial_L \Gamma^I_{KJ} + \Gamma^I_{KM} \Gamma^M_{LJ} - \Gamma^I_{LM} \Gamma^M_{KJ} \,. \label{eq:riemann}
\end{equation}
In generalised geometry, where the transformation property of the connection is (\ref{eq:VarConn}), this object is not a tensor, having an anomalous transformation \cite{Hohm:2011si}
\beq
\Delta_U R^I{}_{JKL} = 2 Y^{MN}{}_{P[K} \partial_{L]} \partial_M U^P \Gamma^I_{NJ} \,.
\label{eq:DeltaRiemann}
\eeq
Instead one is led to define a generalised curvature tensor
\begin{equation}
 \mathcal{R}^I{}_{JKL} = R^I{}_{JKL} + Y^{IM}{}_{LN} R^N{}_{KJM} + Y^{MN}{}_{LP} \Gamma^P_{MK} \Gamma^I_{NJ} \, . \label{eq:sriemann}
\end{equation}

We note immediately that in generalised geometry the Weitzenb\"ock connection is flat, and so we cannot describe the dynamics through the curvature. This is the problem in adopting the Weitzenb\"ock connection. One cannot then form a curvature that will then capture the dynamics.

The torsion of a connection, defined by
\begin{equation}
 T_{JK}{}^I = \Gamma^I_{JK} - \Gamma^I_{KJ}
\end{equation}
is not a generalised tensor. There is however a {\it{generalised torsion}} \cite{Coimbra:2011nw}, denoted $\tau_{JK}{}^I$, defined by
\begin{equation}
 \tau_{JK}{}^I U^J V^K
\equiv \left( \mathcal{L}^\nabla_U - \mathcal{L}^\partial_U \right) V^I\,,
\end{equation}
where $\mathcal{L}^\partial_U$ is the generalised Lie derivative as defined in \eqref{eq:genL1} and $\mathcal{L}^\nabla_U$ is the Lie derivative with all partial derivatives replaced by covariant ones. This gives
\begin{equation}
 \tau_{JK}{}^I = T_{JK}{}^I + Y^{IL}{}_{KM} \Gamma^M_{LJ} \label{eq:tau}
\end{equation}
as the generalised torsion of a connection $\Gamma^I_{JK}$. It is a generalised tensor if the connection obeys (\ref{eq:VarConn}). However, because the Weitzenb\"ock connection behaves only as a connection up to section condition, its generalised torsion is only a tensor up to section condition. For any connection preserving $\eta_{IJ}$, the generalised torsion will be antisymmetric in its lower indices, $\tau_{JK}{}^I = \tau_{[JK]}{}^I$, which follows as a consequence of the compatibility with $\eta$. It is this generalised torsion that we now use to describe the dynamics by constructing the action.

We know that DFT should be invariant under both global $O(d,d)$ and local $H \equiv O(d) \times O(d)$. The Weitzenb\"ock connection \eqref{eq:Wbock} is not invariant under these local $H$-transformations
\begin{equation}
 \Delta_\lambda \Gamma^I_{JK} = {\cal{E}}_{\A}{}^I {\cal{E}}_{\B \, K} \partial_J \lambda^{\A\B}\,,
\end{equation}
However, we may construct a Lagrangian $L$ which is invariant under the local $H$ symmetry and is an \EDIT{$O(d,d)$} scalar. Our DFT action will then be
\begin{equation}
 S = \int dx d\tilde{x} \,  e^{-2d} \, L\,.
\end{equation}

It will be useful to construct a $O(d,d)$ covector using the dilaton $d$ as follows:
\begin{equation}
 \nabla_I d = \partial_I d + \frac{1}{2} \Gamma^K_{KI}\,.
\label{eq:covd}
\end{equation}

We find that the \emph{only} $H$-invariant combination of torsion terms and derivatives of the dilaton is given by
\begin{equation} \label{eq:Lagrangian}
 L = - \frac{1}{12}\EDIT{ \tau_{JK}{}^I} \tau_{MN}{}^L {\cal{H}}_{IL} {\cal{H}}^{JM} {\cal{H}}^{KN} - \frac{1}{4} \tau_{JK}{}^I \tau_{LI}{}^K {\cal{H}}^{JL} - 4 {\cal{H}}^{IJ} \nabla_I d \nabla_J d + 4 {\cal{H}}^{IJ} \nabla_I \nabla_J d \,.
\end{equation}

Inserting the expressions for the torsion and covariant derivative of $d$ we find that we can write \eqref{eq:Lagrangian} as
\begin{equation}
 \begin{split}
  L &= \frac{1}{8} {\cal{H}}^{IJ} \partial_I {\cal{H}}^{KL} \partial_J {\cal{H}}_{KL} - \frac{1}{2} {\cal{H}}^{IJ} \partial_I {\cal{H}}^{KL} \partial_K {\cal{H}}_{JL} +4 {\cal{H}}^{IJ} \partial_I \partial_J d - \partial_I \partial_J {\cal{H}}^{IJ} \\
  & \quad - 4 {\cal{H}}^{IJ} \partial_I d \partial_J d + 4 \partial_I {\cal{H}}^{IJ} \partial_J d + \frac{1}{2} \eta^{IJ} \eta^{KL} \partial_I {\cal{E}}^\A{}_K \partial_J {\cal{E}}_{\A L}  \, .
 \end{split}
\end{equation}
This agrees with the original formulation \cite{Hohm:2010pp} up to the final term which vanishes by the section condition. This final term is only $H$-invariant up to section condition but  as we shall soon see it is crucial in order to match the Scherk-Schwarz reduced theory with the gauged supergravity potential \cite{Grana:2012rr}.

\subsection{Extensions to DFT}
\subsubsection{The RR sector}\label{sec:RRsector}
As we saw in section \ref{secgengeom} there is an isomorphism between spinors in generalised geometry and poly forms.  This is the clue to how RR fields may be included into the DFT  in \cite{Hohm:2011zr,Hohm:2011dv} and also in \cite{Coimbra:2011nw}   similar considerations are used in the non-doubled generalised geometry perspective.  To be more concrete one can realise $Cl(d,d)$ in terms of fermionic creation and annihilation operators that obey 
\be
\{ \psi_i , \psi^j \} = \delta_i^j \ , \quad \psi_i^\dag = \psi^i \ , \quad  \psi_i | 0 \rangle = 0  \ . 
\ee
For instance, an element $R_i{}^j$ of the  $Gl(d)$  subgroup of $O(d,d)$ generating a basis change lifts to 
\be
D(R) = \frac{1}{\sqrt{det R}} \exp \psi^i R_i{}^j \psi_j \ ,
\ee 
in the spin cover.  The factor of $\sqrt{det R}$ that arises is indicative of the isomorphism described in eq.~(\ref{iso}).   Since the invariant metric $\cH$ is also an element of $O(d,d)$ it gives rise to a spinorial representative   we shall denote $\mathbb{S}$ (one may think of this as defined through the invariance properties of the $Cl(d,d)$ gamma matrices).\footnote{Because the time direction is formally doubled in the DFT this causes some subtleties in the definition of $ \mathbb{S}$ which are discussed in \cite{Hohm:2011dv}. }

The Dirac operator decomposes as 
\be
\slashed{\partial}= \Gamma^M \partial_M = \psi^i \partial_i + \psi_i \tilde{\partial}^i \ .
\ee
   One can see from this that in the case of  $\tilde{\partial}\equiv 0$ this Dirac operator acts simply as the exterior derivative.  Then with this in mind one may write a polyform, in this case of RR potentials, as 
\be
\Psi = \sum_p \frac{1}{p!} C_{i_1 \dots i_p} \psi^{i_1 \dots i_p} | 0 \rangle \ . 
\ee
Here we are working in the democratic formalism in which, at the level of the action, p-form potentials and their Hodge duals are included as independent variables but their duality must be imposed by hand separately. Thus the sum in the above runs over $p$ odd for IIA and even for IIB.  To obtain the correct degrees of freedom one imposes a constraint (relating higher p-forms to the Hodge dual of the lower ones) which is given by 
\be
\slashed{\partial}\Psi = - C^{-1} \mathbb{S} \slashed{\partial}\Psi 
\ee
where $C$ is the charge conjugation. Consistency requires that $C^{-1} \mathbb{S}C^{-1} \mathbb{S} = 1$ which restricts the allowed dimensions of the theory (when $d$ is even the allowed cases are $d=2,6,10$ which correspond to the dimensions in which conventional p-form self duality may be imposed).    In the case for which $\tilde{\partial}=0$ for which we can identify $\slashed{\partial}\Psi \sim \sum F^{(p)}$ with $F^{(p)} = d C^{p}$.    Since $\slashed{\partial}$ is nilpotent, we have a DFT generalisation of p-form gauge symmetry
\be
\delta_\Xi \Psi = \slashed{\partial}\Xi
\ee
in which $\Xi$ is a spinor of opposite chirality to $\Psi$. 

The DFT gauge symmetry acts on this spinor according to 
\be
\delta_\xi \Psi = {\cal L}_\xi \Psi  = \xi^M \partial_M \Psi + \frac{1}{2} \partial_M \partial_N \Gamma^M \Gamma^N \Psi \ , 
\ee
which can be though of as the spinorial-Lorentz-Lie derivative supplemented by an additional group theoretic term  (obtained upon symmetrising the indices on the gamma matrices).  Notice that the potentials so defined are not invariant under the B-field gauge transformations --  this is similar to difference between the $A$ and $C$ basis in conventional formalisms.  It was shown in \cite{Geissbuhler:2011mx} that defining ${\cal F} = e^{d} \mathbb{E} \slashed{\partial}\Psi$ where $\mathbb{E}$ is the spinorial representative of the vielbein results in field strengths which, when everything is independent of $\tilde{x}$,  obey the usual Bianchi identities $(d+H\wedge) F= 0$. 

Equipped with the above one can form the correct kinetic terms for the RR fields, given simply by
\be
S = \int dx d\tilde{x} \, e^{-2d} {\cal R}(\cH, d)  + (\slashed\partial \Psi)^\dag \mathbb{S} \slashed\partial \Psi \ . 
\ee
Upon setting $\tilde{\partial}= 0$ one does indeed find that this gives the correct kinetic terms for the RR fields. 
An interesting feature of this approach is that to show gauge invariance  of the action and closure of the algebra one does not appear to invoke the strong constraint in its entirety on the RR sector.  In the frame in which the strong constraint is solved by setting  to zero all dependance  $\tilde{x}$ for the gauge parameters and NS fields,  it remains consistent that the RR potential $\Psi$ is modified to depend linearly on the winding coordinates $\tilde x$  \cite{Hohm:2011cp}.  In the IIA context this can be brought into identification  with the massive IIA theory.  The meaning in IIB remains somewhat mysterious.    To close this section we note that a slightly different  approach to RR fields was presented in \cite{Jeon:2012kd} in which they are packaged not in a spinor of $O(d,d)$ but as a bi-spinor of the local $SO(1,d-1)_L \times SO(d-1, 1)_R$ symmetry in a frame like formalism, which can seem rather natural given the considerations of supersymmetry that we turn to in the next section.

\subsubsection{Fermions and Supersymmetry}
The inclusion of fermions and supersymmetry in DFT has been recently considered   \cite{Hohm:2011nu,Jeon:2011vx,Jeon:2012hp,Jeon:2011sq} with a precedent going back to the seminal work of Siegel \cite{Siegel:1993xq,Siegel:1993th} which was formulated directly in superspace.   The generalised geometry formulation of   \cite{Coimbra:2011nw}  (in which we recall no coordinate doubling takes place) also includes the complete ${\cal N}=2$ supersymmetric completion which was also achieved for DFT in \cite{Jeon:2012hp}.  In the interests of space, and to avoid reproducing technical details, lets only make a brief remark about this. 

They key insight, which is valid in both the string and M-theory constructions, is that whilst scalars take values is a coset $G/H$ the fermions lie in appropriate representations of $H$ which is a local symmetry of the action. In the case at hand we have then a global $O(d,d)$ symmetry and a local $SO(1,d-1)_L \times SO(d-1,1)_R$ symmetry.  This   local symmetry can be manifested by working in a frame like formalism which is needed to couple to fermions.  

For example, the minimally supersymmetric extension  to DFT introduces a  dilatino $\rho$ which is a $MW_+$ spinor of $SO(1,d-1)_L$, a singlet of $SO(d-1,1)_R$ and a singlet of the global $O(d,d)$ together with a gravitino $\psi_m$ which is a $MW_-$ spinor of $SO(1,d-1)_L$, a vector of  $SO(d-1,1)_R$ and a singlet of the global $O(d,d)$.   The supersymmetry parameter in this case is in the opposite chirality representation to the dilatino.  This set up was developed to quadratic order in fermions as described in \cite{Hohm:2011nu} with a full action given in \cite{Jeon:2012hp} (the two approaches differ in their details).   The supersymmetry variations  have the following simple schematic form
\be
\delta_\epsilon d \sim \bar\epsilon \rho \ , \quad \delta_\epsilon E  \sim \bar{\epsilon } \gamma \psi \ , \quad \delta_\e \psi \sim D \epsilon \ , \quad \delta_\e \rho \sim  \slashed{D} \e
\ee
where $E$ represents the a frame fields, and $D$ stands   for an appropriate covariant derivative whose construction and is given in \cite{Jeon:2012hp} .   The supersymmetry algebra so formed closes on supersymmetries, generalised diffeomorphisms, local $SO(1,d-1)_L \times SO(d-1,1)_R$ Lorentz rotations and the fermionic equations of motion. 

In fact one can make the theory supersymmetric without using the section condition by again using the Scherk-Schwarz ansatz \cite{Berman:2013cli}. Again there is an interplay between consistency of the theory and supersymmetry. The conditions for making supersymmetry work are the same as required for consistency of the generalised geometry Courant algebra to close.

\subsubsection{Gauge Fields}
The extension of the doubled approach to gauge fields was already considered by Siegel in \cite{Siegel:1993th,Siegel:1993xq}  and has been recently reconsidered in \cite{Jeon:2011kp} and   \cite{Hohm:2011ex}.

 The approach of \cite{Jeon:2011kp} is to introduce a doubled vector field, which combines the regular gauge field with a set of scalars  to form a  vector in the doubled space 
\be\label{eq:doubledvec}
V_{I} = \binom{\phi^i}{A_i + b_{ij} \phi^j } \ . 
\ee
Then a field strength can be formed using the semi-covariant derivative (i.e. $\nabla = \partial + \hat{\Gamma}$ where the connection is the one in eq. \eqref{eq:Gammadetermined}) according to 
\be
F_{IJ} = \nabla_{I} V_J - \nabla_J V_I  - i [V_I , V_J] \ . 
\ee
Although not covariant in itself under local gauge transformations upon suitable contraction with the projectors defined  in eq.~\eqref{eq:projects} it leads to covariant results.  A YM and DFT gauge invariant action may be obtained as 
\be
\int dx d\tilde x e^{-2d} Tr( P^{IJ} \bar{P}^{KL} F_{IK} F_{JL} ) \ . 
\ee
Expanding into components and setting dependance on the $\tilde{x}$ coordinates to zero reveals that this action contains a Maxwell term in curved space together with kinetic terms  and interactions for the extra scalars that completed the T-duality covariant doubled vector eq.~\eqref{eq:doubledvec}. The result has some similarity (but not exact equivalence) with a topologically twisted Yang-Mills theory. 

An alternative approach was given in  \cite{Hohm:2011ex}. The inclusion of $n$ Abelian gauge fields may be achieved by extending the theory to an $O(d,d+n)$ invariant theory where the inner product becomes 
\be
\label{oddmetric2}
\eta_{IJ}  = \left(\begin{array}{ccc}  0 & \mathbb{1}_d & 0 \\ \mathbb{1}_d & 0 & 0  \\ 0 & 0 & \mathbb{1}_n  \end{array}\right)\, .
\ee
The generalised metric $\cH$ now includes extra components reflecting the dependance on the gauge field; for instance the top-left-most entry becomes $g_{ij} - (c g^{-1} c)_{ij}  + A_i^\a A_{j\a} $ where $c_{ij} = b_{ij} + \frac{1}{2} A_i^\a A_{j\a} $.   Rather elegantly, having redefined $\cH$ and $\eta$ in this fashion the Lagrangian is given by exactly the same one as before in eq.~\eqref{DFT}.  The local DFT gauge symmetry generated by the generalised Lie derivative  $\delta_\xi \cH = {\cal L} _\xi \cH$ when expanded into components gives exactly the correct transformation rule for gauge fields $\delta A^\a_i = L_\xi A^\a_i + \partial_i \Lambda^\a$.   Upon setting $\tilde{\partial} = 0$ the action may be expanded in to components with the result, up to total derivatives as usual, given by 
\be
S  = {1\ov 2\kappa^2}\int_{M_{10}} \sqrt{g}  e^{-2\phi}\left(R + 4 (\partial \phi)^2  - \frac{1}{12}{\hat H^2}{} - \frac{1}{4}F_{ij}^\a F^{ij}_\a  \right) \ , 
\ee 
in which the three-form field strength is modified by a Chern-Simons term $\hat H = H + A^\a \wedge F^\a$.  This indeed looks like an Abelian version of the effective action of the heterotic string.  

The non-Abelian extension was also considered in  \cite{Hohm:2011ex}.  By introducing a  structure constant $f_{MN}{}^K$  it was shown how a non-Abelian deformation of the theory could be constructed in manifestly $O(d,d+n)$ covariant language.  This structure constant obeys standard Jacobi identities but also consistency of the theory required one additional constraint that $f_{MN}{}^K \partial_K = 0$.  The structure constant can be viewed as a spurionic object; it is written in a covariant way however making a particular choice for the structure constant breaks the T-duality symmetry down.   The deformation of the theory also deforms the gauge algebra so that the modified transformation $\delta_\xi V^M = {\cal L} _\xi V^M - \xi^K f_{KL}{}^M V^L $ leaves the action invariant and closes onto a similarly deformed C-bracket.

A somewhat remarkable feature is the extra terms added to the action in this non-Abelian deformation include a scalar potential which is precisely of the form seen in ${\cal N} = 4$ electrically gauged supergravity.  This is further evidence towards a rather intimate relation between gauged supergravity and the doubled formalism which we shall now  explore in more detail. 

\subsection{Relaxing the Strong Constraint: Scherk-Schwarz Reduction and Gauged Supergravity } 
So far we have required that all fields and gauge parameter obey the weak constraint coming from level matching 
\be
\eta^{MN} \partial_M \partial_N \Phi = 0 \ , 
\ee
and the strong constraint on products of fields
\be
\eta^{MN} \partial_M \Phi_1 \partial_N \Phi_2 = 0 \ .
\ee
Together these constraints mean that the theory really only depends on half the coordinates and so, at least locally,  the doubling is really a formal trick that makes manifest $O(d,d)$ symmetry.

However, although the weak and strong constraints are sufficient for the theory to be gauge invariant and the gauge algebra close, they need not be necessary.    One of the most pertinent questions to ask is whether there are consistent ways to relax the strong constraint, that is, can one define the theory for backgrounds that depend on both the regular coordinates of spacetime and the additional dual coordinates?  Answering this in the affirmative could open the door to possibilities far beyond the realm of conventional supergravity.   
 
A breakthrough development  was to consider compactification of the DFT with a Scherk-Schwarz reduction see  \cite{Aldazabal:2011nj,Grana:2012rr,Dibitetto:2012rk,Aldazabal:2013sca,Geissbuhler:2013uka,Geissbuhler:2011mx,Berman:2013cli}.   In this case one finds that upon reduction the DFT gives rise to a lower dimensional gauged supergravities.  What is somewhat remarkable is that this provides a higher dimensional origin for gauged supergravities where the non-geometric fluxes become purely geometric from the point of view of the doubled space.   

This subject has recently been beautifully expounded in great detail in the review article \cite{Aldazabal:2013sca} and to avoid duplication we provide here only a short synopsis in the spirit of \cite{Grana:2012rr}.

To make this dimensional reduction the coordinates of the doubled theory are split into external ones, which we will call $\X$, and  internal ones which we denote as $\Y$.  (For concreteness one may think of the full type II DFT where the 10+10 coordinates of the theory are split into 4 external and 4 dual external coordinates together with  6+6 internal coordinates).   The essence of the approach is to define a duality twist matrix valued in $O(d,d)$ given by 
\be
U^A{}_M(\Y) \in O(d,d) \ . 
\ee
The  Scherk-Schwarz ansatz is that dependance on the internal coordinates in fields and gauge parameters is governed by this twist matrix according to
\be
V^M{}_N(\X, \Y) = (U^{-1})^M{}_A (\Y)  \check{V}^A{}_B(\X)  U^B{}_N(\Y) \ .
\ee
For the dilaton the appropriate ansatz is 
\be
d(\X, \Y) = \check{d}(\X) + \lambda(\Y) \ .
\ee
All dependence on the twist matrix will enter in combinations given by
\be
f_{ABC} = 3 \eta_{D[A} (U^{-1})^M{}_B (U^{-1})^N{}_{C]}  \partial_M U^D{}_N \ , \quad f_A = \partial_M U^M{} _A -  2  (U^{-1})^M{}_A \partial_M  \lambda \ . 
\ee
An immediate requirement is that these \EDIT{be constant}.   One may notice that, and its no coincidence, these \EDIT{objects}  correspond to antisymmetrization and trace of the Weitzenb\"ock connection.   For this presentation, for simplicity we will set \EDIT{ $f_A$ to zero}.  This ansatz should be supplemented with a constraint  that ensures Lorentz invariance of the reduced theory. 

Upon invoking this ansatz the action \footnote{The curvature  entering in eq.~(\ref{grana}) includes a term that vanishes upon the strong constraint which already entered in the  torsion formalism.} 
\be
\label{grana}
S = \int d\X d\Y e^{-2 d} {\cal{R}}(\cH , d ) \ ,  
\ee
one obtains a lower dimensional action
\be
S_{GDFT} \propto \int d\X \, e^{-2 \check{d}} \left( {\cal{R}}(\check{\cH},\check{ d} )  +  {\cal{R}}_f (\check{\cH},\check{ d} ) \right) \ , 
\ee
in which the first term is just the regular DFT action for the lower dimensional theory, and the second term represents   corrections due to the ganging. Ignoring the $f_A$ gaugings (details of which can be found in \cite{Geissbuhler:2011mx}) this correction is of the form  
\be\label{eq:scalarpot}
  {\cal{R}}_f =  -\frac{1}{2} f^A{}_{BC} \check\cH^{BD} \check\cH^{CE}\partial_D \check\cH_{AE}   -\frac{1}{12} f^A_{BC}f^D{}_{EF} \check\cH_{AD} \check\cH^{BE}\check\cH^{CF} - \frac{1}{4} f^A{}_{BC} f^B{}_{AD} \check\cH^{CD } \ , 
  \ee
  which includes gaugings of the kinetic terms and a scalar potential.   
 
 Under this reduction the generalised diffeomorphism symmetry gives rise to a gauged deformation
\be
\delta_{\check\xi} \check V^A =  {\cal L}_\xi V^A  - f^A{}_{BC} \check \xi^B \check V^C \ .  
\ee
Because of the twisting, the C-bracket which appears in the closure of these symmetries also receives a deformation (see also \cite{Hohm:2011ex}) 
\be
[\check \xi_1 , \check \xi_2]^A_f =[\check \xi_1 , \check \xi_2]_C^A - f^A_{BC}\check \xi_1^B\check \xi_2^C \ .  
\ee

The conditions for this theory to be consistent, i.e. closure of the gauge algebra, Jacobiator generating a trivial transformation and invariance of the action, were studied in  \cite{Geissbuhler:2011mx,Grana:2012rr,Geissbuhler:2013uka}.  
The conclusion  is that although  strong and weak constraints on the twist matrix are sufficient for consistency, they are not necessary and can in principle be relaxed.  That this is possible in practice was explicitly demonstrated in  \cite{Dibitetto:2012rk} where consistent gaugings from twist matrices with explicit dependance on both regular and dual internal coordinates were constructed.

The form of the scalar potential in eq.~\eqref{eq:scalarpot} corresponds   to the scalar potential of half maximal gauged supergravities.   Indeed, upon decomposing $\cH$ (which we recall is still a $2d \times 2d$ matrix ) into its internal and external components and demanding that the only non-vanishing components of the gaugings are the ones with internal legs, the identification between gauged supergravity and Scherk-Schwarz reduced DFT can be made complete.  

One might think that this has been a rather convoluted way to obtain gauged supergravity.  After all one could have performed a twisted reduction of the regular type II supergravity in ten dimensions and obtained a gauged theory.  However there exist  lower dimensional gauged supergravities which can not be obtained in this manner.  By allowing for Scherk-Schwarz twists that depend on both the geometric and dual coordinates, DFT provides a higher dimensional origin for exactly these theories!

\subsection{  Related Developments}  
 Space limitations prevent us discussing in any detail a number of   developments related to the contents of this chapter including\footnote{This list is not intended to be exhaustive but we hope it serves as a useful starting point. }: the structure of large (finite) DFT gauge transformations \cite{Hohm:2012gk,Park:2013mpa}, the the matrix theory perspective of non-geometric fluxes  \cite{Chatzistavrakidis:2013qca,Chatzistavrakidis:2012qj};  further perspectives on non-geometric fluxes \cite{Andriot:2012an,Andriot:2012wx} and their relationships to non-commutativity and even non-associativity \cite{Mathai:2004qq,Bouwknegt:2004ap,Mathai:2005fd,Blumenhagen:2010hj,Lust:2010iy,Andriot:2012vb,Blumenhagen:2011ph,Condeescu:2012sp,Mylonas:2012pg}; T-folds, G-structures and supersymmetry  \cite{Gray:2005ea,Grana:2008yw,Grana:2006kf}; brane wrapping rules  \cite{Bergshoeff:2011mh,Bergshoeff:2011zk,Bergshoeff:2011ee}  and the $E_{11}$ \cite{West:2001as}  derivation of Double Field Theory \cite{West:2010ev,Rocen:2010bk} (of which we shall return to later); investigation of the relevant Lie  algebroids \cite{Cavalcanti:2011wu,Baraglia:2011dg,Blumenhagen:2012pc,Mylonas:2012pg, Blumenhagen:2012nt,Blumenhagen:2013aia}.

\section{Duality Symmetric M-theory} \label{sec:Mtheory}
\subsection{M-theory and Eleven-Dimensional Supergravity}

 In 1995 it was realised that the strong coupling limit of the IIA string is an eleven-dimensional theory whose low energy limit is eleven-dimensional supergravity. This unknown mysterious eleven-dimensional theory has become known as M-theory  \cite{Hull:1994ys,Witten:1995ex}. Considering that it originates from a string theory at strong coupling, the properties of M-theory are quite surprising:  the critical dimension is eleven rather than ten and the fundamental objects are  membranes and five-branes rather than strings.

M-theory has a remarkable unifying power: by taking different compactification limits of this single theory one can arrive at all the string theories. The five different versions of string theory, which at first sight appear disparate,  are obtained from M-theory expanded around different vacua.   M-theory provides the overarching umbrella explaining the web of non-perturbative dualities between string theories.  A most elegant realisation of this is that the   IIB $SL(2,\mathbb{Z})$ strong weak duality can be seen as  a simple consequence of M-theory diffeomorphism invariance \cite{Schwarz:1995jq}.  Dimensional reduction also gives rise to the rules and dictionary between the various extended objects of string theory and the membrane and five-brane.   For a review of M-theory and its branes see \cite{Berman:2007bv}.

Despite these tremendous successes,  our understanding of M-theory remains rather incomplete.  We lack even a fundamental description  that allows us to understand the theory beyond its low energy limit.   Although such a formulation of M-theory remains a distant prospect there \EDIT{has been, in recent years, great progress} towards understanding the extended objects of M-theory.  An important breakthrough has been the development of a theory describing the low energy interactions of multiple M2 branes which began with the pioneering work of Bagger and Lamber  \cite{Bagger:2006sk} and of Gustavasson \cite{Gustavsson:2007vu},  culminating in celebrated ABJM model \cite{Aharony:2008ug}.    The signature $N^{3/2}$ degree of freedom scaling has been rigorously  demonstrated for these Chern-Simons matter type theories \cite{Drukker:2010nc}.      Though the theory of interacting M5 branes (a non-Abelian ${\cal N}=(2,0)$ tensor multiplet in six dimensions) remains far from understood,  many of its properties are known and have been beautifully exploited to provide deep new insights about gauge theories.  For instance by considering   M5 branes wrapped on a Riemann surfaces with punctures, Gaiotto  \cite{Gaiotto:2009we} constructed   a large class of novel four-dimensional ${\cal N}=2$ superconformal field theories.  Moreover the Nekarasov partition functions of these have theories have been conjectured in \cite{Alday:2009aq} to be related to correlators in the  Liouville theory  on the corresponding Riemann surfaces.   The six-dimensional M5 theory thus provides a bridge between quantities in a four-dimensional gauge theory and those of a two-dimensional CFT.     Developing the theory of coincident M5 branes remains an important and active avenue of research, interesting recent work in this direction includes \cite{Lambert:2010iw}.

Nonetheless, for what remains of this report we will concern ourselves with just the low energy effective dynamics, namely 11-dimensional supergravity discovered by Cremmer,  Julia and Scherk in 1978 \cite{Cremmer:1978km}.  The action in the bosonic sector is given by
\ba
S_{11} = \frac{1}{2\kappa^2} \int_{M_{11}} \sqrt{g} \left( R - \frac{1}{48} F_{4}^2 \right) - \frac{1}{6} F_4 \wedge F_4 \wedge C_3  \ , \label{11dsugra}
\ea
where the field strength \EDIT{is} defined as $F_4  = d C_3$.   There are 44 bosonic degrees of freedom in the graviton and 84 contained in the three-form which are paired off against the 128 (real) degrees of freedom in a Majorana spin $\frac{3}{2}$ gravitino.  

In 1980, Julia discovered that 11-dimensional supergravity when compactified on tori of various dimensions, exhibited a host of symmetries \cite{Julia}. They are now usually rather plainly referred to as M-theory dualities (or also sometimes hidden symmetries) and it is well known that they are controlled by some Lie group in low dimensions. More specifically,  compactification on a torus of dimension $n$ results in a $D=11-n$ dimensional theory whose symmetry group $G$ is listed in table \ref{GandH}.      For the cases of dimensions smaller than $D=3$ some more complicate objects are involved.    It is believed that in the full M-theory and not just its low energy limit,   an appropriate discrete version of these symmetries survive and these form the  U-dualities of M-theory  \EDIT{\cite{Hull:1994ys}. Let us reiterate a warning given earlier: in the sequel we shall be working with the low energy limit of M-theory, i.e. supergravity, and we will use the phrase U-duality to mean the continuous  ``hidden'' symmetry group  of supergravity.}

\begin{table}[htdp]
\caption{The ``hidden'' symmetry groups $G$ and their maximally compact subgroups $H$ of 11-dimensional supergravity compactified on a torus $T^n$ }
\begin{center}
\begin{tabular}{|c|c|c|c|}
\hline
$D$ & $n$ & $G$ & $H$  \\
\hline
10 & 1 & $SO(1,1)$ & $1$  \\
9 & 2  & $ SL(2)$ &  $SO(2) $\\
8 & 3 & $SL(3) \times SL(2) $& $   SO(3) \times SO(2)$ \\
7 & 4 &  $SL(5)$ & $SO(5) $ \\
6 & 5 & $SO(5,5)$ &$ SO(5)\times SO(5)$  \\
5 & 6 & $E_{6}$ &$ USp(8)$  \\
4 & 7 & $E_{7} $& $SU(8) $\\ 
  3 & 8 & $E_{8}$ & $SO(16)$ \\ \hline
\end{tabular}
\end{center}
\label{GandH}
\end{table}

In 1986, de Wit and Nicolai found some tantalizing clues that these duality symmetries had content beyond what one expects from a theory compactified on a torus. They speculated that the duality groups somehow controlled the theory in a fundamental way, \cite{deWit:1986mz}. The challenge of finding the true meaning of these groups in M-theory has subsequently fascinated many authors resulting in copious publications for example amongst others, \cite{Koepsell:2000xg,Nicolai:2010zz}. 

This motivates two important questions:
\begin{enumerate}
\item How do these symmetries govern the full uncompactified eleven-dimensional theory?
\item Can we make manifest these hidden symmetries? 
\end{enumerate}
In the course of this chapter we shall describe progress in answering both of these questions in the affirmative!   Rather remarkably this path will lead us to a further beautiful insight; the eleven-dimensional supergravity makes a distinction between spacetime (the metric) and fields living on spacetime (the three-form).  Since one expects that both of these fields have common origin in M-theory this divide seems rather uncalled for.   Our approach will lead us to conclude that the metric and the three-forms should themselves be unified into a generalised metric on an extended spacetime. 

In what follows we shall consider not the full eleven-dimensional theory but rather a $1+n$ dimensional analogue given by the same action as eq.~(\ref{11dsugra}) but omitting the Chern Simons term (which can only be defined in eleven dimensions).  This is not to say we are performing a dimensional reduction, quite the contrary: as we include more dimensions we will formulate the theory in a way that makes larger duality groups manifest. No dimensional reduction has been imposed and all the fields can depend on the coordinates associated with the directions in which the duality group acts.

To reinforce this point, consider the case of $n=4$ i.e. the $SL(5)$ duality group.  This is the group that would be obtained upon a $T^4$ reduction of M-theory.   Instead what we shall consider is a $1+4$ dimensional theory where the fields \emph{do} depend on the four spatial coordinates.  We shall show that this proto-M-theory can be reformulated in way that the full $SL(5)$ acts as a manifest symmetry.

\subsection{Generalised Geometry and Extended Spacetime for M-theory}

Recall in the case of DFT that we introduced a doubling of the number of coordinates, thereby giving a geometric realisation of the string winding charges.  In constructing such a theory we found generalised geometry -- in which the number of coordinates is kept the same but the tangent space is doubled to $E= TM\oplus T^\ast M $ -- to be a powerful guiding principle. We then seek a generalised geometry for M-theory in which  $E_{n}$  replaces the role played by $O(n,n)$.   These exceptional generalised geometries have been developed in \cite{Hull:2007zu}  and  \cite{Pacheco:2008ps}.

For the generalised tangent bundle $E= TM\oplus T^\ast M $ we recall that sections consist of one-forms and vectors which are exactly the gauge parameters of B-field gauge transformations and diffeomorphisms.  In the M-theory, and we begin with the case of $n=4$,   we still have diffeomorphisms but now we expect two-form gauge parameters for the three-form field $C_3$.   One is then led to consider the bundle\footnote{In fact bundles of the form $E = TM \oplus \Lambda^p T^\ast M$ were also considered in the mathematical literature \cite{Gualtieri:2003dx}.}, 
\be
E = TM \oplus \Lambda^2 T^\ast M \ . 
\ee
with sections 
\be
U = v + \rho \ . 
\ee
This bundle admits an $SL(5)$ action.  There is an obvious $SL(4)$ action under which vectors and two-forms transform in the usual way (as a $\bf{4}$ and a $\bf{6}$ respectively) -- this represents a 15 parameter subgroup of transformations.    This is supplemented with the action of a three-form $\omega \in \Lambda^3 T^\ast M$ (which has 4 independent components),   which acts as
\be
v+ \rho  \rightarrow  v + (\rho + \iota_v \omega)  \ , 
\ee
and the action of a tri-vector (again with 4 components)  $\beta \in \Lambda^{3}  T M$,  
\be
v+ \rho  \rightarrow  (v +\iota_\rho \beta)  + \rho   \ . 
\ee
These transformations are all natural analogues of the subgroups of $O(d,d)$ encountered previously.  In addition these transformations close on a scaling symmetry giving one extra transformation.  In total we have $\bf{24} = \bf{15} +\bf{ 4} + \overline{\bf{4}} + \bf{1} $ transformations and indeed this corresponds exactly to the decomposition of the adjoint of $SL(5)$ under $SL(4)$.   Thus we find an enhancement of the symmetry with sections of this generalised bundle transforming in the $\bf{10}$ of $SL(5)$.   We may define a Dorfman derivative on this bundle and a Courant bracket which, for this case,   exactly coincide with those of the $T\oplus T^\ast$ geometry defined in eqs.~(\ref{Dorf}) and (\ref{Courant}). 

When we increase the dimension further the picture must be modified.  For $n=5$ one must work with 
\be
E = TM \oplus \Lambda^2 T^\ast M \oplus  \Lambda^5T^\ast M \ . 
\ee
One way of motivating this extension is  that we must now have a geometry that will be able to account for fivebrane modes.  Sections of this bundle now transform as a $\bf{5} \oplus \bf{10} \oplus \bf{1}$ under the geometric $SL(5)$ symmetry.  Following similar arguments to the $n=4$ case one finds that   this symmetry is enhanced to an $E_5 = SO(5,5)$ (or more properly $Spin(5,5)$) underwhich the sections transform as a $\bf{16}$, i.e. a chiral spinor.   The situation for $n=6$ is the same with the bundle $E = TM \oplus \Lambda^2 T^\ast M \oplus  \Lambda^5T^\ast M$ and sections in the fundamental ($\bf{27}$) of $E_{6}$.  For $n=7$ the bundle should be further extended to $E = TM \oplus \Lambda^2 T^\ast M \oplus  \Lambda^5T^\ast M \oplus \Lambda^6 TM$.  This extra factor corresponds to the introduction of a  potential, $C_6$, dual to the three-form.   For $n=8$ the situation becomes radically different, it is thought that the dual graviton field plays a vital role.  We shall not discuss this case further in this report though it represents an important area for future developments.

\begin{table}[htdp]
\caption{The generalised geometries, the coordinate  representation $R_1$, and the section condition representation $R_2$.  }
\begin{center}
\begin{tabular}{|c|c|c|c|c|}
\hline
  $n$ & $G$ &  $E$  &  $R_1$ & $R_2$    \\
\hline
  3 & $SL(3) \times SL(2) $& $  TM \oplus \Lambda^2 T^\ast M$ & $\bf{(3,2)}$ & $\bf{(\bar 3,1)}$  \\
  4 &  $SL(5)$ &  $TM \oplus \Lambda^2 T^\ast M$ & $ \bf{10} $ & $\bf{\bar 5}$ \\
  5 & $Spin(5,5)$ &$   TM \oplus \Lambda^2 T^\ast M \oplus  \Lambda^5T^\ast M  $ &$ \bf{16} $  & $\bf{10}$ \\
  6 & $E_{6}$ &  $TM \oplus \Lambda^2 T^\ast M \oplus  \Lambda^5T^\ast M  $  & $\bf{27} $ & $\bf{\bar 27}$  \\
  7 & $E_{7} $&  $TM \oplus \Lambda^2 T^\ast M \oplus  \Lambda^5T^\ast M \oplus \Lambda^6 TM$ & $\bf{56}$ &$\bf{ 133}$   \\  \hline
\end{tabular}
\end{center}
\label{exceptgeom}
\end{table}

One programme has been the very beautiful reformulation of supergravity, directly in this generalised geometry language \cite{Coimbra:2011ky,Coimbra:2012af}. Here however we shall go one step further, we make the leap that instead of just extending the tangent bundle we should, just as in the case of DFT, introduce extra coordinates that extend spacetime itself.   This is in alignment with the idea used in DFT in the preceding section.   One way to think about this is that we should include an extra coordinate for each point like charge in the reduced theory (although we again emphasise that no dimensional reduction will actually take place) see e.g. \cite{deWit:2000wu}.   For instance consider the $n=4$ example, in which one will have 4 regular coordinates (corresponding to momenta charge) and 6 extra coordinates (which correspond to the ways wrappings of M2 branes in a four-dimensional space).   To be more formal one can directly study the super algebra of eleven-dimensional supergravity
\be
\{ Q_\alpha , \bar{Q}_\beta \} = \Gamma^M_{\a\b} P_M + \frac{1}{2} \Gamma^{MN}_{\a\b} Z_{MN}  + \frac{1}{5!}  \Gamma^{MNPQR}_{\a\b} Z_{MNPQR} \ . 
\ee
Here $P^M$ is the momenta and $Z_{MN}$ and $Z_{MNPQR}$ are central terms corresponding to the presence of M2 and M5 branes.    We look at charges that are point like, i.e. they define either zero-forms or top-forms from the point of view of the external spacetime.   This is summarised in table \ref{pointlike}. 
\begin{table}[htdp]
\caption{The point like charges in lower dimensions  (Roman letters indicate ``internal'' legs and Greek ``external'') }
\begin{center}
\begin{tabular}{|c|c|c|c|c|}
\hline
  $n$ & $G$ &  Point-like charges  & counting   \\
\hline
  3 & $SL(3) \times SL(2) $& $P_m + Z_{mn}$ &  6= 3+3   \\
  4 &  $SL(5)$ &   $P_m + Z_{mn}$   & 10 = 4+ 6 \\
  5 & $Spin(5,5)$ &  $P_m + Z_{mn} + Z_{mnpqr} $ & 16 = 5 +10 + 1   \\
  6 & $E_{6}$ &  $P_m + Z_{mn} + Z_{mnpqr}+ Z_{m \mu \nu \rho \sigma}  $ & 27 +1 = 6 + 15 + 6 +1      \\
  7 & $E_{7} $&   $P_m + Z_{mn} + Z_{mnpqr}+ Z_{  \mu \nu \rho \sigma \tau}  $ & 56 =  7 + 21+21+ 7  \\  \hline
\end{tabular}
\end{center}
\label{pointlike}
\end{table}
This is in broad agreement with the construction of the generalised tangent bundle though we note that in the case of $n=6$ we see the appearance of an extra singlet (whose status in the generalised geometry remains somewhat unclear).   One can continue still further but at $E_8$ one encounters a difference; the number of point like charges is 120 which is not of course an $E_8$ representation, instead it is the antisymmetric \EDIT{two-index tensor representation (the {\bf 120})}  of the maximally compact subgroup $SO(16)$.  Suggestive as the count of point-like central charges is, it appears not to be the ultimate guideline for constructing generalised geometries.  The story is replicated by looking at string like central charges which again fall into representations of $G$ but only for $n\leq6$ and for membrane charges for $n\leq 5$.  

We now turn to the physical fields.  In the $O(d,d)$ DFT case we saw that the $d^2$ components of the metric and NS two-form were combined into a $2d \times 2 d$ matrix $\cH$.  In fact this matrix $\cH$ can be understood to be a representative of the coset $O(d,d)/O(d)\times O(d)$.  The situation here is similar, the number of components contained in the metric, three-form and six form potential exactly equal the dimension of the coset $G/H$ where $G$ are the U-duality groups and $H$ the maximally compact subgroups, i.e. we have for all $n\leq 7$ 
\be
\frac{1}{2}n (n +1) + {n \choose 3} + { n \choose 6} = \dim ( G_{(n)}/ H_{(n)} )  \ . 
 \ee  
The metric, three-form and six form potentials are united in a coset representative which we shall denote by ${\cal M}_{MN}$ which will be a $\dim R_1 \times \dim R_1$ symmetric matrix where we recall that $\dim R_1$ are the number of coordinates in the extended spacetime.

\EDIT{It has been known since the work of de Wit and Nicolai \cite{deWit:1986mz,Nicolai87} who studied the $E_7$ and $E_8$ cases, that fermions of eleven-dimensional supergravity lie in representations of $H$,  the maximally compact subgroup of the U-duality group (more precisely, as recently emphasised in \cite{Coimbra:2011ky,Coimbra:2012af}, they  actually fall in representations of the appropriately defined double cover $\tilde{H}$).}     The appropriate representations are reproduced in table 4.
\begin{table}[htdp]\label{table:spinorrepb}
\caption{The spinor representations  for the gravitino and supersymmetries }
\begin{center}
\begin{tabular}{|c|c|c|c|c|}
\hline
  $n$ & $G$ & $\tilde{H}$   & $\psi$ & $  \epsilon$   \\
\hline
  4 &  $SL(5)$ &   $Spin(5)$    & \bf{16}   &  \bf{4}  \\
  5 & $Spin(5,5)$ &$Spin(5)\times Spin(5)$     &  $\bf{(4,5)} \oplus    \bf{(5,4)}$ & $  \bf{(4,1)} \oplus    \bf{(1,4)}$   \\
  6 & $E_{6}$ & $USp(8)$   & $ \bf{\EDIT{48}} $& $\bf{8}   $   \\
  7 & $E_{7} $&   $SU(8)$  & $ \bf{56}\oplus \overline{\bf{56} }$ &    $\bf{8}\oplus \overline{\bf{8} }  $  \\  \hline
\end{tabular}
\end{center}
\end{table}
Although the inclusion of fermions has been achieved \cite{Coimbra:2011ky,Coimbra:2012af} in the context of the generalised geometry (in which we remind readers no doubling of coordinates takes place - only tangent space is extended),  a comprehensive treatment, at the time of writing, has not been completely extended to the doubled setting.  Recent progress in this direction has been  \cite{Cederwall:2013oaa}.

 We saw that in the case of DFT a key feature of the theory is that there is a constraint (know as the {\it{strong constraint}} or  {\it{physical section condition}})  that reduces the dynamics from the doubled space to a subspace whose dimensionality matches that of physical spacetime. In general, the consistency of the theory requires all the dynamical fields of the theory  obey the  strong constraint.  This in turn means that though the Double Field Theory is a revealing rewriting of ordinary supergravity in which the $O(d,d)$ symmetry is manifest, it is, at least locally, completely equivalent to ordinary supergravity.  However, we also saw that in certain instances this section condition could be relaxed without rendering the theory inconsistent.   We will find both of these features are true in the extended spacetime approach to M-theory of  \cite{Berman:2010is}.  The section condition can be understood as the vanishing of  a certain projection of two derivatives  into a representation which we denote in short hand by 
 \be
 \partial \otimes \partial |_{R_2 }  = 0 \ , 
 \ee
where the representations $R_2$ are given in table \ref{exceptgeom}.   We will see later that indeed this strong constraint can be relaxed in certain circumstances and that this will give the linkage between the  extended spacetime approach to M-theory  and gauged supergravities. 

\EDIT{\subsection{An Overview of the Duality Invariant Construction}}
\EDIT{Before going into detail with a specific example we take the opportunity to summarise the construction of the duality invariant theories giving particular attention to the permissible coordinate dependance. }

\EDIT{We begin first with eleven-dimensional supergravity with its bosonic action given by eq.~\eqref{11dsugra}.  Temporarily,  we denote the full set of eleven spacetime coordinates by $\{\vec{X} \}$. We now make a  split of coordinates   $\{ \vec{X} \} \rightarrow \{t, x^\mu, \vec{y}\}$ which consists of one time coordinate, $t$, $n$ spatial coordinates, $\{x^\mu\}, \mu = 1 \dots n$, and $10-n$ remaining spatial coordinates ${\vec{y}}$.   A conventional dimensional reduction on a torus $T^n$ would result in an $11-n$ dimensional theory obtained by imposing a reduction ansatz that fields do not depend on the coordinates $x^\mu$:
\begin{equation}
\Phi(\vec{X}) \equiv \Phi(t, \vec{y}) \ . 
\end{equation}
This dimensionally reduced theory exhibits the hidden symmetry group $E_{n(n)}$. }

\EDIT{Our ultimate goal would be to reformulate the entire eleven-dimensional supergravity, without making {\em any} dimensional reduction or truncation, in such a way that $E_{n(n)}$ is promoted to a manifest global symmetry.  }

\EDIT{In a  step towards this final goal, let us consider from the outset not the full eleven-dimensional theory  given by eq.~\eqref{11dsugra} but rather a $n+1$ dimensional toy model 
\ba\label{protoM}
S_{1+n} = \frac{1}{2\kappa^2} \int dt d^nx\, \sqrt{G} \left( R - \frac{1}{48} F_{4}^2 \right)  \, .
\ea
The field content is a metric and a three-form which can depend on time and the $n$ spatial coordinates $\{ x^\mu\} $.  There is, of course, no Chern-Simons term here.  We are, in this prototype, completely ignoring the dimensions and coordinates denoted $\{\vec{y}\}$ above.  A dimensional reduction on $T^n$ with an ansatz on fields 
 \begin{equation}
\Phi(t,x^\mu) \equiv \Phi(t) \ . 
\end{equation}
would give rise to a one-dimensional action which exhibits an $E_{n(n)}$  hidden symmetry.   To make our toy-model even simpler, we shall assume that the metric and three-form entering into eq.~\eqref{protoM} are of the form 
\begin{equation}\label{fieldcontent}
G = \left(\begin{array}{c|c} -1 & 0 \\ \hline 0 & g_{\mu \nu}(t, x)  \end{array} \right) \ , \quad C_3 = C_{\mu\nu\rho}(t, x) dx^\mu\wedge dx^\nu \wedge dx^\rho \ ,
\end{equation}
that is to say all fields with temporal legs are set to zero (this is less of a restriction than it may at first seem; for the metric this is equivalent to choosing synchronous gauge in which lapse is set to unity and the shift vector to zero). }

\EDIT{Our more modest immediate goal is the following: {\em to reformulate the $1+n$ dimensional proto-M-theory of eq.~\eqref{protoM} in such  a way that $E_{n(n)}$ becomes a manifest global symmetry without assuming any dimensional reduction ansatz and with fields depending on all dimensions $\{t, x^\mu \}$.} }

\EDIT{To make this possible we will extend the spacetime with  a certain ($n$-dependent) number of additional coordinates that can be thought of, as discussed above and in analogue to the case of DFT, as conjugate to brane winding modes.  This extended space, which will have coordinates $\{\mathbb{X}\}$, supports a linear action of the duality group $E_{n(n)}$.   The metric and three-form of eq.~\eqref{fieldcontent} will be combined into a generalised metric on this extended space. Having done this, we then allow all fields and gauge parameters to depend on not just the physical spacetime coordinates $\{t, x\}$ but on the full gamut of coordinates on this extended spacetime i.e. on $\{ t, \mathbb{X}\}$.   On this extended spacetime we will find a local symmetry encoded by a generalised Lie-derivative which encodes both conventional $n$-dimensional diffeomorphisms and three-form gauge transformations.   Finally, to recover the conventional spacetime and to close the gauge algebra we impose a constraint, the physical section condition, that reduces the theory back to one equivalent to eq.~\eqref{protoM}.}

\EDIT{An evident shortcoming of this proto-M-theory approach is that  a number of dimensions (the $\{\vec{y}\}$ above) are ignored, nonetheless, it is a substantial achievement that  the $E_{n}$ duality group can be made manifest without dimensionally reducing on the $n$-coordinates in which it acts.   The assumptions made in the proto-M-theory approach were designed to simplify the scenario sufficiently to make explicit progress, they are technical in nature and it seems, at least in the authors' opinions, quite possible to relax them.  Indeed, since the publication of the first version of this report on the {\tt arXiv} there has been some great progress \cite{Hohm:2013uia,Hohm:2013vpa} towards doing just this and restoring all the dimensions and making the duality manifest in a full $n \oplus 11- n$ splitting of coordinates.  We anticipate that in the near future this will be completely achieved in full detail, at least for case $n\leq 7$.  }

\EDIT{This construction is best illustrated through concrete examples and since the details vary according to dimension, we focus first on the case of $n=4$.}

\subsection{An Example: $SL(5)$} 
Let us turn now to a concrete example, the case of $n=4$.  \EDIT{That is we consider a proto five-dimensional theory
\ba
S_{5} = \frac{1}{2\kappa^2} \int dt d^4x\, \sqrt{G} \left( R - \frac{1}{48} F_{4}^2 \right)  \, .
\ea
The $E_4 = SL(5)$ duality group acts along four dimensions which we denote by the coordinates   $x^\mu$, $\mu =1 \dots 4$.  In keeping with the work \cite{Berman:2010is},  we will adopt a Hamiltonian approach and after fixing to synchronous gauge as in eq.~\eqref{fieldcontent}  the ten degrees of freedom in the metric are contained in a $4 \times 4$ symmetric matrix $g_{\mu \nu}(x,t)$ which can depend on the spatial coordinates $x$ and time. Similarly the four degrees of freedom in the three-form are contained in  $C_{\mu\nu\rho}(x, t)$.  Since we have fixed to synchronous gauge, covariance in only the four spatial $x$ directions will be maintained.}

\EDIT{We now extend the spacetime} by  introducing an additional six winding coordinates $y_{\mu \nu }= y_{[\mu \nu]}$. Together with the $x^{\mu}$ these for a $\bf{10}$ of  $SL(5)$ which we can express in a covariant way as\footnote{ We will use capital Roman indices, $\mathbb{X}^M,$ to denote the coordinate representation of the generalised spacetime (i.e. the {\bf 10} of $SL(5)$),  lower case Roman indices $X^m$ to denote the fundamental representation of the duality group (i.e. the {\bf 5} of $SL(5)$) and Greek indices $x^\mu$ denote the coordinates of physical spacetime. Barred indices denote flattened/tangent/group indices in the corresponding representation.  } 
\be
\X^M = \X^{[ab]} =  \left\{
\begin{array}{l}
\X^{\m 5}=x^\m,\\
\X^{5\m}=-x^\m,\\
\X^{\m\n}=\frac{1}{2}\eta^{\m\n\a\b}y_{\a\b}.
\end{array}
\right.,
\ee
where $\eta_{\m\n\a\b}$ is the totally antisymmetric symbol $\eta_{1234}=1$, $\eta^{1234}=1$ and the indices $a,b=1\dots5$.   

The the 14 components contained in the metric and three-form fields may be packaged into the $SL(5)/SO(5)$ representative 
\def\cM{{\cal M}}
\begin{equation}
\label{genmet}
\cM_{M  N} =  
\left(\begin{array}{cc}
g_{\m\n}+\frac{1}{2}C_\m^{\phantom{\m}\r\s}C_{\n\r\s} & -\frac{1}{2\sqrt{2}}C_\m^{\phantom{m}\r\s}\eta_{\r\s\a\b}\\
-\frac{1}{2\sqrt{2}}C_\n^{\phantom{m}\r\s}\eta_{\r\s\g\delta} & g^{-1}g_{\g\delta,\a\b} 
\end{array} \right) \ , 
\end{equation}
in which   $g=\det g$ and $g_{\m\n,\a\b} = \frac{1}{2} (g_{\a \m} g_{\b \n}- g_{\a \n} g_{\b \m} )$.    We shall demonstrate later a method to construct these generalised metrics harnessing some more group theoretic power.  \EDIT{We now allow this generalised metric to depend on {\em all} the coordinates of the extended spacetime: 
\be
\cM_{M  N} =\cM_{M  N}(t, \mathbb{X}) \ . 
\ee  }
We now ask a crucial question - how are gauge symmetries encoded in this extended spacetime? It was shown in \cite{Berman:2011jh}   that the diffeomorphism and gauge symmetry of the 3-form potential are a result of reparametrisations of the ordinary space coordinates and winding coordinates, respectively. These form a Courant bracket algebra.  In the extended spacetime we can find an $SL(5)$ covariant version of the derivative that generates generalised diffeomorphism (in much the same way the Dorfman derivative of generalised geometry was promoted to the D-derivative of DFT).  The result is that on a vector we define the generalised derivative
\begin{equation}
{\cal{L}}_\xi V^{ab}=\frac{1}{2}\xi^{cd}\partial_{cd}V^{ab}+\frac{1}{2}V^{ab}\partial_{cd}\xi^{cd}+V^{ac}\partial_{cd}\xi^{db}-V^{bc}\partial_{cd}\xi^{da} \ .
\end{equation}
Generalised Lie derivative is an apt name since the derivative can be expressed as a group theoretic modification to the standard Lie derivative
\beq
\label{sl5deriv}
{\cal L}_X V^M = X^N\pl_N V^M - V^N\pl_N X^M + \e_{a PQ} \e^{a MN} \pl_N X^P V^Q \ ,
\eeq
in which the object $\e_{a M N}$ with mixed indices should be understood as the group invariant taking two $\bf{10}$'s and projecting to the $\bf{5}$ --  it is just the epsilon tensor when we write each of the capital Roman indices of the $\bf{10}$ as an antisymmetric pair of fundamental $\bf{5}$ indices.    When this derivative acts on the generalised metric, one finds upon expanding out into components and setting the derivatives in directions of the winding coordinates to zero, that it does indeed generate diffeomorphism and p-form gauge symmetries.

For the gauge algebra to close we need to invoke the section condition which in this case is given by the projection of two derivatives to the $\bar{5}$ i.e. 
\be
\e^{a MN} \partial_M \bullet \otimes \partial_N \bullet   \equiv \frac{1}{4} \e^{a bc de } \partial_{bc} \bullet \otimes \partial_{de }  \bullet \equiv 0  \ , 
\ee  
in which the bullets denote any field or gauge parameter.  \EDIT{In particular we would require that 
\be
\e^{a MN} \partial_M \bullet   \partial_N \cM_{P Q}(t, \mathbb{X})   = 0  \ .  
\ee }
One way to satisfy this section condition is by setting all derivatives in the winding directions to zero -- it is in this conventional frame  that one reproduces the standard supergravity.

One now needs to provide an action principle encoding the dynamics of the theory.  In this proto-M-theory all fields can depend on all the internal coordinates of the extended spacetime (subject to the section condition), as well as time.   The dynamics for this  generalised metric is thus given by a Hamiltonian  density 
\be
{\cal H} = T + V \ , 
\ee
with the  kinetic term  (where the dots indicate time derivatives)
\be
\label{TBP}
T = -\sqrt{g} \left(  \frac{1}{12} tr (\dot \cM^{-1} \dot \cM ) + \frac{1}{12} (tr (  \cM^{-1} \dot \cM ))^2 \right) \ , 
\ee   
and a potential  
\be
\label{VBP}
\frac{1}{\sqrt{\det{g}} } V = V_1  + V_2 + V_3 + V_4  \ , 
\ee
with 
  \ba
 V_1 =  \frac{1}{12} \cM^{MN} (\partial_{M} \cM^{KL}) (\partial_{ {N}}\cM_{ {K} {L}}) \ ,  &\quad& 
V_2 = - \frac{1}{ 2} \cM^{ {M} {N}} (\partial_{ {N}} \cM^{ {K} {L}}) (\partial_{ {L}} \cM_{ {K} {M}}) \ ,   \\
 V_3 =  \frac{1}{12} \cM^{ {M} {N}} ( \cM^{ {K} {L}} \partial_{ {M}} \cM_{ {K} {L}})( \cM^{ {R} {S}} \partial_{ {N}} \cM_{ {R} {S}})\ , &\quad&
V_4 =  \frac{1}{4} \cM^{ {M} {N}}  \cM^{ {P} {Q}}  ( \cM^{ {K} {L}} \partial_{ {P}} \cM_{ {K} {L}})(   \partial_{ {M}} \cM_{ {N}  {Q}})\ . \nonumber
\ea
\EDIT{ This density should be integrated over the full extended spacetime to obtain the Hamiltonian, so formally we may write  
\be
H = \int dt d^{4+6} \mathbb{X} {\cal H} \ . 
\ee }The powers of square-root of the determinant of the (regular) metric that enter into the above can be written in terms of the determinant of $\cM$ by 
\EDIT{\be
\det g =  (\det \cM)^{-1/2} \, .
\ee}
It may also in fact be absorbed into a rescaling of the generalised metric.\footnote{Upon rescaling the generalised metric by a power of $\det g$ the action takes the same form but the constants in front of each of the four contractions $V_i$ entering into the potential will be altered. Also with a normalisation different to that given in eq.~\ref{genmet} the derivative defined in eq.~\ref{sl5deriv} would also need to be modified to include an appropriate density weight term.}

One can show that this system is invariant under the gauge symmetries generated by the generalised Lie derivative and that upon expansion and setting derivatives in the winding coordinates to zero  it does reproduce exactly the dynamics of the component metric and three-form, and after a lengthy exercise one finds that the potential reduces to the expected form of 
\be 
V= \sqrt{g} \left( R(g) +  \frac{1}{48} F^2 \right) \ .
\ee 
In showing such an equivalence one is required to perform certain integrations-by-parts.  One might be tempted to throw away any resultant boundary terms however in standard gravity we know boundary terms, in particular the York-Gibbons-Hawking term are an important feature in setting up well defined boundary conditions for a variation principle.     A nice feature is that the duality invariant action presented above can be supplemented with a boundary contribution, again with manifest duality invariance, such that when expanded into components and combined with contributions coming from integrations-by-parts  one recovers the  York-Gibbons-Hawking term.    This boundary term is given by \cite{Berman:2011kg}\footnote{A similar result holds for the DFT case.}
\be
S_{bound} = \int_{\partial M}    2 \cM^{MN} \partial_M N_N +  N_N \partial_M  \cM^{MN} \, ,
\ee
where $N_M$ is the normal vector to the boundary in the doubled space.  In the case where we solve the section condition by setting the derivatives in directions of the winding coordinates to zero this is given by 
\be
N_M =  {n^\m  \choose -\frac{1}{\sqrt{2} }C_{\a \b \mu} n^\mu  } \ , 
\ee
where $n^\mu$ is the normal to the boundary in non-extended space. 

\subsection{The Gauge Structure }\label{sec:gaugestruct}
Let us now look beyond the specific $SL(5)$ example presented above.  First we shall discuss the gauge structure which we shall see exhibits a rather  intricate ghost structure.   In conventional geometry diffeomorphisms are generated by the Lie derivative
\be
\delta_u v^m = L_u v^m = [u, v]^m = u^n \partial_n v^m - (\partial_n u^m)  v^n \ . 
\ee
It is useful to view the first term as a transport term and the second as a $\frak{gl}(n)$ transformation matrix  $(\partial_n u^m)$  acting on an object in the fundamental.  In the M-theory context one may look for a similar structure but with the algebra $\frak{e}_{n}$ (together with a real scaling) playing the role of $\frak{gl}(n)$.      That is one is led to consider gauge variation and derivative of the form  
 \be
 \label{Waldramform}
 \delta_U V^M  = {\cal L}_U V^M = U^M \partial_M V^N - \alpha (P_{adj})^M{}_{N}{}^P{}_Q \partial_P U^Q    V^N   + \beta \partial_N U^N V^M \ ,  
 \ee
where $\alpha$ and $\beta$ are constants to be determined and $P_{adj}$ denotes the projection of the tensor product $R_1 \otimes \bar R_1$ to the adjoint representation ($R_1$ is the coordinate representation as given in table \ref{exceptgeom}).  We may refer to this object as a generalised Lie-derivative or \EDIT{in} keeping with the terminology introduced previously a D-derivative.   An alternative view is to just consider adding to the conventional Lie derivative a group theoretic contribution given by the ansatz 
\be\label{eq:genderiv}
 {\cal L}_U V^M =  L_U V^M + Y^{MN}{}_{PQ} \partial_N U^P V^Q  \ , 
\ee
where $Y^{MN}{}_{PQ}$ is to be determined and built from the invariant tensors of the U-duality group.   Just as with DFT, we introduce the C-bracket given by the antisymmetrisation of the D-derivative
\be
[[ U, V]] = \frac{1}{2}  ( {\cal L}_U V - {\cal L}_V U)  \ . 
\ee
The tensor $Y$ can be completely determined by requiring that the D-derivatives close on an algebra given by
\be
[{\cal L}_U  , {\cal L}_V] = {\cal L}_{[[U, V]]}  \ . 
\ee
By performing the commutators one immediately encounters a constraint of the form 
\be
Y^{MN}{}_{PQ} \partial_M \otimes \partial_N = 0 \ . 
\ee
This equation is roughly speaking the section condition.  There are some further conditions on $Y$ detailed in \cite{Berman:2012vc} that completely determine the full form of these tensors to be\footnote{ $\e_{i M N} = \e_{i mn , pq}$ is the $SL(5)$ alternating tensor;   $(\g^i)^{MN}$ are  $16\times16$ MW representation of the $SO(5,5)$ Clifford algebra (they are symmetric and  $(\g^i)_{MN}$ is the inverse of $(\g^i)^{MN}$); $d^{MNR}$ is a symmetric invariant tensor of $E_6$ normalized such that $d^{MNP}\bar{d}_{MNP} = 27$; $c^{MNPQ}$ is a symmetric tensor of $E_7$  and $\epsilon^{MN}$ is the symplectic invariant  tensor of its {\bf 56} representation.  In all cases except $E_{7(7)}$, the tensor $Y$ is symmetric in both upper and lower indices. 
}
\bea
SL(5):  & \quad & Y^{MN}{}_{PQ}= \e^{i MN}\e_{i PQ} \ , \nonumber    \\
SO(5,5): &\quad & Y^{MN}{}_{PQ}  = \frac{1}{2} (\g^i)^{MN} (\g_i)_{PQ} \ ,  \nonumber  \\
E_{6(6)}: &\quad & Y^{MN}{}_{PQ}  = 10 d^{MN R} \bar{d}_{PQR} \ ,\nonumber   \\ 
E_{7(7)}: &\quad & Y^{MN}{}_{PQ}  = 12 c^{MN}{}_{PQ} + \delta^{(M}_P \delta^{N)}_Q + \frac{1}{2} \e^{MN} \e_{PQ } \ . \nonumber  
\eea
The same structure is present in the $O(d,d)$ DFT case for which the tensor takes the form $Y^{MN}{}_{PQ}  = \eta^{MN} \eta_{PQ} $.   In all cases there is a rearrangement that allows one to cast the derivative exactly in the form  of eq.~(\ref{Waldramform}) with a projector to the adjoint.   These results were found for the case of $SL(5)$ in \cite{Berman:2011cg} with the general case given by  \cite{Coimbra:2011ky} and further developed in \cite{Berman:2012vc}.

We require also that the Jacobi identity still holds in a suitable sense.  To see that this is indeed the case one notes that the symmetric part of  ${\cal L}_U V$, 
\be
(( U, V )) = \frac{1}{2}  ( {\cal L}_U V + {\cal L}_V U)  \ . 
\ee
generates a trivial (zero) transformation $  {\cal L}_{((U, V))}=0$. The Jacobiator can be expressed as
\be
Jac( U , V,  W) = \frac{1}{3} (( [[U, V]], W )) + cyclic \ . 
\ee
Then although the Jacobi identity does not hold in itself, when viewed as a gauge transformation the Jacobiator generates zero transformations.

We must now ask how these derivatives account for the gauge parameters of the underlying supergravity.   Diffeomorophisms of the metric account for $n$ parameters,  the gauge parameters associated to $C_3$ provide ${n-1 \choose 2}$ and for $C_6$ we have ${n-1 \choose 5}$.  The counting of gauge parameters for p-forms is slightly subtle because of reducibility; one would at first say that $\delta C_3 = d \Lambda_2$ accounts for ${n \choose 2}$ parameters however this over counts  since $n$ parameters of the form $\Lambda_2 = d \Lambda_1$ need to be subtracted, but this in turn now under counts and we need to add back a single parameter of the form $\Lambda_1 = d \Lambda_0$.    In total the gauge transformations of $C_3$ contribute $ {n-1 \choose 2}= {n \choose 2} - n + 1 $ parameters.  

A novel feature of the generalised D-derivatives introduced above is that they actually  display infinite reducibility or ghosts and ghosts for ghosts.  Lets give an example to see this more clearly and, for the sake of  variety,  we choose $n=5$ where the group is $G= Spin(5,5)$ and the coordinates are in the $\bf{16}$.  The derivative is given by  (contracted spinor indices implicit)
\be
{\cal L}_U V= (U \partial) V^\b +   \frac{1}{8} ( \partial \g^{ab} U) (\gamma_{ab}   V)^\a + \frac{1}{4} (\partial U) V^\a
\ee
One finds that by virtue of the section condition, transformations of the form $U^\alpha = \gamma^{\alpha\beta}_a  \partial_\beta \xi_{(1)}^a$ automatically vanish.  This gives rise to a first order reducibility in the $\bf{10}$.   However for $\xi_{(1)}^\a = (\partial \gamma^a \xi_{(2)})$ , one finds that using an appropriate Fierz identity that  $U^\alpha = \gamma^{\alpha\beta}_a  \partial_\beta \xi_{(1)}^a =  0$ which gives a second order reducibility in the $\bf{16}$.   This continues with a third order reducibility in the $\bf{45}$ and so on.   One can derive a partition function for this reducibility 
\be
Z_5(t) = (1- t)^{-16} (1- t^2)^{10} (1- t^3)^{-16} ( 1- t^4)^{45} \times \dots
\ee
which is of the form $Z =\prod_k (1- t^k)^{A_k}$ where $A_k$ is $(-1)^k$ times the dimension of the $k^{th}$ order reducibility.   Then the total number of gauge parameters is given by the regulated alternating sum $\sum_k A_k$.  One can show that the full partition function in this case can be expressed as 
\be
Z_5(t) = (1- t)^{-11} (1+ t)(1+ 4t + t^2) \ , 
\ee
 and from which one can extract   $\sum_k A_k$ as the exponent of the term that becomes singular at $t=1$. In this case there are $11$  gauge parameters exactly in accord with the conventional counting $11 = 5 + {4 \choose 2}$ described above.   The same approach yields partition functions for $n\leq 7$ which all give the correct counting of gauge parameters.  Intriguingly, this counting  of gauge parameters works even for $n= 8$ even though the algebra does not close correctly. 
 
\subsection{The Section Condition Revisited }
The coordinates of extended spacetime live in a representation $R_1$ of the U-duality group and the section condition is given by projecting out a certain representation $\bar{R}_2$:
\be 
(\partial \otimes \partial) |_{\bar{R}_2} = 0 \ .
\ee 
The section condition reduces the dynamics of the theory to take place on a subspace whose dimension is that of physical spacetime.  Different solutions to the section condition thus relate theories in different duality frames.   We saw above that the section condition was required also to close the gauge algebra.   

However as formulated above the section condition is a rather unwieldy mathematical object.   It would be desirable to have a linear constraint that defines the same subspace.  We are reminded of how a pure-spinor in generalised geometry does exactly this by defining its  maximal isotropic annihilator.   We seek a generalisation of this concept for the U-duality case.   We introduce an auxiliary object $\Lambda$ in some representation $\bar R_3$ that is constrained (much as a pure spinor is constrained) by a relation of the form
\be
\Lambda^2 |_{P} = 0 
\ee
for some representation $P$  with which a linear section condition can be formed by 
\be
\Lambda \otimes \partial |_{\bar R_4} = 0 \ , 
\ee  
where the representation $R_{4}$ is to be determined. 
Let us give two explicit examples of this construction and for details of other cases we refer the reader to \cite{Berman:2012vc}. 

\subsubsection{n=4}
In this case $G= SL(5)$ and we have $R_1= \bf{10}$ and $R_2 = \bf{5}$  and so the section condition is given by 
\be
e^{abcde} \partial_{bc} A \partial_{de} B = 0 \ , \qquad  a = 1 \dots 5  \, .
\ee
Instead we introduce an auxiliary $\Lambda_a$ in the $R_3 = \bf{\bar{5}}$ and consider the linear section condition 
\be
\Lambda_{[a } \partial_{bc]}=0  
\ee
which transforms in the $\bar R_4 = \bf{10}$.   If we choose a particular choice for $\Lambda$, for instance $\Lambda_1 \neq 0$ with other components vanishing we find the linear section condition would imply $\partial_{23}=\partial_{24} = \partial_{25}= \partial_{34} = \partial_{35} =\partial_{45} = 0 $ and then clearly the non-linear section condition is obeyed.  The linear section condition serves to reduce the dynamics to a four-dimensional physical space.  (In this case $\Lambda$ was not subject to any additional ``purity'' constraint.)

\subsubsection{n=5}
In this case $G= Spin(5,5)$ and we have $R_1= \bf{16}$ and $R_2 = \bf{10}$  and so the section condition is given by 
\be
\gamma^{a\a\b}  \partial_{\a} A \partial_{\b} B = 0 \ , \qquad  a = 1 \dots 10\ , \quad \a = 1 \dots 16 \ . 
\ee
We introduce an auxiliary spinor $\Lambda^\a$ in the $R_3= \bf{16}$ and further more invoke a purity constraint (and in this case we really do mean a conventional pure spinor of $Spin(5,5)$ which has 11 independent components).  With this the linear section condition may be expressed as
\be
0 = \Lambda^\a (\gamma^{ab})_\a{}^\b \partial_\b 
\ee
which transforms in the $\bar R_4 = \bf{45}$.  To see that this does the job one makes use of the identification between spinors and forms so that we may write $\Lambda$ as the sum of odd forms and $\partial$ as the sum of even forms:
\be
\Lambda = \phi +  u_{[ab]} + t_{[abcd]} \ , \quad \partial = \partial_{a} + \partial_{[abc]} + \partial_{[abcde]}  \ , \quad     a = 1 \dots 5 \ . 
\ee
The pure spinor constraint on $\Lambda$ determines the four-form piece in terms of the other components.  If we choose  only the zero-form component of $\Lambda$ to be non-zero we find, using the Mukai pairing of forms, that this sets to zero the three-form and five-form components of $\partial$.   What remains are the five components in the direction of the one-form component of $\partial$.  This provides a solution to the non-linear section condition and serves to reduce the dynamics to a five-dimensional physical space.

\subsection{Construction of Generalised Metrics}

The technique of non-linear realisation is rather algebraically involved but let us sketch it with an example omitting the computational details. It has its origins in the work of Borisov and Ogievetsky \cite{Borisov:1974bn}. Here we work with $SL(5)$ but details of the other cases can be found in  \cite{Berman:2011jh}.  The essential trick is to consider not just $SL(5)$ but rather its motion group.   The motion group is the semi-direct product of a group with a representation; the Poincar\'e group is formed by taking the semi-direct product of translations and the Lorentz group.  In this case the motion group of $SL(5)$ will be the semi-direct product of $SL(5)$ with translation generators in the ten-dimensional extended spacetime.  The 24 trace free generators of $SL(5)$ obey 
\be
[M^{a}{}_b , M^{c}{}_d ] = \delta^c_b  M^{a}{}_d -     \delta^a_d M^{b}{}_c   \quad a = 1 \dots 5  \ .  
\ee
The  ten translation generators $P_{ab}= P_{[ab]}$ are normalised so that $Tr(P_{ab} P_{cd} ) = \delta_{ac} \delta_{bd} -  \delta_{ad} \delta_{bc}$ and  have a commutator
\be
[M^{a}{}_b, P_{cd} ] = 2 \delta^a_{[c}  P_{d ] a}   + \frac{2}{5} \delta^a_b P_{cd} \ . 
\ee
It is useful to decompose under $GL(4) \times U(1)$. Under this adjoint splits as  $\bf{ 24 \rightarrow 1_0 + \EDIT{15}_0 + 4_3 + \bar{4}_{-3} }$ which we write as 
\be
M^a_b \rightarrow  K, \  K^i{}_j \ ,  R_{ijk} \ , R^{ijk}  \ , 
\ee
and similarly for the translation generators 
\be
P_{ab} \rightarrow  P_i, \ Z^{ij}  \ . 
\ee
Full details of the commutators in this basis can be found in  \cite{Berman:2011jh}.    One begins by defining the metric on the extended spacetime by introducing a group element 
\be
g_l = e^{x^i P_i} e^{ \frac{1}{\sqrt{2}} y_{jk} Z^{jk} } 
\ee
from which we obtain the left-invariant one-forms 
\be
g_l^{-1} d g_l = dx_i P^i +  \frac{1}{\sqrt{2}} dy_{jk} Z^{jk} = d\X^M P_M
\ee
where we revert to the notation of $\X^M$ being the coordinates in the $\bf{10}$ and $P_M$ the generators of the translations.  This defines the tangent space metric
\be
Tr( g_l^{-1} d g_l \otimes  g_l^{-1} d g_l)  =  \delta_{ij} dx^i dx^j + \delta_{ij, kl} dy_{ij} dy_{kl} \ . 
\ee
To introduce fields we now perform a conjugation by a second group element 
\be
g_E = e^{h_i{}^j  K^i{}_j }  e^{\frac{1}{3!} C_{ijk} R^{ijk} } 
\ee
where $h$ essentially defines the graviton and $C$ the three-form of supergravity.   One finds that 
\be
g_E^{-1}  (g_l^{-1} d g_l) g_E =   {\cal E}_M{}^{A} d\X^M   P_A \ , 
\ee
defines a vielbein for the generalised metric on extended spacetime, namely 
\be
\cM_{MN} = {\cal E}_{M}{}^A {\cal E}_{N}{}^B \delta_{AB} 
\ee 
After some work, using the exact form for the group commutators and various applications of the Baker-Campbell-Hausdorff formula, it can be shown that the metric obtained in this way  coincides with the one presented earlier up to an overall rescaling factor of the $\det g$ (which can be redefined away should one so desire).

This general construction corresponds to the explicit formulations given for the $SO(5,5)$ duality group in  \cite{Berman:2011pe}, the $E_{7}$ formulation of Hillmann which is close related to this work \cite{Hillmann:2009ci}.  Recent progress towards an $E_{8}$ formulation has been reported in \cite{Godazgar:2013rja}.

This is all part of what is known as the $E_{11}$ programme which has been developed over a number of years by Peter West and collaborators. Already in 2003 it was proposed  to consider the non-linear realisation of
$E_{11}$ and its first fundamental represenations \cite{West:2003fc}. This introduced a
 generalisation of  our usual notion of spacetime to include
 additional coordinates and it was equiped with  a corresponding
 generalised vielbein. As the first fundamental representation
 contains all brane charges \cite{West:2004kb,Kleinschmidt:2003jf,West:2004iz,Cook:2008bi} there is a one to one
 correspondence between the coordinates introduced and  brane charges.
 This construction has be used to find all  the gauged  maximal
 supergravities in five dimensions \cite{Riccioni:2007ni}.

 In \cite{West:2010ev}, West constructed  this non-linear realisation, at level zero
 from the IIA viewpoint, and the result was a dynamics that contained
 the NS fields and agreed with that of double field theory.  This
 work was extended \cite{Rocen:2010bk} to level one and the RR sector of IIA
 supergravity. 

In fact, the embedding of the finite exceptional algebras into $E_{11}$ has one specific technical advantage which is that the appropriate measure emerges for the action which is not the case if one uses other realisations.

\subsection{Relation to DFT}
Given their evident similarities it is natural to ask  exactly how do the U-duality invariant theories described above relate to the DFT introduced in the previous chapter.   

Of fundamental importance in the string theory duality web is the relationship between eleven-dimensional supergravity and type IIA supergravity in ten-dimensions  as described in \cite{Witten:1995ex} which we now outline schematically.  We begin with the bosonic sector of eleven-dimensional supergravity consisting of a metric and a three-form $C_3$ whose action is given by 
 \be
S^{(11)} = \frac{ 1}{2 } \int d^{11}x  \sqrt{g} \left( R - \frac{1}{48} F_{4}^2 \right) - \frac{1}{6} F_4 \wedge F_4 \wedge C_3  \ . 
\ee
We reduce to ten dimensions along an   $S^1$ direction,  labelled by the coordinate $z$,  taking an ansatz  
\be
\label{eq:KKansatz}
ds^2_{11} = \tilde{g}_{ij}^2dx^i dx^j + e^{2 \gamma  } (dz+ C_1)^2  \ , \quad C_{3} =  C_3+  B\wedge dz \ . 
\ee 
This results in an action for the massless fields of the form (the ellipsis indicates $C_1$ and $C_3$ terms) 
\be
S^{(10)} = \frac{1}{2 } \int d^{10}x \sqrt{\tilde{g}} e^{\g} \left( R[\tilde{g}]   - 2 (\partial \g)^2 - 2 \tilde{\nabla}^2 \g - \frac{1}{12} H_{ijk}H^{ijk}e^{2 \g} \right) + \dots \ . 
\ee
At first sight this does not look like the action of type II supergravity but it can be brought in to the correct form by performing a Weyl rescaling and identifying the scalar of the KK ansatz in the following manner: 
\be
\tilde{g}_{ij} = e^{-\gamma} g_{ij} \ , \qquad \gamma =  \frac{2}{3}\phi  \ , 
\ee
where $\phi$ is the dilaton.  The result of doing this is to arrive exactly at the action of type IIA supergravity (with $C_1$ and $C_3$ being the relevant RR potentials) 
\ba
S_{\rm  \ IIA} & = & {1\ov 2 }\int_{M_{10}} \sqrt{g} \Bigg[e^{-2\Phi}
\left(R + 4 (\del\Phi)^2 -\frac{1}{12}{H^2}\right)  -\ha\left(  \frac{1}{2}{F_2^2} +\frac{1}{4!}{F_4^2}\right)
\nonumber\Bigg]    - \ha     C_3 \wedge dC_3 \wedge B 
\ .
\label{actmIIA} 
\ea
The relation between the KK scalar $\g$ and the dilaton $\phi$ leads to one of the most famous results of M-theory: it relates the radius of the M-theory circle to the string coupling according to $R_{11} = e^\gamma = e^{\frac{2}{3} \phi} = g_s^{\frac{2}{3}}$.   The Weyl rescaling  also implies that masses of the KK modes are of order $g_s^{-1}$; consistent with their interpretation as D0 bound states. 
 
 Now we turn to the reduction of the  duality invariant M-theory.  We  anticipate that there ought to be a relationship between the $E_{n }$ covariant theory and the $SO(n-1, n-1 )$ DFT. From a representation theory perspective this can be understood by the embedding of the Dynkin diagram of the latter into the former (for the $E_{10}$ perspective see \cite{Kleinschmidt:2004dy} and the $E_{11}$ perspective  \cite{West:2010ev,Rocen:2010bk} ).  Here we take a more pedestrian approach \cite{Thompson:2011uw} that tries to reproduce the above dimensional reduction in the duality invariant context.   Since we do not yet have a U-duality invariant formulation for the full eleven-dimensional theory \EDIT{we must work with the $1+n$ dimensional prototype M-theory described by eq.~\eqref{protoM}.  This introduces some subtleties.    Whilst the metric and two-form sectors of the theory are dimensionally insensitive, the dimensional reduction in the dilaton sector must be treated with some care.  When reducing the $1+n$ dimensional gravity theory of eq.~\eqref{protoM} to $1+(n-1)$ dimensions,  the relation  between the KK scalar and the would-be dilaton needs to be modified  to $(2- \frac{n}{2}) \gamma = - 2 \phi$ such that after the Weyl rescaling the NS sector has a common $e^{-2 \phi }$ pre-factor.}  Also the coefficient of the dilaton kinetic term is sensitive to the dimension.

As ever we illustrate this with a specific example; in this case the reduction from $SL(5)$ invariant M-theory to $SO(3,3)$ invariant DFT.  The extended spacetime of the $SL(5)$ theory is ten dimensional consisting of four regular coordinates and six winding coordinates.  It is evident that one must reduce one regular coordinate (the $S^1$ reduction above) and three winding coordinates so as to arrive at the three+three dimensional extended space of the Double Field Theory.  In essence we use the decomposition of the antisymmetric representation of $SL(5)$   to a vector plus a spinor of $SO(3,3)$ i.e.  ${\bf 10} \rightarrow {\bf 6} + {\bf 4}$. 

For the remainder of this section we will make a small notational change:  to avoid confusion we shall label indices in the vector representation of the extended M-theory spacetime with hatted Roman letters.  The ten coordinates of the $SL(5)$ geometry are  $\X^{\hat{M}}=  X^{ab} = (X^{\mu 5} , X^{\mu \nu}) \equiv (x^{ \mu} ,   \frac{1}{2}\eta^{\m\n\a\b}y_{\a\b} ) $   where we have indicated the decomposition under  $SL(4)$.  We will further decompose the $SL(4)$ indices under $SL(3)$ so that e.g. $X^{\mu 5} = x^\mu = (x^i , z)$.    Correspondingly the derivatives decompose under $SL(3)$ as $\partial_{\hat{M}} = ( \partial_{z 5} , \partial_{i 5 } , \partial_{ij} , \partial_{i z}) $. 

To perform the reduction we will assert that any derivative carrying a ``$z$'' index is set to zero. The coordinates that survive the dimensional reduction are thus given by $\X^M = (x^i, \tilde{x}_i) =  (x^i , y_{z i })$ .   Let us first explore the consequence of this for the section condition  
\be
\e^{a b c d e} \partial_{ b c  } \bullet \partial_{d e } \bullet = 0  \ . 
\ee    
The $a = z$ component of this yields, 
\be
0=\e^{z b c d e} \partial_{ b c  } \bullet \partial_{d e } \bullet =   \e^{z  i j k 5} \partial_{  i j   } \bullet \partial_{ k 5 } \bullet  = \e^{ijk} \partial_{  i j   } \bullet \partial_{ k 5 } \Rightarrow \tilde \partial^i \bullet \partial_i \bullet = 0  
\ee    
which is the corresponding section condition for DFT  after making the identification $\partial_{i 5} \equiv \partial_{i}$ and $\e^{ijk} \partial_{  i j   } \equiv \tilde \partial^k $.   The remaining components of the $SL(5)$ section condition are automatically \EDIT{solved} by virtue of the reduction ansatz. Equally we can consider the reduction of the generalised derivative and show in a similar way  the $SL(5)$ derivative reduces exactly to that of the $SO(3,3)$ DFT \cite{Berman:2011cg}. 

Now let us consider the generalised metric. It is straightforward, but somewhat tedious, to insert the  KK ansatz of eq.~\eqref{eq:KKansatz} into the generalised metric ${\cal M}_{\hat{M} \hat{N}}$ for $SL(5)$ which was given explicitly in eq.~\eqref{genmet}. It suits us to reorder coordinates such that $\X^{\hat{M}} = (\X^{M}; X_{\a})=   (x^{i}, \tilde{x}_{i} ;   z, y_{ij}   ) $ i.e. we shuffle components so that   the top left $6\times 6$ block of ${\cal M}_{\hat{M} \hat{N}}$ now represents the reduced generalised metric in `external' directions and the bottom right $4\times 4$  represents the `internal' metric.   One finds that the generalised metric may be written in the form 
 \be
 \label{MKK}
 M_{\hat{M}\hat{N}} =    \left( \begin{array}{c  | c}  
e^{-\g} {\cal H}_{M N} + e^{2 \g} {\cal C}_{M \a} {\cal G}^{\a \b }{\cal C}_{N \b}  & \quad e^{ 2 \g}   {\cal C}_{M \a} \\
\hline 
  e^{ 2 \g}  {\cal C}_{N \b}  & \quad  e^{2 \g}{\cal G}_{\a \b }
 \end{array} \right)  \ . 
 \ee
Here one immediately recognise the generalised metric of DFT of eq.~\eqref{fibremetric} entering in the top left hand corner.  The metric on the internal space ${\cal G}_{\a \b}$, like ${\cal H}_{MN}$, depends on the NS fields and is given by
\be
{\cal G}_{\a \b }   =  \left(  \begin{array}{cc}
 1 + \frac{1}{2} b_{ij}b^{ij}    & \frac{1}{\sqrt{2}}    b^{ij }  \\
\frac{1}{\sqrt{2}}   b^{kl}  &  \frac{1}{2}  \left(g^{k i }g^{j l} - g^{i l} g^{k j} \right) 
 \end{array} \right)  \ . 
\ee
Just as ${\cal H}$ describes an $O(3,3)/O(3)\times O(3)$  coset representative  we can think of the internal metric as describing the same   but in the spinor representation i.e. an $SL(4, \mathbb{R})/SO(4)$ representative  (the spin cover of $SO(3,3)$ is the real form $SL(4, \mathbb{R})$).\footnote{Actually this is not completely accurate since the internal metric is not uni-modular but has a determinant $g^{-2}$ but, as mentioned previously, there is some freedom to rescale the generalised metric by such a factor. }   The off diagonal entries ${\cal C}_{M \a}$, which carry both a vector and a spinor index, contain all the dependence on the RR fields $C_{1}$ and $C_{3}$.  However, since the RR potentials contain only a spinors worth of degrees of freedom,  ${\cal C}_{M \a}$  is reducible.  

 Furthermore the form of   eq.~\eqref{MKK} resembles a KK ansatz.  Thus we may summarise the key result   with the motto: \emph{The standard KK dimensional reduction ansatz  gets lifted to a  doubled KK reduction in the extended spacetime}.  

The reduction then proceeds by brute force, simply plugging the doubled KK ansatz of eq.~\eqref{MKK} into the action defined in eqs.~\eqref{TBP} and \eqref{VBP} and setting derivatives in the internal directions to zero.  An instructive example of the manipulations involved is that 
\ba
\label{MidM}
 M^{\hat{K}\hat{L}} \partial_{M} M_{\hat{K}\hat{L}}  &=& {\cal H}^{{K L}}\partial_{M} {\cal H}_{K L} + {\cal G}^{{\a \b}}\partial_{M} {\cal G}_{{\a \b}} + 2 \partial_{M}\gamma \, . 
\ea
The factor ${\cal G}^{{\a \b}}\partial_{M} {\cal G}_{{\a \b}}$ gives rise to a term like $tr(g^{-1} \partial_{M} g)$ which the very alert reader will realise is   what is needed to obtain the derivative of the DFT T-duality invariant dilaton.  

 The result of this procedure is that one does indeed recover the NS sector of the DFT given in eq.~\ref{DFT}.  There are two subtleties worth commenting on: firstly as discussed already the fact we are not \EDIT{reducing from eleven (regular) dimensions to ten} gives some different identification between the dilaton and the KK scalar $\gamma$.\footnote{Indeed, for $n=4$ one can see that the relationship between the dilaton and the KK scalar becomes ill-defined.  in this case one can only demonstrate the equality under the assumption that $\gamma = \phi=0$ (the DFT  T-duality invariant dilaton remains however non-zero). }    Secondly in the M-theory approach time remains undoubled and moreover the lapse function has been gauged fixed to unity in eq.~\eqref{TBP}  which must be properly taken into account when performing the Weyl rescaling.    
 
 With regard to the RR sector, one finds  upon dimensional reduction that the only terms with RR fields are again   quadratic in derivatives and have the structure
\be
\label{RRterms2}
V_{RR} \propto  {\cal G}_{\a \b} \partial_M {\cal C}_{K}^\a  \partial_N {\cal C}_{L}^\b  \left(\H^{MN} \H^{KL} - 3 \H^{MK} \H^{NL}  \right) \ . 
\ee
Upon expanding into its constituent undoubled fields this can be seen to correctly package RR terms of the type II theory (restricted to 1+3 dimensions).  The reducibility of  ${\cal C}_{M \a}$ then allows this to be recast to match the formulation of section \ref{sec:RRsector}.   
 
\subsection{Relation to Gauged SUGRA}

The construction of gauged supergravities, that is to say supergravity theories which can accommodate the non-Abelian gauge symmetries of Yang-Mills theories,    has a long history dating back to almost the earliest days of supersymmetry.  A   breakthrough was the discovery in early 1980's by de Wit and Nicolai of the $SO(8)$ gauging of four-dimensional maximal ({\cal N}= 8) supergravity  \cite{deWit:1981eq,deWit:1982ig}.  This gauging can be viewed as a supersymmetry preserving deformation  of the un-gauged theory and, although the global $E_{7(7)}$ of the theory is explicitly broken, the local $SU(8)$ invariance in preserved.   

In general we can consider the process of gauging wherein some subgroup of the global symmetries of a supergravity theory are promoted to gauge symmetries.  In the context of the maximally supersymmetric theories obtained by toriodal reduction of eleven-dimensional supergravity to $D=11-d$ dimensions one may consider   some subgroup of the $G=E_{d(d)}$ symmetry and promote it to a local symmetry in a consistent way preserving all the supersymmetries. 

Importantly, although a given choice of gauging will explicitly break the global $E_{d(d)}$ symmetry, the structure of admissible gaugings are controlled by this global symmetry.  This lies at the heart of the modern ``embedding   tensor'' approach to gauged supergravities. This universal approach introduces a tensor that describes how the gauge generators are embedded into the global symmetry group (for a review see \cite{Samtleben:2008pe}).  Thought of as a \EDIT{spurionic} object, the embedding tensor provides a $E_{d,(d)}$ covariant description of the gauged supergravities. 

It has been argued for some time that the lower dimensional gauged supergravities hint at a higher dimensional origin beyond the standard eleven-dimensional supergravity:   in  \cite{Riccioni:2007ni,Riccioni:2007au}  Riccioni and West considered the application of $E_{11}$ techniques to gauged supergravities  and  the tensor hierarchy considerations of de Wit, Nicolai and Samtleben  \cite{deWit:2008ta} led to the suggestion that gauged supergravities probe M-theoretic degrees of freedom not captured by the conventional supergravity.   
 
 Indeed, we saw in the preceding chapter there was a powerful relationship between the Scherk-Schwarz reduction of the $O(n,n)$ symmetric Double Field Theory and lower dimensional electric gaugings of half-maximal supergravity. The reduction of Double Field Theory permits a higher dimensional origin to gauged supergravities in which so-called  non-geometric fluxes are rendered purely geometric in the extended spacetime.    Here we will outline the same mechanism in the context of the U-duality invariant M-theory. The study of  compactifications of M-theory on twisted tori and the relation with gauge supergravity was also considered in e.g  \cite{DallAgata:2005ff,DallAgata:2005mj,DallAgata:2007fk,DAuria:2005dd,DAuria:2005er,DAuria:2005rv,Hull:2006tp}. Here we shall see how the embedding tensor and local gauge symmetries arise naturally from the generalised Lie derivative of the higher theory and that the dimensional reduction of the action reproduces expected features such as a non-trivial scalar potential. 
 
 Beyond this very elegant description of the gauged supergravities, a key feature of this work is that it represents a consistent relaxation of the section condition.  This is of deep significance; it represents a first step in which the full geometry of the extended spacetime plays an active role and explores the structure of the duality invariant theory beyond   conventional supergravity. 
 
 To make this section pedagogical we shall provide next a short pr\'ecis of the embedding tensor approach to gauged supergravities (see \cite{Samtleben:2008pe} for a more comprehensive treatment and references within) before turning to the Scherk-Schwarz reduction which we shall describe in some detail in the case of $SL(5)$ and then outline the extension to higher rank exceptional groups.

  \subsubsection{The Embedding Tensor and Gauged Supergravity}
\def\Q{\Theta} 
\def\q{\theta} 
\def\mb{\mathbf}
 
 We consider the maximal supergravity theory in  $D=11-d$ with the global symmetry group $ E_{d(d)}$.  In addition, this global symmetry can be complemented with the action of   $\mathbb{R}^+$ which is an on-shell conformal rescaling symmetry  for $D\neq 2$     known as the trombone symmetry.  Thus we shall consider gaugings in which some subgroup $H \subset G = E_{d(d)}  \times \mathbb{R}^+$ is promoted to a gauge symmetry.   Let us denote by $\frak{t}_\alpha$ the generators of $\frak{g}$, the algebra of $G$, and let $M= 1\dots n_v$ label the $n_v$ vector fields of the ungauged theory.   The embedding tensor, $\Theta$,  describes the embedding of $H$ into $G$ and singles out the gauge generators according to 
 \be
 X_{M} = \Theta_{M}{}^\a \frak{t}_\a \ . 
 \ee
 The number of gauge generators is determined by the rank of this embedding tensor and covariant derivatives are obtained by
 \be
 D_\mu = \partial_\mu + g A^M_\mu X_{M} \ . 
  \ee 
  A given choice of $\Theta$ will single out a particular gauged theory and thus break the global symmetry, but we can work without specifying a choice for $\Theta$ and  thereby retaining $G$ as \EDIT{spurionic} symmetry. In addition to defining the gauged covariant derivatives the embedding tensor also determines in a universal way the form of the scalar potential required by supersymmetry. 

There are two vital consistency conditions for the theory.  The first is a closure condition which arises from imposing that the embedding tensor is invariant  under the gauge symmetry.  This reads
  \be
  [X_M, X_M] = X_{MN}{}^P X_P \ , 
  \ee
  in which we have introduced ``structure constants'' which are the embedding tensor in the adjoint $X_{MN}{}^P = \Theta_M^\a (\frak{t}_{\a})_N{}^P$.  It is important to note that this is not a standard closure condition; $X_{MN}{}^P$ need not be   antisymmetric in its lower indices  (and it generally isn't). It is only the projection of $X_{MN}^PX_P$ into the gauge generators that must be anti-symmetric.  The symmetric part of $X_{MN}{}^P$ gives rise to a non-standard Jacobi identity (the Jacobiator does not vanish but again will do so upon contraction into a gauge generator).   The second constraint is a linear one that ensures supersymmetry; in      general  it is given by the vanishing of a certain projector of the embedding tensor.  These results are summarised in table \ref{table:embedding}  
  
  \begin{table}[htdp]\label{table:embedding}
\caption{The allowed representations of the embedding tensor for selected dimensions; the first column of the allowed representations correspond  to trombone gaugings and the final column denotes representations projected out by the supersymmetry constraint.  }
\begin{center}
\begin{tabular}{|c|c|c|c|c|}
\hline
  $d$ & $G$ &  $\Theta$ &  Allowed & Projected \\
\hline
  4 &  $SL(5)$ &   $\bf{10\otimes 24}$    & $\bf{{10} \oplus  15 \oplus \overline{40}}$ &  \bf{175}  \\
  5 & $Spin(5,5)$ &  $\bf{16_s\otimes 45}$     &   $\bf{{16_s} \oplus  144_c }$  &   $\bf{560_s}$   \\
  6 & $E_{6,(6)}$ &  $\bf{27 \otimes 78}$   &$ \bf{{27} \oplus  351} $   &  $\bf{1728}$  \\
  7 & $E_{7,(7)} $&   $\bf{56\otimes 133}$ &   $ \bf{{56} \oplus  912} $    &      $\bf{6480}$    \\  \hline
\end{tabular}
\end{center} 
\end{table}
  
 For instance, in the case of  $D=7$ \cite{Samtleben:2005bp} the vectors of the un-gauged theory lie in the $\bf{10}$, two-forms in the $\bf{5}$ and the scalars parametrise an $SL(5)/SO(5)$ coset.     The gaugings are specified by the embedding tensor $ \Theta_{mn , p }{}^q$ with generators of the gauge group given by  $X_{mn}=\Theta_{mn, b}{}^a T^b_a$.    {\it{A priori}}, $ {\Theta}_{mn , p }{}^q$, lies in the product $\mathbf{10\otimes24}$ which decomposes as 
\begin{equation}
 \Theta_{mn, b}{}^a\in\mathbf{10\otimes24}=\mathbf{10\oplus15\oplus\overline{40}\oplus175}.
\end{equation}
The linear supersymmetry constraint is as projection $\Theta | _{\bf{175}} = 0$ and the surviving components of the embedding tensor may be written as  
\begin{equation}\label{eq:sl5embeddingtensor}
\begin{split}
& \Theta_{mn, b}{}^a=\delta^a_{[m}Y_{n]b}- 2 \e_{mnrs b}Z^{rs,a}- \frac{5}{3} \theta_{kl}(T^a_b)^{kl}_{mn}  +\theta_{mn}\delta^a_b,\\
 \end{split}
\end{equation}
where  $\theta, Y$ and $Z$ are in the $\bf{10, 15}$ and $\bf{\overline{40}}$ respectively and $(T^a_b)^{kl}_{mn}$ are the $SL(5)$ generators in the antisymmetric representation.
  The  components $\theta_{mn}$   in the $\bf{10}$ are those associated with trombone gaugings.   The quadratic closure constraint requires that 
\beq
Z^{m n , p} X_{m n} = 0 \ . 
\eeq        
The scalar potential, at least in the absence of trombone gaugings, is given by
 \beq \label{eq: Vgauged}
V_{gauged} = \frac{1}{64} \left( 2{m}^{a b}  Y_{b c} {m}^{c d} Y_{d a} -   ( {m}^{a b} Y_{a b})^2   \right) + Z^{a b, c} Z^{d e, f} \left( {m}_{a d}  {m}_{b e}  {m}_{c f} - {m}_{a d}  {m}_{b c}  {m}_{e f}     \right) \ . 
\eeq
where $m_{mn}$ is the paramterisation of the scalar coset in the fundamental representation (it may be related to the ${\cal M}_{MN}$ that has been encounter already as ${\cal M}_{MN} = {\cal M}_{mn,pq} = m_{mp}m_{nq} - m_{mq}m_{np}$).     
  
  \subsubsection{Scherk-Schwarz Reduction of the $SL(5)$ U-duality Invariant Theory}
 \def\SSA{Scherk-Schwarz }
We now consider the \SSA reduction ansatz in the $SL(5)$   U-duality invariant theory and its comparison to the seven-dimensional gauged supergravities introduced above \cite{Berman:2012uy}.   

\EDIT{In line with the discussion in section 5.3, we would ideally start with conventional eleven-dimensional spacetime split into the product of some seven-dimensional manifold $M_7$ (with coordinates $x_{(7)}$) and four-dimensional $M_4$ where the U-duality acts.    The later would be the internal manifold of a conventional reduction but in the U-duality invariant framework it is replaced with a ten-dimensional extended space, with coordinates $\X^M$.  We would then be able to define  a seventeen-dimensional theory in which $SL(5)$ acts linearly and is a manifest symmetry.    We then would perform a \SSA reduction in the ten-dimensional extended space  to arrive at a conventional gauged supergravity in seven dimensions. }

\EDIT{
However, the present technical limitations mean that the $SL(5)$ invariant theory we are considering is defined in time plus a ten-dimensional extended space (we repeat that we see no reason why this restriction could not be relaxed in due course).     The result of performing the  \SSA reduction in the ten-dimensional space will be  a one-dimensional action.  Thus we do not directly reproduce all the features, and indeed field content, of gauged supergravity in this way. In particular we don't have the gauge fields which would arise from off-diagonal terms i.e. objects with internal/external mixed  indices.  Although we will not be able to obtain the gauge sector of the theory we will however recover the scalar potential of the theory given in  eq.~\eqref{eq: Vgauged}.      Moreover we will obtain directly expressions for the embedding tensor in terms of the Scherk-Schwarz twist.  }

 The reduction ansatz is that all $\X^M$ dependence of fields and gauge parameters is encapsulated by a $SL(5)$ ``twist'' matrix: 
\be\label{eq:Scherk-Schwarz} 
Q^M(x_{(7)}, \X) = W^M{}_{\bar{A}}(\X)  Q^{\bar{A}}(x_{(7)} )  \, .
\ee
It will be useful in what follows to note that the twist  in the ${\bf 10}$ can be written in terms of a matrix in the fundamental according to $ W^M{}_{\bar{A}} = W^{mn}{}_{\bar a\bar b} =V^{a}{}_{[\bar a} V^b{}_{\bar b]} $.  For clarity we shall place bars over indices on quantities that only depend on $x_{(7)}$ \EDIT{(here we use $x_{(7)}$ to  denote the external coordinates even though, in the prototype M-theory we consider, these consist only of the time direction).} 

Let us first consider the gauge symmetries of the dimensionally reduced theory.  These are obtained by substitution of  the ansatz eq.~\eqref{eq:Scherk-Schwarz} in to the generalised Lie derivative introduced in eq.~\eqref{eq:genderiv}.  The result is that
\be\label{eq:gaugedgenderiv}
({\cal L}_U V)^M = W^M{}_{\bar{A}} \left( ({\cal L}_{\bar{U}} \bar{V})^{\bar{A}}  +  X_{\bar{B} \bar{C} }{}^{\bar{A}} U^{\bar{B}} V^{\bar{C}} \right) 
\ee
where the first term in parenethisis is just the derivative acting on the $x_{(7)}$ dependent part of the field $V$ and the second term encodes the effect of introducing the \SSA twist and indicates the presence of a gauge symmetry.   The would be structure constants have the form
\be\label{eq:structconsts}
 X_{\bar{B} \bar{C} }{}^{\bar{A}} =  X_{\bar{c}\bar{d},\bar{e}\bar{f}}{}^{\bar{a}\bar{b}} = \frac{1}{2} W^{\bar a \bar b}{}_{mn} W^{pq}{}_{\bar{c}\bar{d}} \partial_{pq} W^{mn}{}_{\bar{e}\bar{f}} + \frac{1}{2}\delta^{\bar{a}\bar{b}}_{\bar{e}\bar{f}} \partial_{mn} W^{mn}{}_{\bar c\bar d} +2 W^{\bar a \bar b}{}_{mn } W^{mp}{}_{\bar{e}\bar{f}} \partial_{pq} W^{qn}{}_{\bar c \bar d}    \ . 
\ee
One can see from this that these structure constants are manifestly not anti-symmetric in their lower indices.  We find immediately a first constraint on the form of the \SSA twist: it should be choosen such that the $ X_{\bar{B} \bar{C} }{}^{\bar{A}} $ are indeed constants.

At this stage we can make direct contact with the embedding tensor described in  eq.~\eqref{eq:sl5embeddingtensor} and extract the components in the $\mathbf{10\oplus15\oplus\overline{40}}$ as 
\be
\begin{aligned}
\bf{10}: & \qquad \theta_{\bar{c}\bar{d}} = \frac{1}{10} \left( V^{\bar m}_n \partial_{\bar c \bar d} V^n_{\bar m} - V^{\bar m}_n \partial_{\bar m [ \bar c} V_{\bar{d}]}^m  \right) \ ,  \\
\bf{15}:  & \qquad  Y_{\bar{c}\bar{d}} =  V^{\bar m}_n \partial_{\bar m ( \bar c} V_{\bar{d})}^m  \ ,  \\
\overline{\bf{40}}: &\qquad Z^{\bar{m}\bar{n} , \bar{p} } = -\frac{1}{24}\left( \e^{\bar m \bar n \bar i \bar j \bar k} V_t^{\bar{p}} \partial_{\bar i \bar j}  V^t_{\bar{k}}  
+ V_t^{[\bar m} \partial_{\bar i \bar j} V^{|t|}_{\bar k}\e^{ \bar n ] \bar i \bar j \bar k \bar p }     \right) \ . 
\end{aligned}
\ee
Something rather remarkable about this should be emphasised: we started with a generalised Lie derivative in an extended spacetime whose form was determined by closure and compatibility with U-duality, no extra assumptions about supersymmetry were made.  Here   we find that not only does this Lie derivative give rise to the gauge symmetries associated with the lower dimensional gauged supergravity, only the representations allowed by supersymmetry are present (there is   no component in the $\bf 175$ entering in eq.~\eqref{eq:gaugedgenderiv}).  We may thus conclude that  the U-duality generalised derivative already secretly ``knows'' about supersymmetry.

A second constraint follows from considering the closure conditions.  We recall that prior to twisting the generalised Lie derivative closed on the Courant bracket up to the section condition, i.e. we have a closure rule of the form
\begin{equation}
{\cal L}_{[U_1, U_2]_C} V^M -  [ {\cal L}_{U_1},{\cal L}_{U_2 }]V^{M } = {\cal  F}^{M} \ ,
\end{equation}
where the anomalous term  ${\cal  F}^{M}$ vanished on the section condition.  One way to determine the correct closure constraints in the case at hand would be to explicitly evaluate ${\cal  F}^{M}$ with the gauge parameters obeying Scherk--Schwarz ansatz.   Instead we may simply evaluate the left-hand side of the above relation and use the form of eq.~\eqref{eq:Scherk-Schwarz}. Suppressing all dependance on the coordinates of $M_7$ one obtains the invariance condition
\be
[ X_{\bar{A}}, X_{\bar{B}} ]  = - X_{[\bar{A}\bar{B}]}^{\bar C} X_{\bar C} \ .
\ee
Moreover, from demanding that $X_{\bar{A}\bar{B}}{}^{\bar C}$ be not just constant but an invariant object under the local symmetry transformations one recovers the constraint
\be 
X_{(\bar{A}\bar{B})}^{\bar C} X_{\bar C}  = 0 \ . 
\ee
Let us emphasise one important point here: whilst  it is evident that invoking the section condition is {\em sufficient} to satisfy these constraints it not a {\em  necessary} condition.  Indeed a \SSA reduction permits explicit, albeit constrained, dependance on the extra coordinates of the extended spacetime.

We can now turn to the action defined in  eq.~\eqref{TBP} and eq.~\eqref{VBP}.  \EDIT{ As discussed above, since the U-duality invariant formalism truncates the external manifold to just the time direction,  we don't fully capture all the field content of the gauged supergravity. However we can obtain the scalar potential of the theory and it is this which we shall now outline.   }

We recall the   potential of eq.~\eqref{VBP} was given by
\be
\frac{1}{\sqrt{\det{g}} } V = V_1  + V_2 + V_3 + V_4 
\ee
with 
  \ba
 V_1 =  \frac{1}{12} \cM^{MN} (\partial_{M} \cM^{KL}) (\partial_{ {N}}\cM_{ {K} {L}}) \ ,  &\quad& 
V_2 = - \frac{1}{ 2} \cM^{ {M} {N}} (\partial_{ {N}} \cM^{ {K} {L}}) (\partial_{ {L}} \cM_{ {K} {M}}) \ ,   \\
 V_3 =  \frac{1}{12} \cM^{ {M} {N}} ( \cM^{ {K} {L}} \partial_{ {M}} \cM_{ {K} {L}})( \cM^{ {R} {S}} \partial_{ {N}} \cM_{ {R} {S}})\ , &\quad&
V_4 =  \frac{1}{4} \cM^{ {M} {N}}  \cM^{ {P} {Q}}  ( \cM^{ {K} {L}} \partial_{ {P}} \cM_{ {K} {L}})(   \partial_{ {M}} \cM_{ {N}  {Q}})\ . \nonumber
\ea
In addition to this we may include, with impunity,  any extra terms which vanish when the section condition is invoked.  One such term is given by  
\beq\label{eq:V5}
V_{5} = \epsilon^{a MN} \epsilon_{a PQ}  {\cal E}^A_{R} {\cal M}^{RS} {\cal  E}^B_{S}  \partial_{M} {\cal E}^{P}_A \partial_N {\cal E}^Q_B \ , 
\eeq
where ${\cal E}^A_M$ is a vielbein for the generalised metric ${\cal M}_{MN} = {\cal E}^{A}_M \delta_{AB} {\cal E}^{B}_N$.   This extra term is essential in order to make the linkage with  gauged supergravity One can now insert the \SSA ansatz into these terms and  compare with the literature on seven dimensional gauged supergravity after rewriting the results in terms of the scalar coset representative $m_{mn}$  in the $\bf{5}$ rather than ${\cal M}_{MN}$ in the $\bf{10}$.  If an extra assumption is made that there is no trombone gauging then the contributions from $V_3$ and $V_4$ drop out entirely and one finds that 
\be
V_1 + V_2  - \frac{1}{8} V_5 = V_{gauged} + 2 \partial_{kl} \left(m^{pk} m^{\bar q \bar l} V_{\bar q}^{[q}\partial_{pq} V_{\bar l}^{l]}  \right) \ . 
\ee 
Thus up to a total derivative term that is dropped one does indeed recover the potential for the scalar fields required by gauged supergravity that was defined in eq.~\eqref{eq: Vgauged}.

\subsubsection{Comments on Higher Dimensions}
The $Y$ tensors introduced in section \ref{sec:gaugestruct} govern the universal form   for the generalised Lie-derivative, \be 
 {\cal L}_U V^M =  L_U V^M + Y^{MN}{}_{PQ} \partial_N U^P V^Q  \ . 
\ee
Inserting the \SSA ansatz into the generalised Lie derivative gives rise to the gauge symmetries of the gauged supergravity and, as described for the case $SL(5)$ above, gives an expression for the would-be structure constants (or rather gaugings in the adjoint):
\be\label{eq:embeddingtensor}
X_{ \bar{A}\bar{B} }{}^{\bar C}  = 2 W^{\bar C}{}_M \partial_{[\bar A}  W^M{}_{\bar B]}  +  Y^{\bar C \bar D}{}_{\bar E \bar B} W^{\bar E}{}_M \partial_{\bar D} W^M{}_{\bar A}   \ . 
 \ee 
These are demanded to be constant and gauge invariant (giving rise to the correct closure constraints of gauged supergravity).   One can see that these are not anti-symmetric in the lower indices since in general the final term has a symmetric part.  For cases $n\leq 6$ the relevant tensor for $E_{n(n)}$ can be written schematically as $Y^{MN}{}_{PQ} = d^{a MN} d_{a  PQ}$ where $d$ is an invariant tensor and $a$ is an index in the fundamental of $E_{n}$.  Then a short manipulation shows that 
\be
X_{ (\bar{A}\bar{B}) }{}^{\bar C}  \propto d^{\bar a \bar C \bar D} d_{\bar b  \bar B \bar A} V^{\bar b}_m \partial_{\bar D} V^{m}_{\bar a}  \ .
\ee 
As with the $SL(5)$ case, one finds that only the representations allowed by supersymmetry enter in the above expression for the gaugings.  One may extract the forms of the components of the embedding tensor directly from eq.~\eqref{eq:embeddingtensor}. For instance the trace part of the gauging gives the trombone components of the corresponding embedding tensor.   The exact forms have been extracted for the cases of $E_5$ and $E_6$ in \cite{Berman:2012uy,Musaev:2013rq} and for $E_{7}$ in \cite{Aldazabal:2013mya}.

One can also consider the reduction of the action.  In general one finds that the extra term, which vanishes on the section condition, that needs to be added to the action (c.f. eq.~\eqref{eq:V5}) is given by 
\be
V_5 = Y^{PQ }{}_{MN} \partial_P  {\cal E }^M_{\a}  \partial_Q {\cal E }^{N  \a}  \ , 
\ee
where $  {\cal E }^M_{\a}  $ is the appropriate vielbein  for the generalised metric.   Including this term it can be shown that the correct potential for gauged supergravity emerges upon \SSA reduction.  Further work developing the related geometrical concepts can be found in  \cite{Aldazabal:2013mya,Park:2013gaj,Cederwall:2013naa}.

\section{Conclusions and Discussion}\label{sec:Concs}

This review is of a subject in development. Despite the initial objections to the programme due to the absence of a hierarchy of energy scales, there have been numerous successes. Perhaps most notably is the interpretation of gauged supergravities in terms of Scherk-Schwarz reductions of the extended theory. This then provides the M-theory perspective on gauged supergravities as arising from a single parent theory albeit one with additional novel extra dimensions. 

The doubled $O(d,d)$ case has also proven to be formally much more successful than one might have imagined. The quantum consistency of the theory, something that the faith in string theory is built on, persists for the doubled string. The central charge counting, vanishing of the one-loop $\beta$-function to give the background field equations and the modular invariance of the genus one world-sheet partition function are all very nontrivial checks that the double string passes at the quantum level. What is interesting is that these checks work without the use of the section condition in the background so that they use the whole doubled space (and in the case of the $\beta$-function its doubled geometry). Thus, the world-sheet quantum theory sees the total doubled space \EDIT{albeit} in a very particular way through chiral bosons on its world-sheet.

This theory is the natural environment for the so called non-geometric backgrounds. By allowing nontrivial topology in the extended theory we can produce backgrounds that would be forbidden in usual geometry. These backgrounds emerge in so called exotic branes and in numerous other contexts and may even prove to be interesting phenomenologically. A detailed study of the relevant topological quantities for the extended geometry is not yet complete. Mathematically it would be related to the obstruction theory of the Lie Algebroid associated with the generalised geometry. One would imagine that such obstructions- which in the usual case of a tangent bundle give the obstructions to picking a global section- would provide the {\it{charges}} associated to nongeometries. Related to this, Daniel Waldram has emphasised that it appears the presence of the $O(D,D)$ or exceptional group structure requires the space to be generally parallelisable. Then if the space does not obey the section condition this requirement forces us into the Scherk-Schwarz case. Some preliminary discussion of global aspects is made in \cite{Berman:2014jba}.

The developments for the future will be split between using the formalism that has been developed to provide new perspectives and the further development and understanding of the formalism itself. As an example of the former, one may seek particular solutions to the equations of motion of the extended space and examine their interpretation from the usual spacetime perspective. As an example of the later, the notion of geometry for these theories is something of interest. The torsion formalism provides a very different perspective than we are used to in normal general relativity.  The absence of a description of the higher derivative corrections to string theory- ie. curvature squared terms, makes one feel that something is missing from the formalism itself.

Ultimately, the suspicion is that the theory is tied deeply to the presence of supersymmetry as seen by the physical section condition being identical to the condition for 1/2 BPS states in the theory \cite{Obers:1998fb,West:2012qm,bermanperrynew}. In the extended space all these 1/2 BPS states are massless or tensionless objects \cite{bermanrudolph1,bermanrudolph2}. All the charges and effective masses/tensions arise through the dimensional reduction of the extended theory. This means the extended theory has no real scale (as seen by the $R^+$ trombone symmetry). The implications of this are unclear  but one might hope that some of the physics of string theory beyond the Hagedorn transition might be captured. It was noted in the original work on the Hagedorn transition by Atick and Witten \cite{Atick:1988si} that string theory at temperatures beyond the Hagedorn scale has thermodynamic properties in common with a theory of two times and that the physics beyond Hagedorn should be a spacetime scale invariant theory. Currently, this perspective remains uninvestigated in Double Field Theory.

The role of states that preserve less supersymmetry such as 1/4 BPS states, is unclear. Looking at the simple case of Double Field Theory provides an interesting perspective. The full level matching condition in doubled space is:
\beq  P^I P_I = \alpha' (N_R-N_L) \, .  \label{levmatch} \eeq 
So that for the zero mode sector where $N_R=1, \, N_L=1$  this equation becomes the section condition of the doubled theory which indeed is the doubled space light-cone condition $P^2=0$. Thus the section condition is a cone in doubled momentum space. The reader might be forgiven for thinking that this equation (\ref{levmatch}) is the mass-shell condition, since it relates momentum squared to oscillator number. Indeed in the doubled space the level matching condition takes on exactly the form of  the  mass-shell condition in the doubled space. As soon as we allow other oscillator numbers so that $N_R-N_L \neq 0$ then these states do not lie on the doubled space light-cone. In fact summing over all allowed oscillators will allow us foliate the whole of doubled momentum space. Of course the oscillators are quantised which means the doubled momentum space inherits this quantisation. This may have important implications for the quantisation of Double Field Theory. In the U-duality extended theory a similar structure exists with the 1/2 BPS states again forming a cone in the extended momentum space \cite{Gauntlett:2000ch} and the states with lower supersymmetry filling out the cone and foliating the entire extended momentum space. The  quantisation now comes not from the string quantisation of oscillators but simply from charge quantisation.

Summing over all generalised momenta (perhaps as one would do in a partition function or in a loop diagram) will give a U-duality invariant quantity. This allows the sort of U-duality invariant sums that have appeared in the higher derivative corrections to string theory \cite{Green:2011vz} or in black hole entropies  to become {\it{geometric}} in an extended generalised geometry sort of way. In normal M-theory the $SL(2,\mathbb{Z})$ of IIB comes from the compactification of eleven-dimensional supergravity on a 2-torus. The result of this is that the IIB higher derivative terms come with a piece that has its M-theory origins in a momentum sum around the torus \cite{Green:1997as}. This idea can then be extended to the exceptional groups of U-duality with the derivative corrections in lower dimensions being the generalisation of modular forms to the exceptional groups. These may then be thought of as generalised momentum sums in the extended space. The decision of whether one should include the section condition then is determined by the supersymmetry of the quantity in question. If the quantity should be 1/2 superysmmetric then the section condition must be applied to the momenta in the sum, if it is not then the section condition should not be applied.

There are many more questions that one could consider such as: the full implications for the open string sector; the U-duality geometry if there is such a thing beyond $E_8$; and for $E_8$, the emergence of the dual graviton and its properties. There are so many research directions still open that one expects further new perspectives on string and M-theory will emerge from the formalism in the years to come.

\section{Acknowledgements} 
 
DSB is partially supported by STFC consolidated grant ST/J000469/1. DSB is
also grateful for DAMTP in Cambridge for continuous hospitality.
DCT would like to thank the organisers of the 2012 Modave summer school where he presented a set of lectures that led to the idea of this review.
DCT is supported in part by the Belgian Federal Science Policy Office through the Interuniversity Attraction
Pole P7/37, and in part by the FWO-Vlaanderen through the project G.0114.10N and
by the Vrije Universiteit Brussel through the Strategic Research Program ``High-Energy
Physics'' and  by an FWO postdoc fellowship. We would like to thank foremost Malcolm Perry for extensive collaboration and discussions on most  of the ideas presented here. We have also benefitted from working and having discussions with: Chris Blair, Ralph Blumenhagen, Martin Cederwall, Neil Copland, Gary Gibbons, Hadi and Mahdi Godazgar, Chris Hull, Georgios Itsios, Axel Kleinschmidt, Kanghoon Lee, Emmanuel Malek, Edvard Musaev, Carlos Nunez, Jeong-Hyuck Park, Boris Pioline, Felix Rudolph, Kostas Sfetsos,  Kostas Siampos, Alexander Sevrin, Daniel Waldram and Peter West.

\section{Appendix:  A Chiral Boson Toolbox}
 
 Chiral $p$-form fields may be defined in $D=2(p+1)$ dimensions as potentials with self-dual $(p+1)$-form field strengths.  They are ubiquitous in supergravity and string theory; notable examples being the self-dual Ramond-Ramond five-form in the type IIB theory and the self-dual three-form in the $(2,0)$ tensor multiplet on the world-volume of an $M5$ brane.  Here we will be most interested in the two-dimensional chiral boson which is essential to the duality symmetric doubled formalism.   In this case a chiral boson is simply one that obeys $\partial_- \phi = 0$.   We present a summary here but further details may be found in \cite{Sevrin:2013nca}.  We first review the classical situation and then provide some short discussion of the quantum situation. 

A particular challenge with these chiral fields is to incorporate the self-duality condition at the level of an action whilst at the same time preserving manifest Lorentz invariance.   The difficulties arise primarily because self-duality is a first order differential condition and gives rise to second class constraints.  In the approach of Siegel \cite{Siegel:1983es} a Lagrange multiplier is used to invoke the square of the constraint with the action
\be
S = \int d^2 \s \partial_+ \phi \partial_- \phi  - h_{ +\,+} \partial_- \phi \partial_- \phi \ . 
\ee 
The equations of motion    imply $\partial_- \phi = 0$ and since $h$ drops out of the equations of motion it is a gauge degree of freedom. Indeed this action has a gauge symmetry 
\be\label{siegelgauge}
\delta  h_{ +\,+}  = \partial_+ \epsilon^{-} + \epsilon^{-}  \partial_-  h_{ +\,+}  - h_{ +\,+}  \partial_- \epsilon^{-}   \ , \qquad  \delta \phi =\epsilon^{-}  \partial_- \phi  \ ,
\ee
 which shows that the self-dual part of $\phi$ may be gauged away.  
 A helpful way to view this system is actually as a boson in curved space coupled to gravity in the lightcone gauge, that is to say the Lagrange multiplier $h$ is a component of metric 
 \be
 ds^2 = d\sigma^+ d\sigma^-  +  h_{ +\,+}  d\sigma^+{}^2 \ , 
 \ee
 and the gauge symmetry is simply inherited from the diffeomorphisms.


A Lorentz invariant formulation for chiral $p$-forms was developed by Pasti, Sorokin and Tonin \cite{Pasti:1996vs}.  The essence of this approach is to introduce auxiliary fields that furnish the theory with a gauge invariance.  This  invariance serves to render the anti-self-dual part of the field strength pure gauge. In fact in two dimensions the upshot of this approach is simply to take the Siegel formalism and make use of a Beltrami parametrisation $h_{++} \partial_- a = \partial_+ a$ such that 
\be
\label{PST}
L_{PST} =  \partial_+ \phi \partial_- \phi  - \frac{\partial_+ a}{\partial_- a}  \partial_- \phi \partial_- \phi \ . 
\ee
with the PST gauge symmetry
\be
\delta a = \zeta \, , \quad \delta \phi = \frac{ \partial_- \phi}{\partial_- a} \zeta \, , 
\ee
which descends from the above diffeomorphism symmetry choosing $\zeta = \epsilon \partial_- a $.  More generally one can consider the action  
\be
L_{PST} = \partial_+ \phi \partial_- \phi - \frac{1}{  u_\a u^\a} ( u^\b \cP_\b )^2
\ee
where $u$ is a closed one-form and $\cP = 0$ is the desired constraint.  Locally $u = d a$ the action reduces to eq. (\ref{PST}). 

Finally let us remark that upon gauge fixing either the Siegel approach $h=1$ or the PST approach $\partial_+a = \partial_-a$ or $u_\alpha = \delta^0_\alpha$ one arrives at the Floreanini--Jackiw Lagrangian  \cite{Floreanini:1987as}
\be
\label{FJ}
L_{FJ} =     \partial_- \phi \partial_1 \phi\  . 
\ee
The action corresponding to (\ref{FJ}) has both time translation and spatial translational symmetries with Noether charges related as $H=-P$ which is to be expected for chiral dynamics. The gauge symmetry
\be
\delta_\alpha \phi = \alpha(t) 
\ee
means that the equations of motion that follow from (\ref{FJ}), namely 
\be
\partial_1 \partial_- \phi = 0 
\ee
may be integrated to give solutions which gauge equivalent to the desired chirality constraint $\partial_- \phi = 0$.

The quantisation of 2d chiral bosons is a subtle question about which the historical literature is rather ambiguous and often conflicting.   
The most clear cut case is the Floreanini-Jackiw formalism \cite{Floreanini:1987as}.  Here quantisation can be readily performed, there are no gauge symmetries so nothing needs to be fixed.  This is a first class system with a second class constraint so the passage to the quantum theory can be achieved by means of Dirac brackets. As detailed in \cite{Sonnenschein:1988ug} one finds that the single particle Hilbert space correctly implements the chirality constraint and consists of a continuum of states with $E=k$ and with only positive (i.e. right moving)  $k\geq 0$ excitations.   It is in this formalism that some quantum questions of the doubled formalism have been addressed \cite{HackettJones:2006bp,Berman:2007xn}. 

The case of the Siegel formulation is rather more divisive.   One observation is that either by using Hamiltonian reduction as described in \cite{Faddeev:1988qp} or Dirac procedure as in \cite{Bernstein:1988zd} the Siegel formalism reduces to the Floreanini-Jackiw formalism and quantisation can proceed accordingly.  An alternative approach would be to treat the Siegel gauge symmetry directly using Faddeev-Popov \cite{Imbimbo:1987yt} and BRST methods \cite{Labastida:1987zy}. Doing so finds an immediate problem; that Siegel symmetry is anomalous    
   with an anomaly given by
 \begin{eqnarray}
 \label{anom}
 \delta \Gamma[h_{+\,+}\,] =
 - \frac{c}{12 \pi}  \int d^2 \sigma  \epsilon^{-} \partial_-^3 h_{+\,+}\,   \, .
 \end{eqnarray}
 To counter  this anomaly and to restore the  symmetry at a quantum level, the authors of  \cite{Imbimbo:1987yt}  propose adding a Wess-Zumino term to the action
\begin{eqnarray}
S_{WZ}= \kappa \int d^2\sigma   h_{+\,+}\, \partial^2_- \phi \, . 
\end{eqnarray}  
This extra term explicitly violates the classical gauge symmetry however,   $\kappa$ may be appropriately tuned (and including ghost loops where required),  such that this contribution cancels with that of  eq.~(\ref{anom}) leaving a quantum mechanically gauge invariant system.   Unfortunately this approach has difficulties when then coupled to world sheet gravity.  It was further argued   \cite{Henneaux:1987hz} that such an approach is fundamentally inapplicable because the first class constraint associated to the gauge symmetry of the Siegel action  is the square of a second class constraint.  
 
The quantum aspects of the PST formulation are less discussed, though given the wide use of the formalism rather important to understand.  In \cite{Pasti:1996vs} some classical arguments are presented as to why  the system may be well behaved quantum mechanically.  Firstly  it is observed that  first class constraint for the PST symmetry is {\em not} the square of a second class one and the objection of  \cite{Henneaux:1987hz}  can be avoided.  Since the algebra of the constraints behaves like a Virasoro algebra one can guess that its quantum commutator may acquire a central charge which may be interpreted as the presence of an anomaly in the gauge symmetry.  In  \cite{Pasti:1996vs} it is argued that by appropriately modifying the definition of the constraint, without spoiling the property that it is classically first class, one can engineer a cancellation of the central charge in the quantum constraint algebra.   This argument was made at a rather heuristic level in  \cite{Pasti:1996vs}   and a step to making it more precise was made by Lechner in \cite{Lechner:1998ga} where it was demonstrated explicitly that upon coupling to gravity the only extra contributions to the expected  gravitational anomaly of a chiral boson  \cite{AlvarezGaume:1983ig} were trivial (i.e. removable by local finite counter terms).   One might ask whether this can be made precise without the extra complication of coupling to gravity.  These comments not withstanding there remains an unresolved puzzle as discussed in \cite{Sevrin:2013nca}, since we can  pass from the Siegel formulation to the PST by a Beltrami parametrisation, the results of Polyakov  \cite{Polyakov:1987zb}  can be used to immediately determine the one-loop effective action which still has a non-vanishing variation under the PST symmetry.

\bibliographystyle{JHEP}
\bibliography{PhysReptBibDatabase}

\end{document}